\documentclass[fleqn,usenatbib]{mnras}

\usepackage[T1]{fontenc}
\usepackage{ae,aecompl}
\usepackage{xcolor}

\usepackage{graphicx}	% Including figure files
\usepackage{amsmath}	% Advanced maths commands
\usepackage{amssymb}	% Extra maths symbols

\usepackage{newtxtext,newtxmath}

\def\crate{\dot M_{\rm cool}}
\def\eagle{{\sc eagle}}
\def\surfs{{\sc surfs}}
\def\shark{{\sc Shark}}

\title[Quenching galaxies with \shark\ {\sc v2.0}]{Quenching massive galaxies across cosmic time with the semi-analytic model \shark\ {\sc v2.0}}

\author[C.~D.~P. Lagos et al.]{
\parbox[t]{\textwidth}{
\vspace{-0.5cm}
Claudia del P. Lagos$^{1,2,3}$\thanks{E-mail: claudia.lagos@icrar.org}, Mat\'ias Bravo$^4$, 
Rodrigo Tobar$^{1}$, Danail Obreschkow$^{1,2}$,  Chris Power$^{1,2}$, Aaron S. G. Robotham$^{1,2}$,
Katy L. Proctor$^{1,2}$, Samuel Hansen$^{1}$, \'Angel Chandro-G\'omez$^{1,2}$,
Julian Carrivick$^{1}$}
\vspace*{6pt} \\
$^{1}$International Centre for Radio Astronomy Research (ICRAR), M468, University of Western Australia, 35 Stirling Hwy, Crawley, \\WA 6009, Australia.\\
$^{2}$ARC Centre of Excellence for All Sky Astrophysics in 3 Dimensions (ASTRO 3D).\\
$^{3}$Cosmic Dawn Center (DAWN), Denmark.\\
$^{4}$Department of Physics \& Astronomy, McMaster University, 1280 Main Street W, Hamilton, ON, L8S 4M1, Canada.
\vspace*{-0.5cm}}

\date{Accepted XXX. Received YYY; in original form ZZZ}

\pubyear{2023}

\begin{document}
\label{firstpage}
\pagerange{\pageref{firstpage}--\pageref{lastpage}}
\maketitle

\begin{abstract}
We introduce version {\sc 2.0} of the \shark\ semi-analytic model of galaxy formation after many improvements to the physics included. The most significant being: (i) a model describing the exchange of angular momentum (AM) between the interstellar medium and stars; (ii) a new active galactic nuclei feedback model which has two modes, {a wind and a jet} mode, with the {jet} mode tied to the jet energy production; (iii) a model tracking the development of black hole (BH) spins; (iv) more sophisticated modelling of environmental effects on satellite galaxies; and (v) automatic parameter exploration using Particle Swarm Optimisation. We focus on two timely research topics: the structural properties of galaxies and the quenching of massive galaxies. For the former, \shark~v2.0 is capable of producing a more realistic stellar size-mass relation with a plateau marking the transition from disk- to bulge-dominated galaxies, and scaling relations between specific AM and mass that agree well with observations. For the quenching of massive galaxies, \shark~v2.0 produces massive galaxies that are more quenched than the previous version, reproducing well the observed relations between star formation rate (SFR) and stellar mass, and specific SFR and BH mass at $z=0$. \shark\ v2.0 produces a number density of massive-quiescent galaxies $>1$~dex higher than the previous version, in good agreement with JWST observations at $z\le 5$; predicts a stellar mass function of passive galaxies in reasonably good agreement with observations at $0.5<z<5$;  and environmental quenching to already be effective at $z=5$.
\end{abstract}


%%%%%%%%%%%%%%%%%%%%%%%%%%%%%%%%%%%%%%%%%%%%%%%%%%

%%%%%%%%%%%%%%%%% BODY OF PAPER %%%%%%%%%%%%%%%%%%

\section{Introduction}

Our current theory of galaxy formation and evolution is intimately linked to the growth of structures in the universe, which are thought to form hierarchically. The current preferred cosmological model is the $\Lambda$ cold dark matter ($\Lambda$CDM), in which the cosmic web evolves, for the most part, by the effect of gravity. Cosmological simulations of galaxy formation attempt to follow the formation of galaxies as the cosmic web forms from the very early to the local universe, providing a wide range of predictions across cosmic time, environment and galaxy populations (see reviews of \citealt{Somerville15,Vogelsberger20}).

Among the most popular tools to simulate the formation of galaxies in a cosmological context are hydrodynamical simulations and semi-analytic models (SAMs). Both have pros and cons. Hydrodynamical simulations have the main advantage of solving for the evolution of baryons and DM simultaneously, avoiding some key simplifications made in SAMs regarding the symmetry of galaxies and halos and the relevant baryonic components of each. The most important advantages of SAMs are their speed, ability to thoroughly explore the parameter space to understand potential degeneracies between physical models and parameters, and as a result, the very large cosmological boxes they can be run on, still covering a very large dynamic range in both halo and stellar mass of the produced galaxies \citep{Baugh06,Benson10}. Compared to the current generation of cosmological hydrodynamical simulations, SAMs can push to about two to three orders of magnitude below the stellar mass resolution of hydrodynamical simulations for the same cosmological volume. For this reason, SAMs remain an essential part of the toolkit in the quest of understanding galaxy formation and evolution. 

In \citet[hereafter L18]{Lagos18c}\defcitealias{Lagos18c}{L18}, we introduced the \shark\ (v1.1) SAM, an open source, flexible and highly modular SAM. \shark\ has been extensively used for many applications (some of which are listed in $\S$~\ref{sharkdesign}). One characteristic of both SAMs and hydrodynamical simulations is the continuous development of the physical models included to either introduce new physics into the models, make more physical assumptions, and/or improve the agreement with observations once tensions have been identified. \shark\ is no exception and since \citetalias{Lagos18c}, the model has seen continuous improvements on many of the physical models included, where we recognised too simplistic assumptions were made, and to address areas of tension that have been identified between the model and observations. In this paper, we introduce a new version of \shark\ (v2.0), after significant development of the physical models, which include, but are not limited to, more physical models tracking the properties of supermassive black holes (BHs) and active galactic nuclei (AGN) feedback, the angular momentum evolution of galaxy components, and environmental effects that affect the evolution of satellite galaxies. In this paper, we focus on two areas of tension that have been identified between \shark\ v1.1 and observations: a stellar mass-size relation that approximates a single power-law, while observations display a clear plateau in the size-mass relation associated with the transition from disk- to bulge-dominated galaxies \citep{Lange15}; and an overall scarcity of massive-quiescent galaxies at $z\gtrsim 2$ compared with observations. More details about these two areas of tension are presented in $\S$~\ref{sharkdesign}.

 The second tension above is particularly interesting as it has been reported across many cosmological simulations of galaxy formation \citep{Gould23,Valentino23}. This problem has worsened thanks to the exquisite observations of the James Webb Space Telescope (JWST). These have revealed that massive-quiescent galaxies 
 %overall scarcity of massive-quiescent galaxies in the early universe is a problem that has been reported  across many cosmological simulations of galaxy formation. The One of the most interesting ones relates to the existence of massive-quiescent galaxies in the early Universe, which thanks to the James Webb Space Telescope (JWST) have been recently characterised 
 are relatively common at $z>3$ (e.g. \citealt{Carnall23,Valentino23,Nanayakkara22,Long23}) and more so than previous observational inferences had indicated \citep{Carnall20,Weaver22,Gould23}. These galaxies typically have stellar masses in excess of $10^{10}\,\rm M_{\odot}$ and number densities $\gtrsim 10^{-5}\,\rm Mpc^{-3}$. Some of these galaxies have signs of having ceased their star formation (a.k.a. quenched) recently ($\approx 100$~Myr), while others are consistent with older stellar population ages ($\sim 1$~Gyr; \citealt{Glazebrook23}). The latter may imply these massive galaxies to have formed at $z>10$, potentially posing a problem to structure formation in $\Lambda$CDM \citep{Boylan-Kolchin23}. 

%show that current galaxy formation simulations, including \shark\ (as presented in \citetalias{Lagos18c}), tend to produce a number density of these massive-quiescent galaxies that is too low compared with current observational constraints. 
Although there are still many effects that could lead to systematic errors in the inferred stellar masses, redshifts and star formation rates (SFRs) of massive-quiescent galaxies in the observations, we focus here on how the predictions around this population changes from \shark\ v1.1 and v2.0 after the significant revisions of the model. We generalise the problem to understanding the quenching of massive galaxies across cosmic times. We leave for future work the critical assessment of systematic biases in the inferred properties of massive-quiescent galaxies.

%reducing the potential tension with current simulation predictions.

%Recent challenges for simulations based on observations.

%Aim of this paper.
This paper is organised as follows. $\S$~\ref{sharkdesign} briefly describes the \shark\ model and the $N$-body DM only simulations we use. $\S$~\ref{LargeModifications} describes in detail the significant modifications made to \shark\ in this new version 2.0 relative to v1.1, and the parameters we adopt for the default \shark\ v2.0 model. Minor modifications are presented in Appendix~\ref{AppendixMinorModifications}. $\S$~\ref{basicresults} presents key results on the abundance of galaxies across cosmic time and scaling relations, including structural relations in $\S$~\ref{structuresec}, in the local universe. We also present a supplementary material with additional comparisons with observations to show that previous areas of agreement between \shark\ and observations remain so. 
$\S$~\ref{basicresults2} focuses on the quenching of galaxies in \shark\ v2.0 and compares with v1.1 to reveal how much more efficient quenching is across cosmic time. The reader interested solely in the problem of massive-quiescent galaxies at $z\gtrsim 2$ can skip to $\S$~\ref{sec:quench2}, where we cover this in detail.
$\S$~\ref{conclusions} presents our main conclusions. 
 
\section{The semi-analytic model \shark}\label{sharkdesign}

\shark, hosted on GitHub\footnote{\href{https://github.com/ICRAR/shark}{\url{https://github.com/ICRAR/shark}}}, takes into account physical processes that we think are critical in shaping the formation and evolution of galaxies. These are (i) the collapse and merging of DM halos; (ii) the accretion of gas onto halos, which is modulated by the DM accretion rate; (iii) the shock heating and radiative cooling of gas inside DM halos, leading to the
formation of galactic disks via conservation of specific angular momentum of the cooling gas; (iv) star formation (SF) in galaxy disks; (v) stellar feedback from the evolving stellar populations; (vi) chemical enrichment of stars and gas; (vii) the growth via gas accretion and merging of BHs; (viii) heating by AGN; (ix) photoionization of the intergalactic medium; (x) galaxy mergers
driven by dynamical friction within common DM halos which can
trigger starbursts (SBs) and the formation and/or growth of spheroids; (x) collapse of globally unstable disks that also lead to SBs and the formation and/or growth of bulges. \citetalias{Lagos18c} included several different models for gas cooling, AGN feedback, stellar and photo-ionisation feedback, and star formation. 

The model presented in \citetalias{Lagos18c}, has been tested thoroughly across several publications and shown to reproduce several observed relations, including: 
the optical colour bimodality and how this depends on stellar mass \citep{Bravo20}; the atomic hydrogen 
(HI)-halo mass relation and HI clustering \citep{Chauhan20,Chauhan21}; the panchromatic emission of galaxies from the far-ultraviolet (FUV) to the far-infrared (FIR) across cosmic time \citep{Lagos19,Lagos20,Chen23}; the redshift distribution of bright FIR galaxies and the redshift evolution of their number density \citep{Lagos19,Casey21,Long22}, among many other successes. In addition, \shark\ has been used to make predictions for the gravitational wave signal from  binary stellar \citep{Rauf23} and supermassive \citep{Curylo22} black holes.

Some areas of tension with observations have also been identified, such as a weak downsizing signal and quenching timescales that are independent of stellar mass \citep{Bravo23}; a number density of passive galaxies that appears to be too low at $z\ge 3$ \citep{Long22,Gould23}; a stellar size-mass relation that is very close to a single power-law and massive-end of the stellar mass function (SMF) that is too shallow compared to observations (Proctor et al. in preparation). Proctor et al. (in preparation) crucially show that some of these limitations were not easily solved by modifying the parameters of the model and instead were inherent to the physical models included in \shark. 

The identification of these tensions in addition to the desire to continue to improve the model have led us to the continuing improvement and inclusion of new models within \shark. These new models are described in detail in $\S$~\ref{LargeModifications}. We have solidified these changes into a new version of \shark\ v2.0 available on GitHub. Before describing the new models, we introduce the suite of $N$-body simulations on top of which \shark\ runs in $\S$~\ref{surfs}.

\subsection{The \surfs\ simulations: halos catalogues and merger trees}\label{surfs}

%\begin{table}
%        \setlength\tabcolsep{2pt}
%        \centering\footnotesize
%        \caption{Simulation parameters of P-Millennium.}
%        \begin{tabular}{@{\extracolsep{\fill}}l|cc|p{0.45\textwidth}}
%                \hline
%                \hline
%            %Name & L210N1536\\
%    \hline
%    Box size [$\rm cMpc/h$] & 542.16 \\
%%    Number of particles &  $5040^3$\\
%    Particle Mass [$\rm M_{\odot}/h$]& $1.06\times 10^8$\\
%    %Softening Length [$\rm ckpc/h$] & 4.5\\
%    \hline
%        \end{tabular}
%        \label{simus}
%\end{table}

\begin{table}
        \setlength\tabcolsep{2pt}
        \centering\footnotesize
        \caption{Simulation parameters of the \surfs\ run used in this paper.}
        \begin{tabular}{@{\extracolsep{\fill}}l|cc|p{0.45\textwidth}}
                \hline
                \hline
            Name & L210N1536\\
           \hline
   Box size [$\rm cMpc/h$] & 210 \\
   Number of particles &  $1536^3$\\
   Particle Mass [$\rm M_{\odot}/h$]& $2.21\times10^8$\\
   Softening Length [$\rm ckpc/h$] & 4.5\\
   \hline
       \end{tabular}
       \label{simus}
\end{table}

%We run \shark\ over the P-Millennium $N$-body simulation, introduced in \citet{Baugh19}. The cosmological parameters correspond to a total matter, baryon and $\Lambda$ densities of $\Omega_{\rm m}=0.307$, $\Omega_{\rm b}=0.04825$ and $\Omega_{\rm  L}=0.693$, respectively, with a Hubble parameter of $H_{\rm 0}=h \, 100\,\rm Mpc\, km\,s^{-1}$ with $h=0.6777$, scalar spectral index of $n_{\rm s}=0.9611$ and a power spectrum normalization of $\sigma_{\rm 8}=0.8288$. Technical specifications of the simulation are presented in Table~\ref{simus}.  P-Millennium has $271$ snapshots containing the full particle and halo data.

%Merger trees and halo catalogues were constructed using {\sc SUBFIND} \citep{Springel01} and merger trees were built using D-halos \citep{Jiang14}. Halos with $\ge 20$ particles are included in the catalogues, giving a minimum halo mass of $2.12\times 10^9\,\rm M_{\odot}$. 

We run \shark\ over the \surfs\ suite of $N$-body, DM only simulations \citep{Elahi18}. Most of the \surfs\ runs have cubic volumes of $210\,\rm cMpc/h$ on a side, and span a range in particle number, currently up to $8.5$ billion particles, and adopt a $\Lambda$CDM \citet{Planck15} cosmology. The cosmological parameters correspond to a total matter, baryon and $\Lambda$ densities of $\Omega_{\rm m}=0.3121$, $\Omega_{\rm b}=0.0491$ and $\Omega_{\rm  L}=0.6879$, respectively, with a Hubble parameter of $H_{\rm 0}=h \, 100\,\rm Mpc\, km\,s^{-1}$ with $h=0.6751$, scalar spectral index of $n_{\rm s}=0.9653$ and a power spectrum normalization of $\sigma_{\rm 8}=0.8150$. \surfs\ was produced using a memory lean version of the {\sc gadget2} code on the Magnus supercomputer at the Pawsey Supercomputing Centre. In this paper, we use the L210N1536 simulation, with the specifications of Table~\ref{simus}. \surfs\ produces $200$ snapshots for each simulation, typically having a time span between snapshots in the range of $\approx 6-80$~Myr.

Merger trees and halo catalogs were constructed using the phase-space finder {\sc VELOCIraptor} \citep{Elahi19VR,Canas18} and the halo merger tree code {\sc TreeFrog}, developed to work on {\sc VELOCIraptor} \citep{Elahi19TF}. We refer to \citetalias{Lagos18c} for more details on how the merger trees and halo catalogs are constructed for \shark, and to \citet{Poulton18,Elahi19VR,Elahi19TF,Canas18} for details of {\sc VELOCIraptor} and {\sc TreeFrog}. 

%From the D-halos catalogues, 
From the {\sc VELOCIraptor} catalogues, 
we take the halo and subhalo masses, and halo virial radii. Other properties of halos are calculated as described in $\S$~4.2 in \citetalias{Lagos18c}. The existence of central and satellite subhalos in the 
%{\sc SUBFIND} 
{\sc VELOCIraptor} 
catalogues leads to \shark\ galaxies existing in 3 different types: type = 0 is the 
central galaxy of the central subhalo, while every other central
galaxy of satellite subhalos are type = 1. If a subhalo merges onto another one and it is not the main progenitor, it is treated as defunct.
All the galaxies of defunct subhalos are made type = 2 and transferred to the list of galaxies of the central subhalo of their descendant host halo (see $\S$~4.1 in \citetalias{Lagos18c} for more details).

\section{New baryon physics models and technical updates in \shark}\label{LargeModifications}

The modifications presented in this paper have been released in the public repository of \shark\ in the version {\sc v2.0}. Below we introduce the physical models that signify large improvements from our previous installment of \shark. In addition, we have implemented small changes to other existing physical models as described in Appendix~\ref{AppendixMinorModifications}. 

\subsection{ISM-stars angular momentum exchange}\label{newjmodel}

%EXTEND SHARK TO INCLUDE HALO'S AM FROM COOLING PART OF THE HALO TO TEST WHAT HAPPENS 
%IF THAT IS INCLUDED.

In this new version of \shark\ we include different levels of complexity for the 
%To estimate the disk scale radii, $r_{\rm s}$, 
%we follow the 
exchange of specific angular momentum between the cooling gas, interstellar medium (ISM) and stellar disk. 
Here, the specific angular momentum will be referred to as $j\equiv J/M$. 
The default in \citetalias{Lagos18c} assumes the cooling gas to have the same $j$ of the DM halo,

\begin{equation}
	j_{\rm cool} = \frac{J_{\rm h}}{M_{\rm halo}},
\end{equation}

\noindent where $J_{\rm h}$ is the 
halo's angular momentum, which is calculated from the mass and halo's spin
parameter, following \citet{Mo98},

\begin{equation}
        J_{\rm h}= \frac{\sqrt{2}\,G^{2/3}}{(10\,H(z))^{1/3}} \lambda_{\rm DM}\, M^{5/3}_{\rm halo}.
\label{jhalo}
\end{equation}

\noindent $j_{\rm cool}$ is then input in the set of ordinary differential equations (ODEs) 
that control the exchange of angular momentum 
(Eqs.~(\ref{eqn:sff:j})-(\ref{eqn:sff:jf})). Section~4.2 of \citetalias{Lagos18c} describes how $\lambda_{\rm DM}$ is obtained and Appendix~\ref{AppendixMinorModifications} presents a minor modification to the calculation of  $\lambda_{\rm DM}$ introduced in \shark\ v2.0.

The gaseous and stellar disks also 
 exchange angular momentum at a rate $\dot{J}_{\rm g,s}$. 
In its simplest form, 

\begin{equation}
	\dot{J}_{\rm g,s} = \psi \,j_{\rm cold}, 
        \label{AMexchange1}
\end{equation}

\noindent where $\psi$ is the instantaneous SFR and  
$j_{\rm cold}$ (not to be confused with $j_{\rm cool}$) is the specific angular momentum of the gaseous disk. 
This is the default assumption made in \citetalias{Lagos18c} and in most SAMs, 
except for those that follow disks as a set of annuli that can evolve independently 
(e.g. \citealt{Stringer07}; \citealt{Stevens16}). 

\citet{Mitchell18} showed that in the cosmological hydrodynamical simulation \eagle\ \citep{Schaye14,Crain15,McAlpine15}, 
the stellar $j$ was systematically lower than in the GALFORM SAM at fixed stellar mass. 
\citet{Mitchell18} suggested that a key modification required by GALFORM to evolve the specific angular momentum of galaxies more realistically was to consider the fact that stars form from molecular gas only. The latter  
tends to be more concentrated in the centres of galaxies relative to the total ISM \citep{Lagos11}. This implies that stars will be systematically forming 
from low specific angular momentum in comparison to the total ISM. 
\citet{Mitchell18} showed that this physical effect is behind the overly large sizes of galaxies in GALFORM.
Thus, a more sophisticated model should include the fact that stars form from molecular gas, which resides in the high column density regions of disks, and is more concentrated than the total gas disk. 
%\citet{Mitchell18} show that this physical effect 
%is behind the overly large sizes of GALFORM. 

Here, we implement a more realistic exchange between the specific angular momentum of the ISM and the stars 
following the lessons from \citet{Mitchell18}. This is done by 
calculating $\dot{J}_{\rm g,s}$ as 

\begin{equation}
	\dot{J}_{\rm g,s} = 2\,\pi\,  \int_{0}^{\infty} v(r)\,r\, \Sigma_{\rm SFR}\,{\rm d}r,
	\label{AMexchange2}
\end{equation}

\noindent where $\Sigma_{\rm SFR}$ depends on the assumed SF law and $v(r)$ is the circular velocity radial profile. 
We include a boolean parameter in \shark,  
{\tt angular$_{-}$momentum$_{-}$transfer}, which set to {\tt true} means 
the stellar disk angular momentum is calculated as Eq.~(\ref{AMexchange2}), and 
to {\tt false} uses Eq.~(\ref{AMexchange1}) instead. Thus, the results of \citetalias{Lagos18c} 
can be fully recovered with the latter option. 
%Throughout this paper we will refer to 
%the model of \citet{Lagos18c} as \shark-default, and 
%the \shark\ model variation with exactly the same parameters as \shark-default
%except for {\tt angular$_{-}$momentum$_{-}$transfer = true} as 
%\shark-AM. gIn the appendix we show some of the diagnostic relations 
%\citet{Lagos18c} used to calibrate the free parameters of the model to show that the 
%\shark-AM model does not lead to significant changes and thus 
%we do not require recalibration.

The half-mass gas and stellar disk sizes are then calculated as 
$r_{\rm gas} = f_{\rm norm}\, j_{\rm cold} / V_{\rm circ}$ and 
$r_{\star} = f_{\rm norm}\, j_{\star} / V_{\rm circ}$. Here, we set 
$f_{\rm norm} = 0.839$, which is the value in an idealized exponential disk \citep{Guo11}.
%following the relation between 
%$r\,v_{\rm circ}$ and $j$ that \citet{Swinbank17} reported for the 
%{\sc EAGLE} simulations. Note that the value of $f_{\rm norm}$ is slightly smaller than 
%the idealized value ($0.835$) adopted by \citet{Guo11} and \citet{Zoldan18}.

Before presenting the way we solve for the evolution of the galaxy angular momentum split by components, we look at the assumptions we make. From the definition  $J=j\,M$, with $J\equiv |\vec{J}|$ and $j\equiv |\vec{j}|$, it follows that  ${\rm  \partial} J/{\rm \partial}t= M\,{\rm  \partial} j/{\rm \partial}t  + j\,{\rm  \partial} M/{\rm \partial}t$. The two terms represent angular momentum transfer by torques and advection, respectively. We assume that internal to galaxies and from the gas cooling down to galaxies, ${\rm  \partial} j/{\rm \partial}t = 0$ (i.e. no torques), and hence ${\rm  \partial} J/{\rm \partial}t=j\,{\rm \partial} M/{\rm  \partial }t$. This is a simplification, as cosmological hydrodynamical simulations have shown torques can be important between the gas cooling and the galaxy (e.g. \citealt{Stevens16b}). 
The second important assumption is that all the components in the galaxy and gas cooling have angular momentum vectors that are aligned. This is not necessarily the case. \citet{Contreras17} showed that halos have an angular momentum vector that can evolve significantly in direction under the presence of high mass mergers. Quieter 
assembly histories lead to smaller changes of the direction of the vector. This in principle could lead to misalignments between the halo and the galaxy, as 
shown by \citet{Lagos15c}. 
The two assumptions above are made to simplify the problem, but we will consider in the future models that relax these assumptions and study the effect that would have on the scaling relations analysed in $\S$~\ref{structuresec}.
%However, here we do not consider this effect.

Simultaneously to the mass and metal exchange (see Eqs.~($49$) to ($58$) in \citetalias{Lagos18c}), in the case of SF in disks, 
we solve for the angular momentum exchange between 
these components following the assumptions above, as follows:

\begin{eqnarray}
	\dot  J_{\star}     &=&  (1-R)\, \dot{J}_{\rm g,s}  \label{eqn:sff:j} \\
	\dot  J_{\rm cold}  &=& \crate\, j_{\rm cool}- (1-R+\beta_{\star})\, \dot{J}_{\rm g,s} \\
	\dot  J_{\rm cold,halo}   &=& - \crate\, j_{\rm cool}\\
	\dot  J_{\rm hot,halo}   &=& \dot{m}_{\rm out,\star}\,  j_{\rm out} - \dot{m}_{\rm ejec}\, j_{\rm out}\\
	\dot  J_{\rm ejec}   &=& \dot{m}_{\rm ejec}\, j_{\rm out}. \label{eqn:sff:jf}
\label{eqn:sflc}
\end{eqnarray}

\noindent Here, $\dot{J}_{\rm g,s}$ is as described in Eq.~(\ref{AMexchange2}), $\dot{M}_{\rm cool}$, $\dot{m}_{\rm out,\star}$ and $\dot{m}_{\rm ejec}$ are the gas cooling rate (Eq.~(5) in \citetalias{Lagos18c}), outflow rate due to star formation and ejection rate from the halo, respectively; $\beta_{\star}=\dot{m}_{\rm out,\star}\,\psi^{-1}$, where $\psi$ is the instantaneous SFR, is the wind mass loading, and $R$ is the fraction of mass recycled to the ISM (from stellar winds and supernovae; see \S~4.4.6 in \citetalias{Lagos18c} for details).
In the case of the hot halo and ejected gas mass components, the angular momentum growth depends on the 
specific angular momentum of the outflowing gas. This in principle allows for outflows to affect the angular momentum of the disk in a differential fashion, which would be the case 
if the outflow rate was an explicit function of radius (as it has been proposed by detailed stellar feedback models, 
e.g. \citealt{Creasey12,Hopkins12,Lagos13}). However, in \shark\ v2.0 we assume $j_{\rm out} = j_{\rm cold}$, and leave for future work the exploration of how outflow rates that change radially can impact the angular momentum structure of galaxies.
%%%%%%%%%%%%%%%%%%%%
%Note that the form in Eqs.~\ref{eqn:sff:j}-\ref{eqn:sflc} implicitly assume that the angular momentum vectors 
%of the halo, cooling gas, ISM and galaxy stars are all aligned. This is not necessarily the case. \citet{Contreras17} showed 
%that halos have an angular momentum vector that can evolve significantly in direction under the presence of high mass mergers. Quieter 
%assembly histories lead to smaller changes of the direction of the vector. This in principle could lead to misalignments between the halo and the galaxy, as 
%shown by \citet{Lagos15c}. However, here we do not consider this effect.

Note that implicitly in Eqs.~(\ref{eqn:sff:j})-(\ref{eqn:sff:jf}) there is a third assumption, which is that the relaxation time of the hot halo gas is such that after a SF episode, it regains the same specific angular momentum as the DM halo. This assumes that any change in the hot halo gas $j$ due to outflows is negligible.  

\subsection{Dynamical friction leading to galaxy mergers}\label{poultonmodel}

Inherent limitations of halo and subhalo finders generally imply that subhalos cease to be tracked when they reach too few particles, or when they become indistinguishable from the 3D or 6D density background, depending on whether a 3D or 6D structure finder is used. This does not imply that galaxies in those subhalos should be immediately merge onto the central galaxy, as in some cases subhalos stop being tracked at large distances from the halo's centre \citep[typically 0.5--1 times the virial radius; see][]{Poulton2019}. For SAMs, this means that we have to resort to analytic calculations of the dynamical friction timescale to have a more realistic proxy for when a satellite galaxy should merge onto the central.

Recent work by \citet{Poulton2020} tracks (sub)halos that are identified in dense environments in $N$-body simulations, and provides estimates for their orbital properties. By comparing the merger timescale of these simulated subhalos with various analytical prescriptions, they are able to identify regimes where previously published analytical prescriptions do not provide a realistic estimate of merger timescales. Specifically, they find that in systems with relatively large host-to-subhalo mass ratios, these analytic approximations all systematically underpredict the merger timescale. This has important implications for both the satellite population and the high-mass galaxy population reported in SAMs, given the importance of mergers in the mass assembly of large galaxies that is expected from a hierarchical model of structure formation \citep{Robotham14}. 
%An under-abundance of satellite galaxies surviving at $z=0$ could significantly affect the stellar mass function, for example.

Follow-up work presented in \citet{Poulton2021} provided a new analytic model to compute the dynamical friction timescale ($\tau_{\rm{merge}}$) of satellite galaxies that much more accurately captures the merger timescales seen over a wide dynamic range in halo masses. In this work we implement this new dynamical friction timescale in \shark. The default model presented in \citetalias{Lagos18c} adopted the dynamical friction timescale of \citet{Lacey93}, which compared to \citet{Poulton2021} merges small galaxies too quickly.

\citet{Poulton2021} formulate the merger timescale in two regimes, based on their result that $\tau_{\rm{merge}}$ is more strongly dependent on position for subhalos outside of the virial radius of the host than for subhalos inside the virial radius. The merger timescale is calculated as

\begin{eqnarray}
  \tau_{\rm{merge}} = \begin{cases} 
        5.62\sqrt{\frac{R_{\rm vir,host}}{GM_{\rm encl,host}(r)}}r^{0.8}R_{\rm peri}^{0.2} & \text{for } r< R_{\rm{vir,host}}, \\
        5.62\frac{R_{\rm vir,host}}{\sqrt{GM_{\rm vir,host}}}r^{0.3}R_{\rm peri}^{0.2} & \text{for } r\geq R_{\rm{vir,host}}, \\
    \end{cases}  
\end{eqnarray}

\noindent where $r$ is the position of the subhalo relative to the halo's centre, $R_{\rm peri}$ is the pericentric distance from the host centre, $M_{\rm encl, host}(r)$ is the mass of the host enclosed within a radius $r$ assuming an \citet{Navarro97} profile, and $M_{\rm vir,host}$ is the virial mass of the host halo. These properties are all determined from the underlying $N-$body simulation. In \shark\ this model can be adopted by setting {\tt merger\_model = poulton20}. Note that the default in \citetalias{Lagos18c} was {\tt merger\_model = lacey93}.
%) in the input parameter file.

\subsection{AGN feedback: Wind and jet mode feedback}\label{newagnmodel}

The default model presented in \citetalias{Lagos18c} adopted the \citet{Croton16} AGN feedback model. That model in itself was inspired by \citet{Croton06}, which included only a radio mode of AGN feedback. The heating power of AGN feedback in the radio mode of \citet{Croton06} was calculated using the bolometric luminosity of AGN. For the latter, only the BH accretion rate coming from the hot-halo mode was considered (Eq. $10$ in \citealt{Croton06}). However, BHs at any one time can be growing by multiple channels; in particular, the accretion rate onto the BHs can be much higher than that coming from the hot-halo mode only if the galaxy is experiencing a SB (which in \shark\ can be either triggered by galaxy mergers or disk instabilities). There is no physical reason for why in these cases only the fractional contribution to the accretion rate from the hot-halo mode should be included in the AGN heating power calculation. In \shark\ v2.0, we implement a new model for AGN feedback that includes two modes: a jet and a wind mode, and uses the physical properties of the BH to calculate a jet power and a {  AGN}-driven outflow rate and velocity. 

{  Fundamentally, our model differs from previous ones in that feedback in previous works is modelled as dependent on the source of the accreting gas (e.g. \citealt{Croton06,Bower06,Lagos08,Somerville08,Fontanot20}), while in our model, it depends on the BH properties. Especifically, the mode of feedback depends on the type of BH accretion disk expected for different accretion rates. This is closer to the approach taken in cosmological hydrodynamical simulations, where the BH accretion rate is used to classify accretion disks between thin and thick disks (e.g. \citealt{Sijacki07,Dubois12,Weinberger17}).}

Before describing our new AGN feedback model, we describe how the three fundamental properties of BHs, their mass, accretion rate and spin, are calculated.

\subsubsection{BH mass, accretion rate and spin}\label{BHnohairproprs}

In addition to BH-BH mergers, in which we instantaneously add the masses of the BHs that are merging, BHs can also grow via gas accretion in two different modes. The BH accretion rate due to hot-halo mode is calculated as in \citetalias{Lagos18c},

\begin{equation}
\dot{m}_{\rm BH,hh} = \kappa\,\frac{15}{16} \pi\,G\,\mu\,m_{\rm p}\,\frac{\kappa_{\rm B}\,T_{\rm vir}}{\Lambda(T_{\rm vir},Z_{\rm hot})}\,m_{\rm BH},
\label{acchh}
\end{equation}
\noindent where $m_{\rm BH}$ is the BH mass, $G$ is the gravitational constant, $\mu$ is the atomic weight, $m_{\rm P}$ the proton mass, $\kappa_{\rm B}$ is Boltzmann's constant, $T_{\rm vir}$ the virial temperature of the halo, $Z_{\rm hot}$ the halo gas metallicity, $\Lambda(T_{\rm vir},Z_{\rm hot})$ the cooling function, and $\kappa$ a free parameter. For the case of BH growth during SBs, we follow \citetalias{Lagos18c},

\begin{equation}
\delta\,m_{\rm BH,sb} = f_{\rm smbh}\,\frac{m_{\rm gas}}{1+(v_{\rm smbh}/V_{\rm vir})^2},
\label{bhgrow_sbs}
\end{equation}

\noindent where $m_{\rm gas}$ and $V_{\rm vir}$ are the cold gas mass reservoir of the SB and the virial velocity of the halo, respectively. $f_{\rm smbh}$ and $v_{\rm smbh}$ are free parameters. As in \citetalias{Lagos18c}, we estimate the BH accretion rate in this mode assuming that the accretion timescale scales with the
bulge dynamical timescale, $\tau_{\rm acc,sb}=e_{\rm sb}\,r_{\rm bulge}/v_{\rm bulge}$, where $e_{\rm sb}$ is an e-folding parameter. The accretion rate during SBs is thus,

\begin{equation}
\dot{m}_{\rm BH,sb} = \frac{\delta\,m_{\rm BH,sb}}{\tau_{\rm acc,sb}}.
\end{equation}

\noindent The total accretion rate onto the BH at any one time is $\dot{m}_{\rm BH}=\dot{m}_{\rm BH,hh} + \dot{m}_{\rm BH,sb}$. The BH accretion disk structure is expected to be a strong function of the accretion rate. We thus define a normalised accretion rate based on the Eddington luminosity, $L_{\rm Edd}$,

\begin{equation}
    \dot{m} = \frac{\dot{m}_{\rm BH}}{\dot{m}_{\rm Edd}},
    \label{eddmacc}
\end{equation}

\noindent where $\dot{m}_{\rm Edd}=L_{\rm Edd}/(0.1\, c^2)$, and $c$ is the speed of light. 

When $\dot{m}>\dot{m}_{\rm ADAF}$ {  (with ADAF referring to Advection Dominated Accretion Flow; \citealt{Rees82}), and $\dot{m}_{\rm ADAF}$ being a parameter}, the accretion disk is expected to be thin and efficiently cool; this regime is commonly referred to as thin-disk (TD; \citealt{Shakura73}).
If instead, $\dot{m}<\dot{m}_{\rm ADAF}$, the accretion disk is expected to be unable to cool efficiently by radiation due to the energy generated by viscosity {  (i.e. the accretion disk is in the ADAF regime)}. It is broadly assumed the transition between TD and ADAF happens at $\dot{m}_{\rm ADAF}=0.01$. The ADAF regime according to \citet{Mahadevan97}, can be further subdivided into two regimes: the lower accretion rate ADAF regime ($\dot{m}<\dot{m}_{\rm crit,visc}$), in which heating of the electrons is dominated by viscous heating, and a 
higher accretion rate ADAF regime ($\dot{m}_{\rm crit,visc}<\dot{m}<\dot{m}_{\rm ADAF}$), in which the ion-electron heating dominates the heating of the electrons. On the other extreme, we classify AGN as super-Eddington (SE) if $\dot{m}>\eta$, with $\eta\sim 1$. In summary, four BH accretion regimes are defined:

\begin{itemize}
    \item SE: $\dot{m}>\eta$,
    \item TD: $\dot{m}\geqslant \dot{m}_{\rm ADAF}$ and $\dot{m}\le \eta$,
    \item ADAF$_\mathrm{high}$: $\dot{m}_{\rm crit,visc}<\dot{m}<\dot{m}_{\rm ADAF}$,
    \item ADAF$_\mathrm{low}$: $\dot{m}\le \dot{m}_{\rm crit,visc}$.
\end{itemize}

\noindent {  Even though AGN are allowed to be in the SE regime in \shark\ v2.0, in practice this rarely happens and such AGN never dominate the number density at any luminosity (Bravo et al. in preparation).}

The dimensionless BH spin vector, ${\bf a}$, is defined as ${\bf a} \equiv {\bf J}_{\rm  BH} /J_{\rm max} = c\, {\bf J}_{\rm BH} / G\, m^2_{\rm BH}$, where ${\bf J}_{\rm BH}$ is the angular momentum vector of the BH. To calculate ${\bf a}$, we implement three different models, which in \shark\ v2.0 can be selected by setting the variable {\tt spin$_{-}$model}. Below, we refer to the norm of the spin vector as $a=|{\bf a}|$. The list of models are presented below:

\begin{itemize}
    \item {\tt constant}. This corresponds to the simplest assumption of a constant spin. The radiation efficiency of a BH depends on the radius of the last stable orbit, which in itself depends on the BH spin. The value adopted for the constant spin in \shark\ v2.0 is $0.67$, which is equivalent to assuming a constant radiation efficiency of $0.1$ \citep{Bardeen72}. (Note that $a=0$ gives $\epsilon=0.057$ and $a=1$ gives $\epsilon = 0.42$; see Eqs.~(\ref{epsilonTD})-(\ref{EqZ2})).
    \item {\tt volonteri07}. This is a simple scaling relation inspired by the spin-BH mass relation found by \citet{Volonteri07}, who presented a model of BH accretion from a warped disk. The authors found that on average more massive BHs have a higher spin. We use the average relation they found and scale the spin directly with the BH mass as
\begin{equation}
    a = 0.305 \, {\rm log_{10}}\left(\frac{m_{\rm BH}}{\rm M_{\odot}}\right) - 1.7475.
\end{equation}
\noindent We limit $a$ to be in the range $[0,1]$ in this model.
\item {\tt griffin19}. This model is the full implementation in \shark\ v2.0 of the sophisticated BH spin development model of \citet{Griffin19}. {  This model is an update of the BH spin model presented in \citet{Fanidakis10}.} The model follows the changes in BH spin produced by BH-BH merger and gas accretion. For the latter, \citet{Griffin19} presented three different models, which we also implemented in \shark. These are the prolonged accretion, self-gravitating accretion disk and warped accretion disk. The user can choose between these three models by setting the variable {\tt accretion$_-$disk$_-$model}. Details of these models are presented in Appendix~\ref{AppendixSpin}. Note that in this model $a$ can take values in the range $[-1,1]$. This model includes the full implementation of the warped accretion disk model described in \citet{Volonteri07}, rather than a fit to their results as adopted by {\tt spin$_{-}$model=volonteri07}.
\end{itemize} 

With the BH accretion rate and spin defined, we can calculate the AGN bolometric luminosity. We first define the radiative accretion efficiency for a thin accretion disk, $\epsilon_{\rm TD}$,

\begin{equation}
    \epsilon_{\rm TD} = 1-\sqrt{1- \frac{2}{3\, \hat{r}_{\rm lso}}}, \label{epsilonTD}
\end{equation}

\noindent where $\hat{r}_{\rm lso}$ is the radius of the last stable circular orbit in units of
the gravitational radius of the BH, $r_{\rm G}\equiv G\, m_{\rm BH}/c^2$. We calculate  $\hat{r}_{\rm lso}$ following \citet{Bardeen72}
\begin{equation}
\hat{r}_{\rm lso} = 3 + Z_2 \pm \sqrt{(3 - Z_1)(3 + Z_1 + 2\, Z_2)}, \label{riso_eq}
\end{equation}
\noindent with the negative (positive) sign corresponding to the case when the angle between the BH spin and the accretion disk is less (larger) than $90$~degrees. We refer to these two cases as co- and counter-rotation, respectively. $Z_1$ and $Z_2$ are define as,
\begin{equation}
Z_1 = 1 + (1 - a^2)^{1/3} \left[(1 + a)^{1/3} + (1 - a)^{1/3} \right],
\end{equation}

\noindent and
\begin{equation}
Z_2 = \sqrt{3 a^2 + Z^2_1}.\label{EqZ2}
\end{equation}

\noindent Note that BHs can be counter-rotating only in the {\tt griffin19} spin model. 

We compute the AGN bolometric luminosity following \citet{Griffin19} for the thin disk and ADAF regimes, and following \citet{Griffin20} for the super-Eddington regime:

\begin{equation}
    L_{\rm bol} =\begin{cases} 
    \eta\, \left(1 + {\rm ln}\left(\frac{\dot{m}}{\eta}\, \frac{\epsilon_{\rm TD}}{0.1}\right)\right)\,L_{\rm Edd},\,\text{in  SE} \\
    \epsilon_{\rm TD}  \dot{m}_{\rm BH}\, c^2,\,\text{in TD}\\
    0.2\,\epsilon_{\rm TD}  \dot{m}_{\rm BH}\, c^2 \left(\frac{\dot{m}}{\alpha^2_{\rm ADAF}}\right)\, \left(\frac{\beta}{0.5}\right) \, \left(\frac{6}{\hat{r}_{\rm lso}}\right),\,\text{in ADAF$_\mathrm{high}$}\\
    0.0002\,\epsilon_{\rm TD}  \dot{m}_{\rm BH}\, c^2 
    \left(\frac{\delta_{\rm ADAF}}{0.0005}\right)\, \left(\frac{1-\beta}{0.5}\right) \, \left(\frac{6}{\hat{r}_{\rm lso}}\right)
    ,\,\text{in ADAF$_\mathrm{low}$}
    \end{cases} \label{Lboleq}
\end{equation}

\noindent Here, $\alpha_{\rm ADAF}$ is the viscosity parameter in the ADAF regime, $\delta_{\rm ADAF}$ is the fraction of viscous energy transferred to the electrons, which has a value between $0.1$ and $0.5$ (see review of \citealt{Yuan14}), and $\beta$ is the ratio of gas pressure to total pressure (i.e. the sum of gas pressure and magnetic pressure). Following \citet{Griffin19}, $\beta=1-\alpha_{\rm ADAF}/0.55$, 
  $\alpha_{\rm ADAF}=0.1$ and $\delta_{\rm ADAF}=0.2$. 
With these parameters defined, we introduce the boundary between the two ADAF regimes \citep{Griffin19},

\begin{equation}
    \dot{m}_{\rm crit,visc} = 0.001\,\left(\frac{\delta_{\rm ADAF}}{0.0005}\right)\, \left(\frac{1-\beta}{\beta}\right)\, \alpha^2_{\rm ADAF}.
\end{equation}

\subsubsection{AGN jet mode feedback}\label{radiomodemodel}

In \shark\ v2.0, the {  jet} mode of AGN feedback is assumed to only occur when halos have reached a quasi-hydrostatic equilibrium (i.e. halos have developed a hot gaseous halo). To evaluate this, we use the \citet{Correa18} criterion, in which a cooling and heating terms are calculated and compared. These terms are computed per halo as follows,

\begin{eqnarray}
  \Gamma_{\rm cool}(M_{\rm halo},r)  &=& M_{\rm hot}\, \frac{n_{\rm H}(r)\,\Lambda(T_{\rm vir},Z_{\rm hot})}{\mu \, m_{\rm p}},\label{gamma_cool}\\ 
   \Gamma_{\rm heat}(M_{\rm halo}) &=& \frac{3\,k_{\rm B}\, T_{\rm vir}}{2\,\mu\,m_{\rm p}}\, \frac{\Omega_{\rm b}}{\Omega_{\rm m}}\dot{M}_{\rm halo}\,\left(\frac{2}{3}\,f_{\rm hot} + f^{\rm halo}_{\rm acc,hot}\right).\,\,\,\,\,\,\,\,\,
\end{eqnarray}

\noindent Here, $M_{\rm halo}$ is the virial mass, $M_{\rm hot}$ is the halo gas mass, $n_{\rm H}(r)$ is the gas volume density of hydrogen atoms at $r$, $\Omega_{\rm b}$ and $\Omega_{\rm m}$ are cosmological parameters, $\dot{M}_{\rm halo}$ is the matter accretion rate onto a halo, $f_{\rm hot}$ is the fraction of gas in the halo relative to the universal baryon fraction ($\equiv M_{\rm hot}/(\Omega_{\rm b}/\Omega_{\rm m}\,M_{\rm  halo})$) and $f^{\rm halo}_{\rm acc,hot}$ is the fraction of the accreting gas that is hot. The latter fractions were parametrised as a function of the halo mass by \citet{Correa18} as follows,

\begin{eqnarray}
      f_{\rm hot} &=& 10^{-0.8 + 0.5\,x - 0.05\,x^2},\\
      f^{\rm halo}_{\rm acc,hot} &=& \frac{1}{e^{-4.3\,(x+0.15)} + 1},\\
\end{eqnarray}

\noindent with $x$ defined as, 
\begin{equation}
      x = {\rm log}_{10} \left(\frac{M_{\rm halo}} {10^{12}\,\rm M_{\odot}}\right).
\end{equation}
    
\noindent The matter accretion rate onto halos is calculated using \citet{Dekel09},

\begin{equation}
    \frac{\dot{M}_{\rm halo}}{\rm M_{\odot}\, Gyr^{-1}} = 0.47 \, \left(\frac{M_{\rm halo}}{10^{12}\,\rm M_{\odot}} \right)^{0.15}\, \left(\frac{1+z}{3}\right)^{2.25} \, \frac{M_{\rm halo}}{\rm M_{\odot}}.
\end{equation}

\noindent In principle, we can measure $\dot{M}_{\rm halo}$ directly from the {\sc VELOCIraptor} and {\sc TreeFrog} catalogues. However, there are cases of mass swapping between merger branches which can lead to sudden large changes in the halo mass. To avoid such big discontinuities in mass, we opt to use the fitting function above. Chandro-G\'omez et al. (in preparation) present a detailed analysis of the frequency of mass-swapping events for different simulations, subhalo finders and tree builders, and the effect they can have on galaxies in \shark.

A halo is considered to be capable of forming a hot halo when $\Gamma_{\rm heat} > \Gamma_{\rm cool}$. Under this condition, the accumulated shock-heated gas at the virial radius gains the necessary pressure through external shock-heating to overcome the energy loss from radiative cooling. In \shark, we assume the relevant density in Eq.~(\ref{gamma_cool}) to be the gas density of the halo at $R_{\rm vir}$ where the shocks for the accreting gas are expected to happen, which we approximate as 
$n_{\rm H}(R_{\rm vir})=200\,\rho_{\rm crit}/\mu\,m_{\rm P}$, where $\rho_{\rm crit}$ is the critical density of the universe.
%mean hot gas density of the halo, which for an isothermal sphere is $n_{\rm H}(R_{\rm vir})=M_{\rm hot}/(4\,\pi\,R^3_{\rm vir}\,\mu\,m_{\rm P})$. 
To allow flexibility in \shark, we include the parameter $\Gamma_{\rm thresh}$, so that halos with $\Gamma_{\rm cool}/\Gamma_{\rm hot} < \Gamma_{\rm thresh}$ are considered to have formed a hot halo. Appendix~\ref{sec:hothalo} shows at which halo mass, most of the halos comply with $\Gamma_{\rm cool}/\Gamma_{\rm hot} < \Gamma_{\rm thresh}$.

In the default model presented in \citetalias{Lagos18c}, the heating rate of the AGN in the radio mode was computed from the AGN bolometric luminosity produced by the hot-halo mode accretion rate only \citep{Croton06},

\begin{eqnarray}
    \dot{m}_{\rm heat} &=&  \frac{L_{\rm hh}}{0.5\,V^2_{\rm vir}},\\
    L_{\rm hh} &=& 0.1\,\dot{m}_{\rm BH,hh}\, c^2.
\end{eqnarray}

\noindent In the new {  jet} AGN feedback model, we compute the jet power (summed over both jets, assuming jets to be symmetrical) following \citet{Meier02},

\begin{equation}
    \frac{Q_\mathrm{mech}}{\rm erg\, s^{-1}}=
    \begin{cases}
        2.5\cdot10^{43}\left(\frac{M_\mathrm{BH}}{10^9M_\odot}\right)^{1.1}\left(\frac{\dot{m}}{0.01}\right)^{1.2}a^2\text{ if $\dot{m}\ge\dot{m}_{\rm ADAF}$,}\\
        2\cdot10^{45}\left(\frac{M_\mathrm{BH}}{10^9M_\odot}\right)\left(\frac{\dot{m}}{0.01}\right)a^2\text{ if $\dot{m}<\dot{m}_{\rm ADAF}$.}
    \end{cases}
    \label{eq:Lmech}
\end{equation}

\noindent In our new model, a fraction $\kappa_{\rm jet}$ of $Q_{\rm mech}$ is used to offset $\dot{M}_{\rm cool}$, so that the new cooling luminosity is reduced by $\kappa_{\rm jet} \, Q_{\rm mech}$. If $L_{\rm cool} < \kappa_{\rm jet} \, Q_{\rm mech}$ the cooling flow is completely shut off. Otherwise, the cooling rate is defined as

\begin{equation}
    \dot{M}^{\prime}_{\rm cool} = \dot{M}_{\rm cool}\, \left(1-\frac{\kappa_{\rm jet} \, Q_{\rm mech}}{L_{\rm cool}}\right).\label{kappa_radio}
\end{equation}
\noindent Here, $\kappa_{\rm jet}$ is a free parameter controlling how efficient the jet power is in heating the halo gas. Note that in principle, $\kappa_{\rm jet}$ can be $>1$ if the jets are efficient in producing buoyant bubbles that lift up gas, as recently suggested by \citet{Husko23}.
%reduced by $(L_{\rm cool} - \kappa_{\rm radio} \, Q_{\rm mech})/L_{\rm cool}$, where $L_{\rm cool}$ is the cooling luminosity.
%, where $L_{\rm cool} = 5\, k_{\rm B}\, T_{\rm vir} \dot{M}_{\rm cool} / 2\,\mu\,m_{\rm p}$.
%\begin{equation}
%    \dot{m}_{\rm cool} =  (L_{\rm cool} - \kappa_{\rm radio} \, Q_{\rm mech})\kappa_{\rm radio}\,%\frac{Q_{\rm mech}}{0.5\,V^2_{\rm vir}},
%    \label{mheating}
%\end{equation}

%\noindent where $\kappa_{\rm radio}$ is a free parameter of order unity. 
Note that in this model, both BH accretion modes, hot-halo and SB, contribute to the mechanical power of jets, and hence {  jet} mode feedback can happen regardless of the source of gas accretion.

In \shark\ v2.0, satellite galaxies can continue to accrete gas that is cooling from the hot halo they have retained (as it is gradually stripped; see $\S$~\ref{RPSmodel}). This means that satellite galaxies can also undergo AGN {  jet} mode feedback. The way this operates is the same as above. However, in situations where $L_{\rm cool} < \kappa_{\rm jet} \, Q_{\rm mech}$ in a satellite galaxy, we assume that the excess power, $P_{\rm excess}=\kappa_{\rm jet} \, Q_{\rm mech} - L_{\rm cool}$, can be used to heat up the hot gas reservoir of the central subhalo. If this is the case, then the effective heating power affecting the gas cooling onto the central galaxy is larger than $\kappa_{\rm jet} \, Q_{\rm mech}$ of the central galaxy by $\sum_{i} P_{\rm excess,\it i}$, where $i$ are all the satellite galaxies with an excess power $>0$. In the practice $P_{\rm excess}$ makes little difference to the heating rate offsetting gas cooling of central subhalos.

\subsubsection{AGN wind mode feedback}\label{QSOfeedbackmodel}

We consider outflows in {  AGN galaxies} to be driven by direct radiation pressure on dust grains, as originally introduced by \citet{fabian1999}.
For this, we follow the description of the production of outflows of \citet{murray2005} and complement that with the calculations of \citet{ishibashi2015} for the outflow velocities produced in the case of radiation pressure on dust grains. In this model we use gas metallicity and gas content as proxies of dust mass (as described below).

\citet{murray2005} discussed the role of radiation pressure on dust in driving galactic-scale winds through momentum deposition.
\citet{murray2005} assumed an isothermal sphere profile, and argue that in the optically-thick limit and from the momentum equation of the gas, one can derive a critical luminosity at which the effective gravity is reduced by the momentum deposition of the radiation. This critical luminosity for an isothermal sphere can be written as,

\begin{equation}
    \frac{L_\mathrm{crit}}{3\cdot10^{46}\ \mathrm{erg}\ \mathrm{s^{-1}}}\approx 10\, f_{\rm gas} \left(\frac{\sigma}{200\ \mathrm{km}\ \mathrm{s^{-1}}}\right)^4,
\end{equation}

\noindent where $f_{\rm gas}$ is the gas fraction of the bulge and $\sigma$ its velocity dispersion (which in this case we equate to the stellar velocity dispersion). 
If the BH's bolometric luminosity is in excess of this critical value, $L_\mathrm{bol}>L_\mathrm{crit}$, the net motion of the gas in the bulge would be outwards (i.e., outflow). {  For galaxies with stellar masses $>10^{10}\,\rm M_{\odot}$ that are undergoing a starburst in \shark\ v2.0, $L_{\rm crit}$ typically takes values $10^{45}-5\times 10^{46}\,\rm erg\,s^{-1}$, meaning that only bright QSOs are expected to be powerful enough to drive outflows.}

If an outflow is produced, and following \citet{murray2005}, it should be capable of sweeping up the ISM gas that is outside the sublimation radius.
However, the latter tends to be very small ($<100$~pc) and hence, we consider the total ISM gas mass in the bulge, $m_\mathrm{gas,b}$, to participate in the outflow.
We take as the relevant timescale for the outflow to be the Salpeter time (i.e. the time to double the mass of the BH), defined as

\begin{equation}
    \tau_\mathrm{Salp}=43\Gamma^{-1}\ \mathrm{Myr}.
\end{equation}

\noindent where $\Gamma\equiv L_\mathrm{bol}/L_\mathrm{Edd}$.
With this timescale, we calculate the outflow rate to be 

\begin{equation}
    \dot{m}_\mathrm{out,AGN}=\frac{m_\mathrm{gas,b}}{\tau_\mathrm{Salp}}.\label{outflowrateBH}
\end{equation}

\noindent The reason why we consider the Salpeter time to be the relevant timescale here is because this timescale assures us that the BH will grow to values that are comparable to local BHs as measured in \citet{mcconnell2013}, and because the Salpeter timescale is of a similar magnitude as the duration of SBs. It also ensures that enough stars are produced in the galaxies in \shark\ to lie in the Faber-Jackson (\citealt{murray2005,power2011}) and BH-bulge mass relations. Following the arguments presented in \citet{Nayakshin09}, using $\tau_{\rm Salp}$ in Eq.~(\ref{outflowrateBH}) can potentially underestimate the impact of {  wind} feedback in galaxies with small bulges (those with a stellar velocity dispersion $<150\,\rm km\,s^{-1}$), but it is a fair representation of the timescale at which BHs grow in more massive bulges. Because {  the wind-mode} feedback is expected to be significant in massive galaxies only, we consider this assumption to be reasonable.

We estimate the terminal velocity of the outflow following \citet{ishibashi2015} in the limit of radiation pressure on dust grains and assuming that the whole ISM content is contained in an expanding shell:

\begin{eqnarray}
    v_\mathrm{out,AGN} &\approx& 320\ \mathrm{km}\ \mathrm{s^{-1}} \left(\frac{L_\mathrm{bol}}{10^7L_{\odot}}\right)^{1/2}\left(\frac{\kappa_\mathrm{UV}}{10^3\ \mathrm{cm}^2\ \mathrm{g}^{-1}} \right)^{1/4}\nonumber \\ 
    &&\cdot\,\left(\frac{m_\mathrm{gas,b}}{M_{\odot}}\right)^{-1/4}.
\end{eqnarray}

\noindent Here, $\kappa_\mathrm{UV}$ is the UV opacity.
We use the \citet{ishibashi2015} approximation $\kappa_\mathrm{UV}=10^3\,f_\mathrm{dg,MW}\ \mathrm{cm}^2\ \mathrm{g^{-1}}$, with $f_\mathrm{dg,MW}$ being the dust-to-gas mass ratio relative to the Milky-Way value.
Assuming a constant metallicity-dependent dust-to-gas mass ratio, we can then write $\kappa_\mathrm{UV} = 10^3\,(Z_\mathrm{gas,bulge}/Z_{\odot})\, \mathrm{cm}^2\ \mathrm{g}^{-1}$, where $Z_\mathrm{gas,bulge}$ is the fraction of mass in metals relative to the total gas mass in the bulge. 
%We have also tested a different velocity prescription, following \citet{king2019},

%\begin{equation}
%    \frac{v_\mathrm{out,QSO}}{1230\ \mathrm{km}\ \mathrm{s^{-1}}} \approx \left(\frac{\sigma}{200\ \mathrm{km}\ %\mathrm{s^{-1}}}\right)^{2/3}l^{1/3}.
%\end{equation}

%\noindent where $\sigma$ is the velocity dispersion of the galaxy bulge and $l=L_\mathrm{QSO}L_\mathrm{QSO,Edd}^{-1}$ the ratio between the QSO and Eddington luminosities. This lead to consistently lower outflow velocities, which already are low compared to observed outflows \citneed, so for this work we decided on the model by \citet{ishibashi2015}.
%\comment{MATIAS}{A plot comparing both models could be nice.}

Similar to the description of SB-driven outflows in \citetalias{Lagos18c}, we define an excess energy of the {  AGN}-driven outflow that can be used to eject gas from the halo as,

\begin{equation}
    E_\mathrm{excess,AGN}=\epsilon_\mathrm{wind}\,\frac{v^2_\mathrm{out,AGN}}{2}\,\dot{m}_\mathrm{out,AGN}.\label{energyexcess}
\end{equation}

Here, $\epsilon_\mathrm{wind}$ is a free parameter in which we enclose variations of geometry and other simplifications of our modelling.
We then define the gas ejection rate due to these {  AGN-driven} outflows as:

\begin{equation}
    \dot{m}_\mathrm{ejec,AGN}=\frac{E_\mathrm{excess,AGN}}{0.5\,V^2_\mathrm{circ}}-\dot{m}_\mathrm{out,AGN}
\end{equation}

\noindent where $V_\mathrm{circ}$ is the circular velocity of the halo. 
This can be reduced to

\begin{equation}
    \dot{m}_\mathrm{ejec,AGN}=\left(\epsilon_\mathrm{AGN}\,\frac{v^2_\mathrm{out,AGN}}{v^2_\mathrm{circ}}-1\right)\,\dot{m}_\mathrm{out,AGN}.
\end{equation}

To include the {  wind-mode} feedback, we modify the differential equations (Eqs. $49-58$ in \citetalias{Lagos18c}) that control the evolution of the stellar ($M_\star$), cold gas ($M_\mathrm{cold}$), hot halo gas ($M_\mathrm{hot}$) and ejected gas ($M_\mathrm{ejec}$) masses, and their respective metals ($M^Z_\star$, $M^Z_\mathrm{cold}$, $M^Z_\mathrm{hot}$ and $M^Z_\mathrm{ejec}$), as follows:

\begin{eqnarray}
\dot{M}_{\star} &=& (1-R) \psi \label{eq:sff} \\
\dot{M}_{\rm cold} &=& \crate-(1-R+\beta_\star+\beta_{\rm AGN})\psi \\
\dot{M}_\mathrm{cold,halo} &=& -\crate \\
\dot{M}_\mathrm{hot,halo} &=& (\dot{m}_\mathrm{out,\star}+\dot{m}_\mathrm{out,AGN})\nonumber \\
    && -(\dot{m}_\mathrm{ejec,\star}+\dot{m}_\mathrm{ejec,AGN}) \\
\dot{M}_\mathrm{ejec} &=& \dot{m}_\mathrm{ejec,\star} \\
\dot{M}_\mathrm{lost} &=& \dot{m}_\mathrm{ejec,AGN}\\
\dot{M}_{\star}^Z &=& (1-R)Z_\mathrm{cold}\psi \\
\dot{M}_\mathrm{cold}^Z &=& \crate Z_\mathrm{cold,halo}\nonumber\\
    && +(p-(1+\beta_{\star}+\beta_\mathrm{AGN}-R)Z_\mathrm{cold})\psi \\
\dot{M}_\mathrm{cold,halo}^Z &=& -\crate Z_\mathrm{cold,halo}\\
\dot{M}_\mathrm{hot,halo}^Z &=& \dot{M}_\mathrm{hot,halo}Z_\mathrm{cold}\\
\dot{M}_\mathrm{ejec}^Z &=& Z_\mathrm{cold}\dot{m}_\mathrm{ejec,\star}\\
\dot{M}_\mathrm{lost}^Z &=& Z_\mathrm{cold}\dot{m}_\mathrm{ejec,AGN}\label{eq:sfflast}
\label{eq:sflc}
\end{eqnarray}

\noindent where 
%$\beta_{\star}\equiv\dot{m}_\mathrm{out,\star}\psi^{-1}$ and
$\beta_\mathrm{AGN}\equiv\dot{m}_\mathrm{out,AGN}\,\psi^{-1}$ is the mass loading due to {  wind-mode} feedback, $Z_\mathrm{cold}\equiv M_\mathrm{cold}^ZM_\mathrm{ cold}^{-1}$ and $Z_\mathrm{cold,halo}\equiv M^Z_\mathrm{cold,halo}M_\mathrm{cold,halo}^{-1}$ are the metallicities of cold gas in the ISM and the cold gas in the halo (the part actively cooling), respectively, and 
%$\phi$ the instantaneous SFR, 
%$\dot{M}_{\rm cool}$ the cooling rate, 
$p$ the metal yield. $R$, $\psi$ and $\beta_{\star}$ were defined when introducing Eqs.~(\ref{eqn:sff:j})-(\ref{eqn:sflc}).
%, and $R$ the fraction of mass recycled to the ISM (from stellar winds and SNe). 
Note that $\beta_{\rm AGN}$ is only positive in the case of SBs and where $L_\mathrm{bol}>L_\mathrm{crit}$, and hence during star formation in the disk, there is no {  wind-mode} feedback. When gas is ejected from the halo due to {  wind-mode} feedback, we assume this gas is lost and never reincorporated into the halo. {  This, however, barely impacts the baryon content of halos, as galaxy halos with masses $\gtrsim 10^{12}\,\rm M_{\odot}$ in \shark\ v2.0 have baryons fractions close to the universal baryon fraction, in agreement with 
%The latter is consistent with results from 
cosmological hydrodynamical simulations  \citep{Wright24}.}
%, which show that once gas is lost from halos due to AGN feedback, it does not come back at later times. This happens regardless of the AGN feedback model implemented. The main difference between AGN feedback models is how much gas can escape the halo in the first place.} 
{  The assumption about AGN-driven outflows from the halo being lost is different from stellar feedback-driven outflows in \shark\ v2.0, where the gas can be reincorporated following a timescale that depends on halo properties (see Eq.~18 in \citetalias{Lagos18c}).}

\subsection{Ram pressure stripping of the halo and ISM gas}\label{RPSmodel}

\shark\ v1.1 included a treatment of instantaneous gas stripping of the hot halo of satellite galaxies. As soon as galaxies became satellites, their hot halo gas was instantaneously removed and transferred to the hot halo gas of the central galaxy of the host halo. Here, we include a treatment of ram pressure stripping (RPS) of both the hot halo and ISM gas of satellite galaxies as described below. 

For the RPS of the hot halo, we follow \citet{Font08}, which follows the RPS criterion found by \citet{McCarthy08} in their hydrodynamical simulations, for a spherical distribution of gas. The halo gas beyond a radius $r_{\rm sat}$ (measured from the centre of the satellite galaxy) is removed if the ram pressure at that position exceeds the binding energy the gas feels due to the satellite's gravitational potential, 

\begin{equation}
\rho^{\rm cen}_{\rm halo,gas}(R) \, v^2_{\rm sat} > \alpha_{\rm RPS} \frac{G\,M_{\rm sat}(r_{\rm sat}) \, M^{\rm sat}_{\rm halo,gas}}{8\, r^{\rm sat}_{\rm vir}\,\, r^3_{\rm sat}},\label{RPShalo}
\end{equation}

\noindent where $M_{\rm sat}(r_{\rm sat})$ is the total mass enclosed in $r_{\rm sat}$, $\rho^{\rm cen}_{\rm halo,gas}(R)$ is the central galaxy's halo gas density at the position of the satellite galaxy relative to the halo centre, $R$, and $v_{\rm sat}$ is the velocity of the satellite in the frame of the host halo. Both $R$ and $v_{\rm sat}$ are the subhalo's position and velocity for satellites type 1 (see below for treatment of satellites type 2). The parameter $\alpha_{\rm ram}=2$ in \citet{McCarthy08}. With the aim of allowing for flexibility in the code, the latter is left as a free parameter, which in the default \shark\ v2.0 is set to $\alpha_{\rm RPS}=1$. Note that $M^{\rm sat}_{\rm halo,gas}$ and $r^{\rm sat}_{\rm vir}$ are the halo gas mass and virial radius the satellite galaxy had right before it became a satellite. The latter assumes that the gas is stripped from the satellite's halo outside-in without affecting the density internal to the stripping radius. 
We find $r_{\rm sat}$ by assuming equality in Eq.~(\ref{RPShalo}) and strip away all the gas that is at radii $>r_{\rm sat}$ that has not yet been stripped. 

For the satellite's ISM, we assume a similar model for RPS, and remove all the gas outside $r$, with $r$ being the radius at which
\begin{equation}
    \rho^{\rm cen}_{\rm halo,gas}(R) \, v^2_{\rm sat} = 2\pi\,G \Sigma_{\rm gas}(r)\,\left[\Sigma_{\rm gas}(r) + \Sigma_{\star}(r)\right]\label{RPSISM}
\end{equation}

\noindent where $\Sigma_{\rm gas}$ and $\Sigma_{\star}$ are the ISM's and stellar surface densities at $r$. Note that the latter has contributions from both the disk and bulge components. The ISM in the disk and bulge follows an exponential profile, with half-mass radii sizes $r_{\rm gas,disk}$ and $r_{\rm gas, bulge}$. The stellar component of the disk, also follows an exponential profile with half-mass radius, $r_{\star, \rm disk}$, while the bulge stars follow a Plummer profile, with half-mass radius $r_{\star, \rm bulge}$. In \shark\ v2.0,  $r_{\rm gas,disk}$ and $r_{\star, \rm disk}$ are calculated self-consistently, following the description in $\S$~\ref{newjmodel}, while $r_{\rm gas, bulge}=r_{\star, \rm bulge}$ and are calculated as described in $\S\, 4.4.12$ of \citetalias{Lagos18c}. The same profiles are used to calculate $M_{\rm sat}(r_{\rm sat})$ in Eq.~(\ref{RPShalo}). 

For the RPS of the ISM, we make the same assumption as for the RPS of the halo, and assume the gas and stellar profiles internal to $r$ in Eq.~(\ref{RPSISM}) are unchanged under RPS, and hence we keep track of the stripped ISM, so that we can incorporate it when computing the gas surface density in Eq.~(\ref{RPSISM}). We also assume that the galaxy radii do not change due to RPS. The latter is a sensible assumption if RPS acts on short timescales (shorter than a dynamical relaxation timescale). 

When a type 1 satellite becomes type 2 (its host subhalo disappears from the subhalo catalogue), we assume that any left over hot gas is instantaneously transferred to the central subhalo's hot gas, but the ISM continues to experience RPS following the equations above.

\subsection{Tidal stripping of gas and stars}\label{TDSmodel}

In addition to the presence of dynamical friction, which ultimately leads to galaxy mergers, the tidal stripping of these infalling satellite galaxies is also an important environmental process. 
%The latter contributes to the growth of stellar halos of galaxy groups or clusters of galaxies. 
Tidally stripped stellar material on the outskirts of halos typically does not contribute to observational estimates of the central galaxy stellar mass and instead is associated to an independent component, which we referred to as ``intra-halo stellar mass''.  This can have an important impact on comparing observations of the SMF with model predictions. Recent work using cosmological hydrodynamical simulations estimates that up to half of the stellar mass bound to a central subhalo can be attributed to stellar halos, rather than the central galaxy (e.g. \citealt{Canas20,Proctor23}).

%e tidally stripped stellar mass from satellites is a mass budget that ultimately does not contribute to the central galaxy, and hence it can have an important effect on the high-mass end of the SMF, with about half of the stellar mass being locked in stellar halos rather than the central galaxy (e.g. \citealt{Canas20,Proctor23}).
%which is dominated in number by high-mass satellites in dense environments.

We include in \shark\ v2.0 a new physical model to describe the tidal effects felt by satellite galaxies from their central component. This implementation is based on the work of \citet{Errani2015}, who use high resolution $N$-body simulations of a Milky-Way mass halo to track the evolution of tidal material that can be associated with dwarf spheroidal galaxies. 
Although \citet{Errani2015} did their analysis specifically for a Milky-Way halo, the problem is highly self-similar and it mostly depends on the fractional mass loss rate experience by an object orbiting in an \citep{Navarro97} profile.
\citet{Errani2015} parameterise the evolutionary tracks of the stripped material 
%in terms of the mass the satellite subhalo had before infall, $m_{\rm h,0}$, and the current mass, $m_{\rm h}$, enclosed within a half-stellar mass radius of the satellite galaxy. 
%halo mass with respect to the mass the halo had before infall, $m_h$, enclosed at the half-light radius, $r_h$. 
%The below fitting function is used to fit to the evolutionary tracks 
following the work of \citet{Penarrubia2008},

\begin{equation}
    \frac{M_{\star,\rm TS}}{M_{\star,0}} = \frac{2^\alpha x^\beta}{(1 + x)^\alpha}, 
\end{equation}
where $x=M_{\rm sh}/M_{\rm h,0}$, $M_{\rm h,0}$ is the total halo mass the now satellite subhalo had at infall, $M_{\rm sh}$ is the current subhalo mass, 
$M_{\star,\rm TS}$ is the amount of stellar mass that has been tidally stripped from the satellite galaxy that is hosted by the subhalo in question, and $M_{\star,0}$ is the stellar mass the galaxy had at infall. In principle the halo masses above should not be the total current and at infall ones, but the mass enclosed within the half-light radii of the galaxy currently or at infall, respectively. However, \citet{Errani2015} found that the half-light radius of the galaxy is barely affected by tidal stripping, and hence one can assume a constant half-light radius to halo's scale radius ratio. Here we adopt the parameters $\alpha = 3.57$ and $\beta = 2.06$, corresponding to a ``cuspy'' \citet{Navarro97} profile and half-light radius to halo's scale radius ratio $=0.2$.
%halo mass enclosed within a half-stellar mass radius of the now satellite subhalo, at some point in its orbit. 
%$f(x)$ gives the fraction of stellar material that is transferred from the satellite galaxy to the stellar halo of the central galaxy due to tidal effects, with respect to the mass the galaxy had before infall, $M_{\star}/M_{\star,0}$. 
%In the above equation, we adopt values of $\alpha = 3.57$ and $\beta = 2.06$, corresponding to a ``cuspy'' \citet{Navarro97} profile and $r_{\rm star}/a = 0.2$. 
These values correspond to the largest effect tidal stripping can have on the stellar content of a satellite galaxy according to \citet{Errani2015}. This allows us to test the maximal effect tidal stripping can have on our galaxy population. We also implement a minimum remaining halo mass fraction of 1\% (where the fraction gives the difference between the subhalo mass at the current snapshot and the subhalo mass at infall; this is a parameter in \shark, named {\tt minimum\_halo\_mass\_fraction}) to ensure that a halo can only lose at most 99\% of its mass. 
%Note that \citet{Errani2015} found that the half-light radius of the galaxy is barely affected by tidal stripping, and hence here we assume tidal stripping only affects the stellar mass of the galaxy. 
This model can be used by setting {\tt tidal\_stripping = true} in the input parameter file.

We implement this in a way that the disk, which has the larger radius, is stripped first, followed by the bulge. If the galaxy has a cold gas reservoir, we strip the amount of gas that is outside a radius $r_{\rm str}$ beyond which a fraction of stellar mass $M_{\star,\rm TS}/M_{\star,0}$ has been stripped.

\begin{table*}
    \centering
        \caption{List of new models and parameters included in \shark\ v2.0. The parameters presented in \textcolor{purple}{purple} are those that are varied in {\tt optim} when searching for a best fit to the $z<0.1$ SMF. We present a suggested range for the parameters and in parentheses the value adopted in our best-fitting model. The relevant equation/section for each model or parameter is presented in the third column. Parameters that were introduced in  \citetalias{Lagos18c} refer to the equations in that paper.}
    \label{tab:parameters}
    \begin{tabular}{c|c|c}
    \hline
    \hline
    Parameter & Suggested value range (adopted) & Equation/Section \\
    \hline
    \hline
     Physical Process & AGN Feedback and BH growth & \\
    \hline
     model & bower06, croton15, lagos22 (lagos22) & $\S$~\ref{newagnmodel}\\
     spin$_{-}$model & constant, volonteri07, griffin19 (griffin19)  & $\S$~\ref{BHnohairproprs} \\
     accretion$_{-}$disk$_{-}$model & prolonged, selfgravitydisk, warpeddisk (warpeddisk) & Appendix~\ref{AppendixSpin} (only for spin$_{-}$model=griffin19)\\
     wind$_{-}$feedback & true or false (true) & $\S$~\ref{QSOfeedbackmodel}\\
         {\color{purple}$\kappa$} &  {\color{purple}$10^{-5}-10^2$ (10.31)} & {\color{purple}Eq.~(\ref{acchh})}\\
         $f_{\rm smbh}$ &  $10^{-5}-10^{-1}$ ($10^{-2}$) & Eq.~(\ref{bhgrow_sbs})\\
         $e_{\rm sb}$ & $0.5-50$ (15) & $\S$~\ref{BHnohairproprs}\\
         $\dot{m}_{\rm ADAF}$ & 0.01 & $\S$~\ref{BHnohairproprs}\\
         $\delta_{\rm ADAF}$ &  0.1-0.5 (0.2) & Eq.~(\ref{Lboleq})\\
         $\alpha_{\rm ADAF}$ &  $0.05-0.5$ (0.1) & Eq.~(\ref{Lboleq})\\ 
         $\eta$ & 1-10 (4) & $\S$~\ref{BHnohairproprs}\\
         $\alpha_{\rm TD}$ &  $0.05-0.5$ (0.1) & Appendix~\ref{AppendixSpin}\\
         {\color{purple}$\kappa_{\rm jet}$} &  {\color{purple}$10^{-1}-3$ ($0.023$)} & {\color{purple}Eq.~(\ref{kappa_radio})}\\
         {\color{purple}$\Gamma_{\rm thresh}$} & {\color{purple}$0.01-100$ (10)} & {\color{purple}$\S$~\ref{radiomodemodel}}\\
         $\epsilon_{\rm wind}$ & $0.1-100$ (10) & Eq.~(\ref{energyexcess})\\
 \hline
 \hline
 Physical Process & Tidal and Ram-pressure Stripping & \\
 \hline
 tidal$_{-}$stripping & true or false (true) & $\S$~\ref{TDSmodel}\\
 gradual$_{-}$stripping$_{-}$ism  & true or false (true) & $\S$~\ref{RPSmodel}\\
 gradual$_{-}$stripping$_{-}$halo  & true or false (true) & $\S$~\ref{RPSmodel}\\
 $\alpha_{\rm RPS}$ & 0.1-10 (1) & Eq.~(\ref{RPShalo})\\
 minimum$_{-}$halo$_{-}$mass$_{-}$fraction & $10^{-5}-10^{-1}$ ($10^{-2}$) & $\S$~\ref{TDSmodel}\\
 \hline
 \hline
 Physical Process & Dynamical Friction & \\
 \hline
 merger$_{-}$timescale$_{-}$model & lacey93 or poulton20 (poulton20) & $\S$~\ref{poultonmodel}\\
 \hline
 \hline
 Physical Process & Chemical Enrichment & \\
 \hline
 evolving$_{-}$yield & true or false (true) & Appendix~\ref{evolvingyield} \\
 \hline
 \hline
  Physical Process & Star Formation & \\
 \hline
 {\color{purple}$\nu_{\rm SF}$} & {\color{purple}$0.2 - 1.7\,\rm Gyr^{-1}$ ($1.49\,\rm  Gyr^{-1}$)} & {\color{purple}Eq.~(7) in \citetalias{Lagos18c}}\\
  {\color{purple}$\eta_{\rm burst}$} &  {\color{purple}1-20 (15)} &  {\color{purple}$\S$~4.4.3 in \citetalias{Lagos18c}}\\
{\tt angular$_{-}$momentum$_{-}$transfer} & true or false (true) & $\S$~\ref{newjmodel}\\
\hline
\hline
Physical Process & Stellar Feedback & \\
\hline
{\color{purple}$\beta$} & {\color{purple}$0.5-5$ (3.79)} & {\color{purple}Eqs.~(25)-(28) in \citetalias{Lagos18c}}\\
{\color{purple}$\beta_{\rm min}$} & {\color{purple}$0.01-1$ (0.104)} & {\color{purple}Appendix~\ref{StellarFeedbackChanges}}\\
$v_{\rm hot}$ & $50-500\,\rm km\,s^{-1}$ ($120\,\rm km \, s^{-1}$) & Eqs.~(25)-(28) \citetalias{Lagos18c}\\
\hline
\hline
Physical Process & Reincorporation & \\
\hline
{\color{purple}$\tau_{\rm reinc}$} & {\color{purple}$1-30$~Gyr (21.53\,Gyr)} & {\color{purple}Eq.~(30) in \citetalias{Lagos18c}}\\
{\color{purple}$\gamma$} & {\color{purple}$-3\,\rm to\,0$ (-2.339)} & {\color{purple}Eq.~(30) in \citetalias{Lagos18c}}\\
{\color{purple}$M_{\rm norm}$} & {\color{purple}$10^9-10^{12}\,\rm M_{\odot}$ ($1.383\times 10^{11}\,\rm M_{\odot}$)} & {\color{purple}Eq.~(30) in \citetalias{Lagos18c}}\\
\hline
\hline
    \end{tabular}
\end{table*}

\subsection{Automatic parameter exploration}

We perform an automatic search for a suite of best-fitting parameters by using a Particle Swarm Optimisation (PSO) python package, {\tt optim}, introduced in Proctor et al. (in preparation). We here use the $z<0.1$ SMF of \citet{Li09} based on SDSS as our primary constraint. In addition to this, {\tt optim} can also use as constraints the SMF at $z=0.5,\, 1$ and $2$ (from \citealt{Weaver22}), the cosmic star formation rate density (CSFRD) of \citet{Driver17}, the HI mass function of \citet{Jones18} and the total stellar-size mass relation of the Galaxy and Mass Assembly (GAMA) survey (see Appendix~\ref{SecSizesGAMA}). Proctor et al. (in preparation) introduce in detail {\tt optim} and these constraints. Note that {\tt optim} can be easily used as a standalone tool, and hence be adapted to work with other SAMs. 
{  We decide not to use empirical AGN properties, such as their luminosity function (LF), to constrain the parameters of the model. This is because we empirically find that the model tends to produce an AGN LF that agrees with observations. This will be discussed in detail in Bravo et al. (in preparation).}

Following Proctor et al. (in preparation), we only vary the following parameters: 

\begin{itemize}
    \item AGN feedback parameters: $\kappa$ (Eq.~\ref{acchh}), $\kappa_{\rm jet}$ (Eq.~\ref{kappa_radio}), and $\Gamma_{\rm thresh}$ ($\S$~\ref{radiomodemodel}). These parameters control the efficiency of gas accretion onto the BH in the hot-halo mode, the efficiency of coupling between the mechanical power of the AGN and the hot halo of the galaxy, and the cooling-to-heating specific energy ratio threshold above which hot halos form, respectively.
    \item Star formation parameters: $\nu_{\rm SF}$ (Eq. (7) in \citetalias{Lagos18c}), which controls the conversion efficiency between the surface density of molecular gas and SFR; {  and the efficiency boost in SBs, $\eta_{\rm burst}$ ($\S$~4.4.3 in \citetalias{Lagos18c}), which controls how much higher the star formation efficiency is during SBs, $\eta_{\rm burst}\nu_{\rm SF}$.} 
    \item Stellar feedback parameters: $\beta$ (Eqs. (25)-(28) in \citetalias{Lagos18c}) and $\beta_{\rm min}$ (Appendix~\ref{AppendixMinorModifications}). These parameters control the dependence of the outflow mass loading on the circular velocity of the galaxy, and the minimum value the mass loading can take, respectively.
    \item Disk instability parameters: $\epsilon_{\rm disc}$ (Eq. (35) in \citetalias{Lagos18c}), which controls the threshold of the stability parameter, below which disks are considered globally unstable to collapse.
    \item Gas reincorporation parameters: $\tau_{\rm reinc}$, $\gamma$ and $M_{\rm norm}$ (Eq.~(30) in \citetalias{Lagos18c}), control how quickly gas that has been removed from halos due to stellar feedback can be reincorporated into the halo. Once gas is reincorporated is available for cooling.
\end{itemize}

Other parameters are left fixed, following the values in Table~\ref{tab:parameters}. The ranges in which we vary the parameters above and the best-fitting values are presented in Table~\ref{tab:parameters}. {  Parameters that are not mentioned in Table~\ref{tab:parameters} but appear in Table~3 in \citetalias{Lagos18c} have been left unchanged from those adopted in \citetalias{Lagos18c}.}

Even though we only use the $z=0$ SMF as a numerical constraint for {\tt optim}, we visually inspected other results of the model to ensure we were not seriously compromising their agreement with observations by using the $z=0$ SMF only. For example, we inspected gas scaling relations, the mass-metallicity relation and the $z<2$ CSFRD and made sure they look sensible. Some of these results are presented in the supplementary material. 

Note that the above does not mean that we change the parameters from the best-fitting ones, but that we used the visual inspection of other results to draw reasonable prior ranges in some of the parameters that we input to {\tt optim}. An example of that, is that we limited the {\tt stable} parameter (see Table~\ref{tab:parameters}) to a maximum of $1.2$. Higher values lead to unphysically large numbers of bulge-dominated galaxies.

\subsection{Other technical updates}

Many small bug fixes and improvements have been made in the \shark\ codebase
since version 1.1.0 was released. Below, we highlight the most important ones:

\begin{itemize}
    \item The memory footprint of \shark{} has been considerably reduced
    by reorganising some of its internal structures,
    with improvements of about $20$\%, depending on the input data.
    \shark{} is usually memory bound,
    and thus this reduction allows more \shark{} instances to run
    in the same set of resources.
    \item Executions are now fully reproducible. By default a random seed is used, but one can be given. The seed used by each run is logged and stored in the output files.
    \item Performance has been improved in two fronts:
    first, we now perform better load balancing when running in multi-threaded mode, reducing walltime; secondly, an overall reduction of small, temporary memory allocations
    allow \shark{} to run more streamlined.
    \item Improved infrastructure:
    dropped requirement for the HDF5 C++ libraries,
    moved our automatic per-commit checks from Travis CI to GitHub actions,
    added support for more systems and compilers (e.g. MSVC in Windows), more runtime information is written (timing, memory usage, etc).
\end{itemize}

\section{Characteristics of the galaxy population: abundances and scaling relations}\label{basicresults}

\begin{figure*}
\begin{center}
\includegraphics[trim=1mm 1mm 1mm 1mm, clip,width=0.99\textwidth]{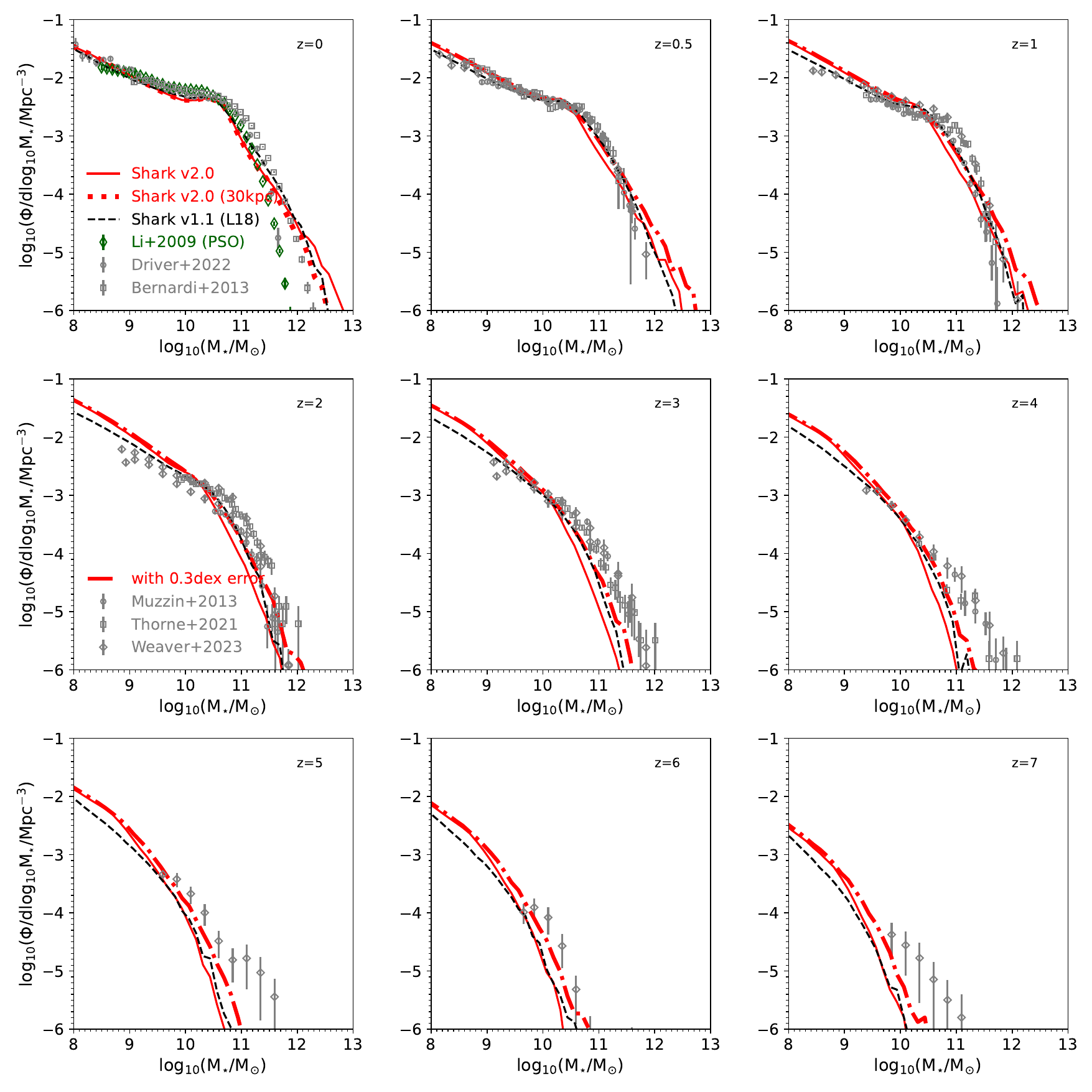}
\caption{Galaxy SMF in \shark\ v2.0 (solid red line) and the default model in \citetalias{Lagos18c} (dashed black line), from $z=0$ to $z=8$, as labelled. We show observations from \citet{Li09, Bernardi13,Driver22} at $z=0$, \citet{Muzzin13,Thorne21,Weaver22} at $0.5\le z\le 4$, and from \citet{Weaver22} only at $z\ge 5$, as labelled. \citet{Li09}, shown in green in the top-left panel, is the constraint used in {\tt optim}. \shark\ v2.0, we also show the SMF using the stellar mass in galaxies enclosed in an aperture of $30$~kpc (red dotted line, shown only at $z=0$), and when we apply a random error in the stellar mass of $0.3$~dex at $z\ge 1$ (red dot-dashed line).}
\label{SMF}
\end{center}
\end{figure*}

\begin{figure*}
\begin{center}
\includegraphics[trim=12mm 12mm 15mm 15mm, clip,width=0.75\textwidth]{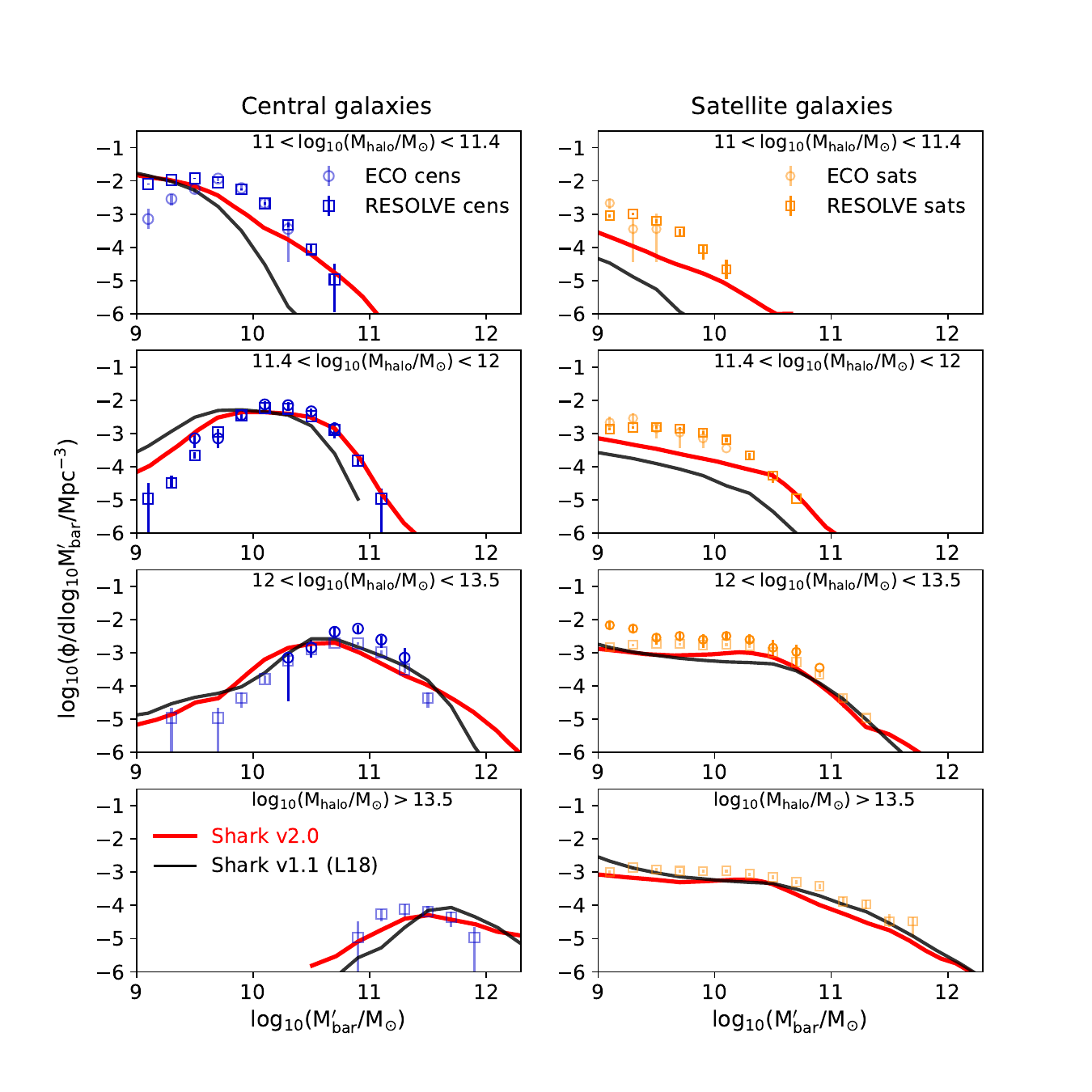}
\caption{Galaxy baryon mass function in \shark\ v2.0 (red lines) and the default model in \citetalias{Lagos18c} (black lines), as labelled, at $z=0$ for central (left) and satellite (right) galaxies in $4$ bins of host halo mass, as labelled in each panel. Observations from the galaxy surveys RESOLVE and ECO from \citet{Eckert16} are shown as symbols for central galaxies (blue symbols) and satellites (yellow symbols). The baryon mass here is defined in the same way as in \citet{Eckert16}, $\rm M{^\prime}_{\rm bar} = M_{\star} + 1.4\,M_{\rm HI}$, with $M_{\rm HI}$ being the atomic hydrogen mass of the galaxy.}
\label{BMF}
\end{center}
\end{figure*}

In this section we discuss fundamental observed properties of galaxies, comparing the results of the new \shark\ v2.0 model using the best-fitting parameters of Table~\ref{tab:parameters} with the default model presented in \citetalias{Lagos18c}. 
%Angular momentum, sizes evolution.

\subsection{Galaxy properties: abundances}

\subsubsection {The stellar mass function}

Fig.~\ref{SMF} shows the SMF from $z=0$ to $z=7$ comparing with observations. For $z=0$ we show the observational estimates of \citet{Li09} which were used for the parameter tuning, and those of \citet{Bernardi13,Driver22} to show the large systematic uncertainties that permeate the SMF, especially at the high-mass end. 
%By construction, \shark\ v2.0 provides a good fit to the observations of \citet{Li09}, better than was achieved in \citetalias{Lagos18c}. This is not only due to the improved automatic parametrisation of the model, but also to the improved physics. Proctor et al (in preparation) show that when using \shark\ v1.1 in tandem with the PSO {\sc optim} package, we fail to provide a fit as good at the one found here for \shark\ v2.0. This boils down to the inability of the default AGN feedback model adopted in \citetalias{Lagos18c} to produce a steep-enough high-mass end of the SMF. 
{  By construction, \shark\ v2.0 provides a good fit to the observations of \citet{Li09}, similarly to \shark\ v1.1. However, as shown hereafter, \shark\ v2.0 significantly outperforms its predecessor in more advanced metrics, such as kinematic scaling relations ($\S$~\ref{sec:scalingrels}) and estimators of quenching ($\S$~\ref{basicresults2}). As a technical improvement to \shark\ v1.1, where model parameters have been tuned by visual inspection, we are now fitting the parameters with an automated and statistically robust method (Proctor et al., in preparation).}

At $z\ge 0.5$, \shark\ v2.0 tends to predict a steeper low-mass end of the SMF compared to \citetalias{Lagos18c}, due to the adopted parameters leading to a weaker dependence of the mass loading on the circular velocity of the galaxy (see Fig.~\ref{fig:ml_vcirc}). At the high-mass end, \shark\ v2.0 produces a steeper break of the SMF than \citetalias{Lagos18c}, and lower-mass galaxies at fixed number density.
%, in better agreement with the observations, at least at $z=0.5,\,1$. 
At $z\ge 2$, the high-mass end in \shark\ v2.0 produces too low number densities of massive galaxies, $>10^{11}\,\rm M_{\odot}$, but this is easily alleviated if we assume stellar masses have uncertainties which are Gaussian-distributed with a width of $\approx 0.3$~dex (see dot-dashed red lines). Uncertainties of that magnitude are quite common in deriving stellar masses due to the required assumptions (e.g. SFH, metallicity history, initial mass function, among others) - see \citet{Marchesini09} for a thorough list of uncertainties and \citet{Robotham20} for a recent example of the uncertainties associated to $z\approx 0$ stellar masses.

At $z\ge 5$, the SMF starts to behave more like a power-law rather than a Schechter function, lacking a clear stellar mass break. This is accentuated when we include random errors to the \shark\ stellar masses. Observations appear to behave similarly, with a clear stellar mass break present only up to $z\approx 3-4$. Note also that the low-mass end of the SMF becomes increasingly steeper with increasing redshifts in \shark\ (both v2.0 and \citetalias{Lagos18c}). The observations appear to follow a similar trend. Better observational constraints at the low-mass end, however, are needed to confirm this trend. 

\subsubsection{The baryon mass function of satellite and central galaxies}

\begin{figure}
\begin{center}
\includegraphics[trim=3mm 2mm 2mm 15mm, clip,width=0.48\textwidth]{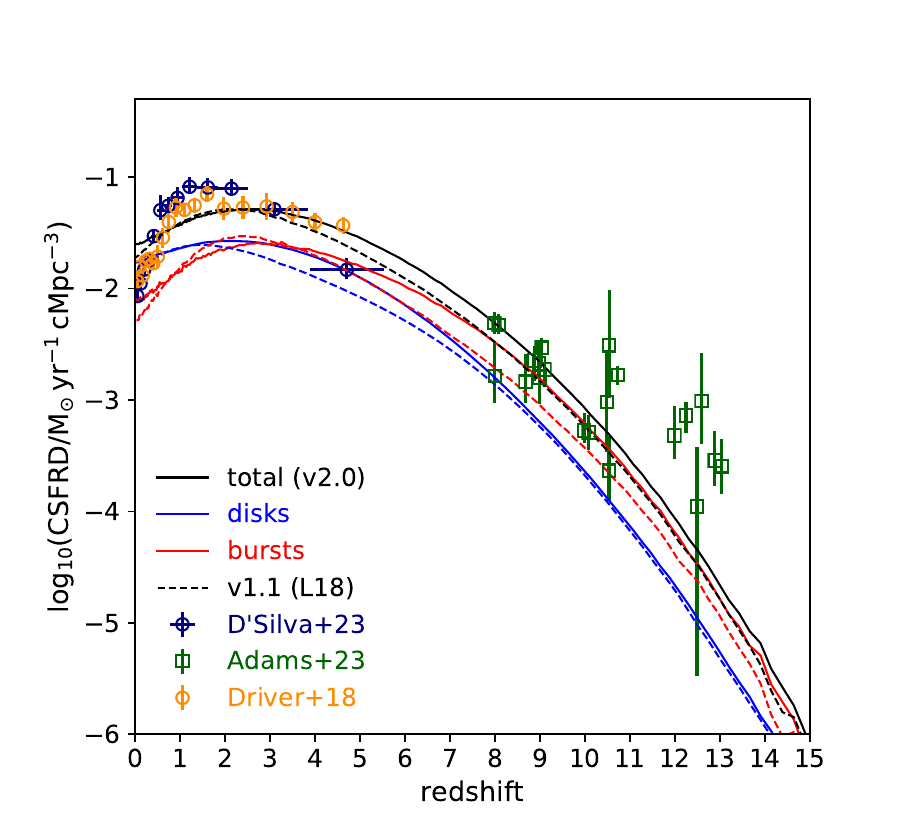}
\includegraphics[trim=3mm 2mm 2mm 15mm, clip,width=0.48\textwidth]{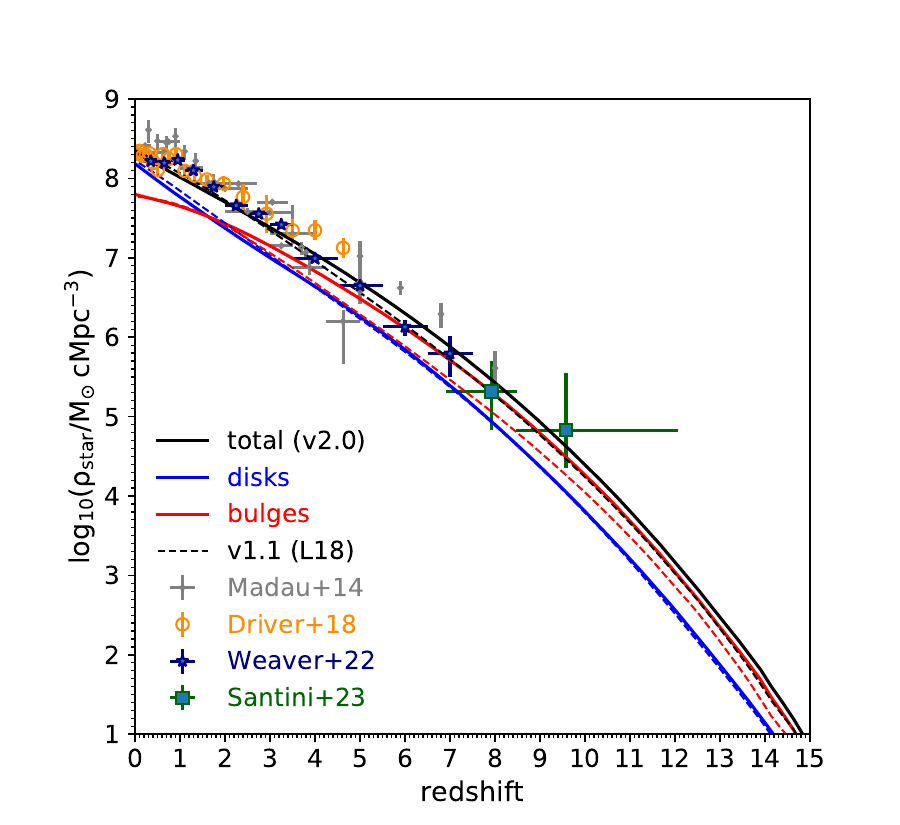}
\caption{Cosmic SFR (top panel) and stellar mass (bottom panel) density as a function of redshift. We show the total contribution of all galaxies in black, galaxy disks in blue and galaxy bulges in red, for the \shark\ v2.0 default model (solid lines) and the default model in \citetalias{Lagos18c} (dashed lines). In the top panel we show the observations from \citet{Driver17,D'Silva23} and the compilation of \citet{Adams23}, as labelled. The latter includes results from \citet{Oesch18,Harikane23,Bouwens23XDF,Bouwens23JWST,Donnan23,Perez-Gonzalez23} and their own measurements, most of them using the JWST. In the bottom panel we show observational constraints from \citet{Madau14,Driver17,Weaver22,Santini23}. Data from \citet{Madau14} has been re-scaled to a Chabrier IMF.}
\label{CosmicSFR}
\end{center}
\end{figure}

Fig.~\ref{BMF} shows the baryon mass function in bins of halo mass, separating centrals and satellite galaxies at $z=0$. To allow for a fair comparison with the observations of \citet{Eckert16}, we define the baryon mass as the sum of the stellar mass and the HI mass times $1.4$. The latter factor was introduced by \citet{Eckert16} to account for the Helium contribution. Note that \citet{Eckert16} ignored the contribution of molecular and ionised gas to the baryon mass. This is a reasonable assumption at $z=0$ given the small molecular gas mass fractions in gas-rich galaxies (which tend to be low-mass galaxies) and the overall low gas fractions in the regime where molecular and atomic gas make a comparable contribution to the gas content of galaxies (i.e. massive galaxies) \citep{Catinella18}. The observations of \citet{Eckert16} were presented for two complementary surveys RESOLVE and ECO; the former is more sensitive but of smaller area than the latter. We show the results from these two surveys using different symbols.

In low-mass groups, $11<{\rm log}_{10}(M_{\rm halo}/\rm M_{\odot})<11.4$ and $11.4<{\rm log}_{10}(M_{\rm halo}/\rm M_{\odot})<12$ {  (top and second rows of panels)}, we see that \shark\ v2.0 is able to produce a much higher number density (by an order of magnitude) of satellite galaxies across the dynamic range probed by the observations compared to \citetalias{Lagos18c}. This is in large part driven by the inclusion of the new dynamical friction timescale of \citet{Poulton2021}, which leads to much longer dynamical friction timescales in low-mass galaxies than what is obtained by using \citet{Lacey93}, allowing them to survive for longer and exist in bigger quantities. 
%Another important contributor to this difference are the parameters adopted for the stellar feedback leading to an overall lower efficiency.
In addition, \shark\ v2.0 recovers a baryon mass function for central galaxies that matches the high-mass end of the observations much better than \citetalias{Lagos18c}. This has to do with the modified parameters of stellar feedback, which make it overall less efficient in the new model. 

The overall higher number density of satellite galaxies in \shark\ v2.0 compared to \citetalias{Lagos18c} is still present in halos of masses $12<{\rm log}_{10}(M_{\rm halo}/\rm M_{\odot})<13.5$ but the difference is much smaller than that seen at lower halo masses. This is because the \citet{Poulton2021} dynamical friction timescales get closer to those of \citet{Lacey93} for more massive satellites. In the highest halo mass bin, ${\rm log}_{10}(M_{\rm halo}/\rm M_{\odot})>13.5$, satellite galaxies in \shark\ v2.0 display a sharper high-end cut off than \citetalias{Lagos18c}, which agrees better with observations. For central galaxies, the main difference is the slightly higher baryon masses produced by \shark\ v2.0 in the  $12<{\rm log}_{10}(M_{\rm halo}/\rm M_{\odot})<13.5$ halo mass bins. Observational errors are quite large in this regime, so within the uncertainties, both \shark\ v2.0 and \citetalias{Lagos18c} agree with observations reasonably well. Note that here we have not included any potential confusion between the tagging of central and satellite galaxies, which tends to be rather large, especially in low-mass halos (see discussion in \citealt{Bravo20}).

\subsection{Star formation rate and stellar mass cosmic density evolution}

Fig.~\ref{CosmicSFR} shows the cosmic SFR and stellar mass density evolution from $z=0$ to $z=15$. Observational constraints from \citet{Madau14,Driver17,D'Silva23,Adams23,Weaver22,Santini23} are shown. For the CSFRD, \shark\ v2.0 predicts higher densities at $z\gtrsim 3$ than \citetalias{Lagos18c}, driven by both the contribution from SBs and star formation in disks being higher in the new model than \citetalias{Lagos18c}. 
%The difference between \shark\ v2.0 and \citetalias{Lagos18c} for the CSFRD in SBs is appreciably larger than the difference in the CSFRD in disks. 
For the CSFRD in disks, we see that at $z\gtrsim 8$, \shark\ v2.0 and \citetalias{Lagos18c} predict very similar levels, but the differences remain for the CSFRD in SBs from $z\approx 3.5$ even up to $z=15$. The overall qualitative trend of SBs dominating the CSFRD at $z\gtrsim 3$ that was presented in \citetalias{Lagos18c} is maintained in \shark\ v2.0. The new model compares more favourably with observations in general, giving a higher CSFRD at $z\gtrsim 4$, closer to 
%
%closer to $0.1\,\rm M_{\odot}\,yr^{-1}\,cMpc^{-3}$, and a higher CSFRD at higher redshifts that are closer 
to current JWST estimates beyond $z=8$ within the uncertainties. At $z\lesssim 4$ \shark\ v2.0 and v1.1 have similar CSFRDs. 

The overall higher CSFRD at high redshift in \shark\ v2.0 compared to \citetalias{Lagos18c} naturally leads to a cosmic stellar mass density (CSMD) in the early universe that is higher in \shark\ v2.0 than \citetalias{Lagos18c}. It is interesting that just the SB contribution to the CSFRD and CSMD in \shark\ v2.0 is similar or even higher than the total in \citetalias{Lagos18c} at $z\gtrsim 7$. This shows that these predictions are still subject to large variations and that better constraints from the high redshift universe are needed to narrow down the parameter space that is plausible in the early universe.  

%analysing the entire GAMA and COSMOS surveys, and from the JWST are also included. 
%supplementary material: HI, H2 MF, LFs as shown in Lagos et al. 2019.

\subsection{Galaxy scaling relations}\label{sec:scalingrels}

\subsubsection{The stellar-halo mass relation}
\begin{figure}
\begin{center}
\includegraphics[trim=5mm 4.5mm 3mm 4mm, clip,width=0.499\textwidth]{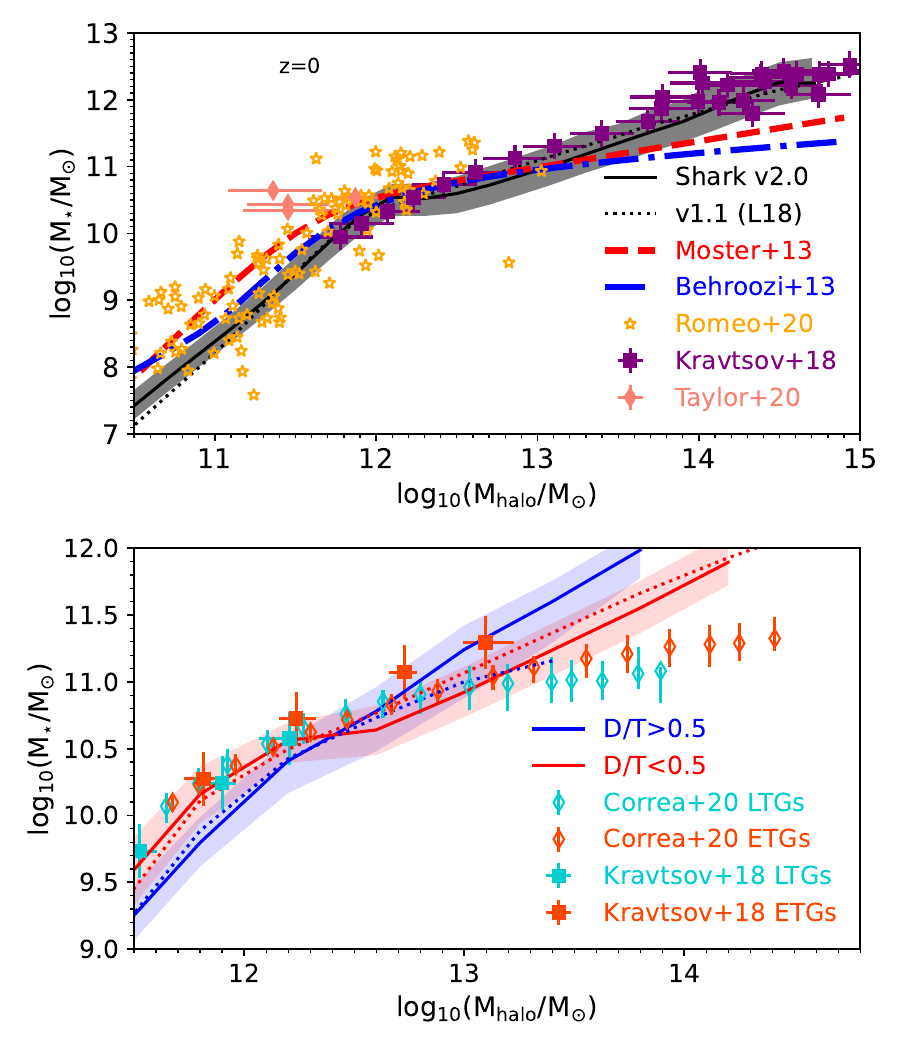}
\caption{{\it Top panel:} The stellar-halo mass relation for central galaxies in \shark\ v2.0 (solid black line) and \citetalias{Lagos18c} (dashed black line) at $z=0$. Lines show the median relation for bins with $\ge 10$ galaxies. For \shark\ v2.0 we also show the $16^{\rm th}-84^{\rm th}$ percentile ranges as the shaded region. For reference, we show the empirical constraints from \citet{Moster13,Behroozi13}, and the more direct observational constraints from \citet{Kravtsov18,Taylor20,Romeo20}, as labelled. {\it Bottom panel:} As in the top panel but splitting galaxies between bulge- (red) and disk-dominated (blue) galaxies, using a disk-to-total stellar mass ratio (D/T) of $0.5$. Galaxies above (below) the threshold are considered disk-(bulge-)dominated. Observational constraints from \citet{Kravtsov18} and \citet{Correa20} for late- and early-type galaxies, LTGs and ETGs, respectively, are also shown, as labelled.} 
\label{SMHM}
\end{center}
\end{figure}

The top panel of Fig.~\ref{SMHM} shows the stellar-halo mass relation for central galaxies at $z=0$ for \shark\ v2.0 and \citetalias{Lagos18c}. At the low-halo mass end, $\rm log_{10}(M_{\rm halo}/M_{\odot})<12$, \shark\ v2.0 produces slightly higher stellar masses at fixed halo mass than \citetalias{Lagos18c}, in better agreement with empirical constraints from abundance matching. This is the direct result of the stellar feedback parameters adopted in \shark\ v2.0 compared to \citetalias{Lagos18c} (see Table~\ref{tab:parameters}). At $\rm 12<log_{10}(M_{\rm halo}/M_{\odot})<14$, \shark\ v2.0 produces smaller stellar masses at fixed halo mass than \citetalias{Lagos18c}, while at higher halo masses both models converge to similar values. The difference seen is the result of the new AGN feedback model leading to a more efficient quenching of central galaxies in massive halos in \shark\ v2.0 (discussed in detail in \S~\ref{basicresults2}). Note that this is not just due to the choice of parameters, but the capability of the new AGN feedback model to more efficiently quench massive galaxies overall. Proctor et al. (in preparation) show that for the AGN feedback, the implementation introduced in \citetalias{Lagos18c} was not capable of getting a very sharp SMF break at the high-mass end, while the new model allows for that. 

Both \shark\ v2.0 and \citetalias{Lagos18c} produce a stellar-halo mass relation at $\rm log_{10}(M_{\rm halo}/M_{\odot})>12$ that is steeper than the empirical relations of \citet{Moster13,Behroozi13}. However, compared to the more direct estimates of \citet{Kravtsov18}, \shark\ compares favourably, especially at the galaxy cluster regime. \citet{Kravtsov18} argued that the difference with the empirical relations of \citet{Moster13,Behroozi13} is due to those studies using SMFs that are severely surface brightness-limited, while the \citet{Kravtsov18} data reaches lower surface brightness values, impacting the recovered stellar mass. We also show the observational measurements of \citet{Taylor20} combining weak lensing with stellar mass measurements from the GAMA survey. These again are significantly different than the empirical relations of \citet{Moster13,Behroozi13}, showing that there are still many systematic uncertainties in the stellar-halo mass relation which are usually not captured in the presented errorbars. Individual local Universe galaxies (primarily of late-type morphology) compiled by \citet{Romeo20} are also shown, depicting the significant galaxy-to-galaxy scatter of the relation.

The bottom panel of Fig.~\ref{SMHM} shows the stellar-halo mass relation of central galaxies in \shark\ v2.0 and v1.1 for disk- and bulge-dominated galaxies, separately. We define these populations using a disk-to-total stellar mass threshold of $0.5$. In both \shark\ versions, and at fixed halo mass, bulge-dominated galaxies have higher stellar masses than disk-dominated ones at $M_{\rm halo}<10^{12.5}\,\rm M_{\odot}$. This is due to bulge-dominated galaxies forming earlier than disk-dominated galaxies and having had more starbursts during their lifetimes. 
%Note that selecting galaxies with increasing disk-to-total stellar mass ratios leads to even lower stellar masses at fixed halo mass (compare the solid and dashed blue lines in the bottom panel of Fig.~\ref{SMHM}). 
%For bulge-dominated galaxies, changing the exact threshold to define them makes a negligible difference. 
At $M_{\rm halo}>10^{12.5}\,\rm M_{\odot}$, the trend reverses in \shark\ v2.0 but not in v1.1, with disk-dominated galaxies in v2.0 being more massive at fixed halo mass than bulge-dominated galaxies. This happens because AGN feedback in v2.0 is not efficient enough in the disk-dominated galaxies of these high halo masses, so they form stars very efficiently and end up more massive than the bulge-dominated ones. In \shark\ v1.1, the efficiency of AGN feedback is more directly tied to the halo mass than in the new model, leading to the differences seen at $M_{\rm halo}>10^{12.5}\,\rm M_{\odot}$. 
%Also note that \shark\ v2.0 produces a negligible number of disk-dominated galaxies at $\rm log_{10}(M_{\rm halo}/M_{\odot})>12.7$, so the medians for that population stop at that halo mass scale.

Interestingly, \citet{Correa20} reported 
%the opposite trend for 
that in the \eagle\ cosmological hydrodynamical simulations, disk-dominated galaxies have a {\it higher} stellar masses than bulge-dominated galaxies at all halo masses. 
We find that the difference with the trend in \shark\ v2.0 is driven by disk instabilities\footnote{{  In \shark, the treatment of disk instabilities is described in $\S$~4.4.8 in \citetalias{Lagos18c}. In short, a stability parameter is computed with the properties of the galaxy disk, and if this is below a given threshold, then the disk of that galaxy is assumed to be fully destroyed, with all the gas and stellar content being moved to the bulge.}} in the model. If we completely switch off disk instabilities, we find that the resulting disk-dominated galaxies have on average higher stellar masses than bulge-dominated ones at fixed halo mass (not shown here), trend that qualitatively agrees with \eagle. This shows that the exact treatment of disks during these episodes of global instabilities has an important impact on the resulting stellar-halo mass relation. This agrees with the conclusions of \citet{Romeo20} who argue that disk gravitational instabilities regulate the stellar-to-halo mass relation in galaxies in the halo mass range of their sample (see orange stars in Fig.~\ref{SMHM}).

Observations still have uncertainties that are too large to be able to confidently verify the conflicting predictions of the stellar-halo mass relation by galaxy morphology. In fact, \citet{Correa20} show, using data from the Sloan Digital Sky Survey (SDSS), that depending on how stellar masses and morphologies are measured, one can get different offsets between bulge- and disk-dominated galaxies in the stellar-halo mass relation. Future surveys, such as the 4MOST WAVES \citep{Driver19WAVES}, will provide exquisite measurements of halo masses down to $\rm log_{10}(M_{\rm halo}/M_{\odot})\approx 12$ in the local Universe, which will allow a robust diagnosis of the predictions discussed here.

%measurements with better measurements of both stellar and halo mass are 
\begin{figure}
\begin{center}
\includegraphics[trim=5mm 5mm 4mm 4mm, clip,width=0.44\textwidth]{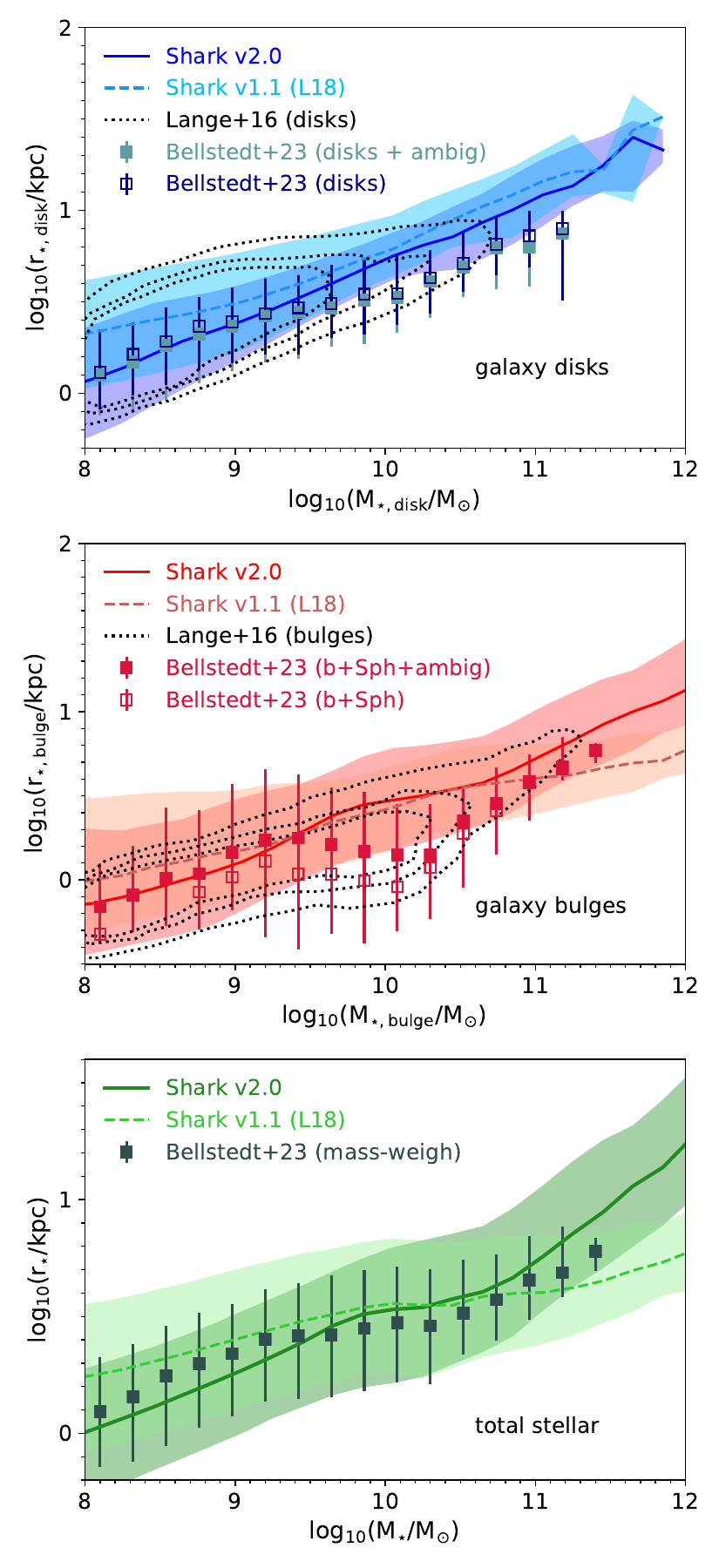}
\caption{Size-stellar mass relation for the disk (top panel) and bulge (middle panel) components of galaxies, and for the entire galaxy in the bottom panel, at $z=0$, for \shark\ v2.0 (solid lines) and \citetalias{Lagos18c} (dashed lines). In the top two panels, 
we plot disk (bulge) half-stellar mass radii vs. disk (bulge) stellar mass, as labelled, while the bottom panel shows the galaxy half-stellar mass radius vs total stellar mass. We show bins with $\ge 10$ objects.
Lines with the shaded regions show the medians and $16^{\rm th}-84^{\rm th}$
percentile ranges, respectively.
%Solid and dashed lines show \shark\ v2.0 and \citetalias{Lagos18c}, respectively. 
Dotted lines in the top two panels show the $50^{\rm th}$, $68^{\rm th}$ and $90^{\rm th}$ percentile regions of the GAMA observations of \citet{Lange16}, while the symbols show the more recent GAMA results of \citet{Robotham22,Bellstedt23} (see Appendix~\ref{SecSizesGAMA} for details). For the latter we show the medians as symbols, and the  $16^{\rm th}-84^{\rm th}$ percentile ranges as errorbars.
%, after improvements in the photometry and the profile/SED fitting of galaxies. For the bulges, we show the size-mass relation of \citet{Bellstedt23} when we include bulges, spheroids, and include or ignore galaxies classified as ``ambiguous" (see text for details).
The bottom panel shows the mass-weighted stellar size-mass relation of  \citet{Robotham22,Bellstedt23} (see Appendix~\ref{SecSizesGAMA} for how this was computed).}
%For the latter the symbols show the medians, and the errorbars the $16^{\rm th}-84^{\rm th}$
%percentile ranges.}
\label{Sizesz0}
\end{center}
\end{figure}

\begin{figure*}
\begin{center}
\includegraphics[trim=5mm 5mm 1.5mm 1.5mm, clip,width=1\textwidth]{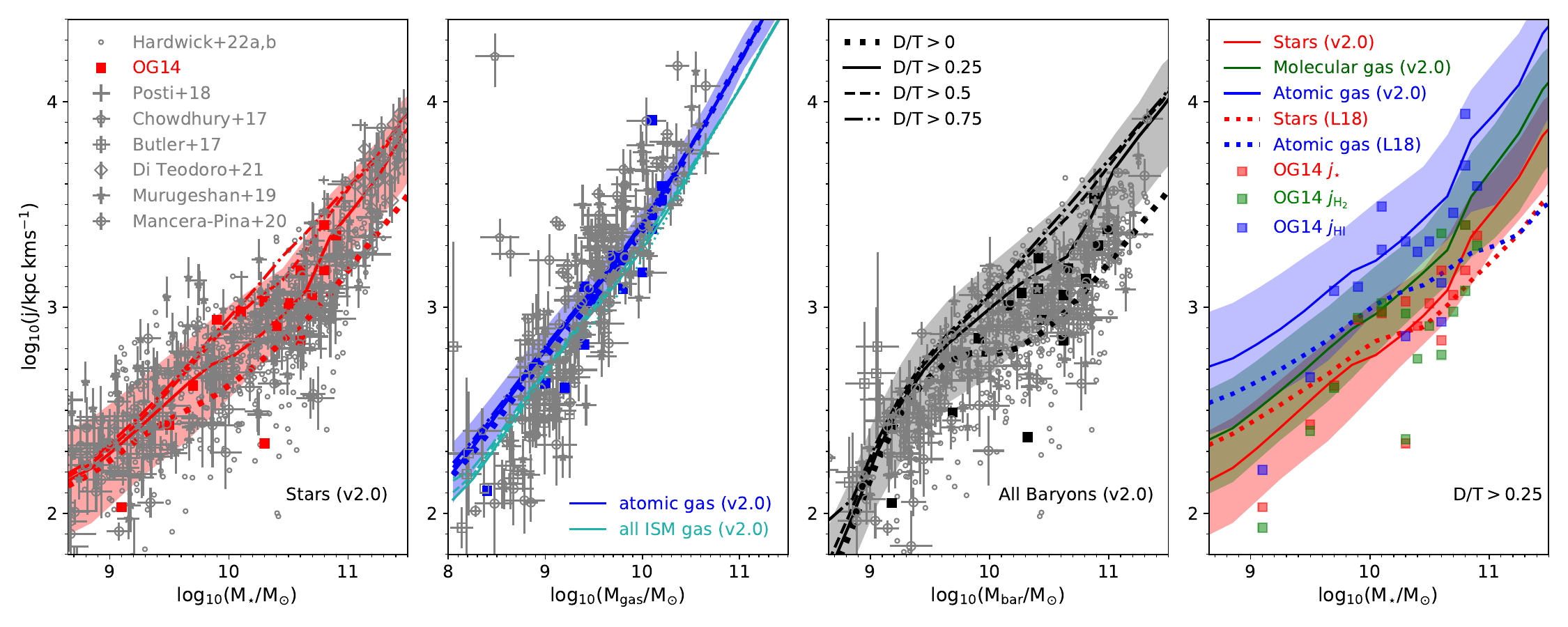}
\caption{sAM of the stars as a function of stellar mass (left panel), sAM of the total ISM and atomic gas a function of total ISM mass (middle-left panel), sAM of all baryons in the galaxy vs baryon mass (middle-right panel), and sAM of atomic, molecular gas and stars vs stellar mass (right panel), for galaxies in \shark\ v2.0 at $z=0$. In the first three panels we show the median relations for galaxies with varying disk-to-total stellar mass ratios, as labelled in the third panel. The shaded region shows the $16^{\rm th}-84^{\rm th}$ percentile ranges for galaxies with $\rm D/T>0.5$ only. 
The fourth panel shows the median sAM-stellar mass relations for galaxies with $\rm D/T>0.25$. In this panel we also show the median relations of \shark\ v1.1 (\citetalias{Lagos18c}) for $j_{\star}$ and $j_{\rm HI}$ (in this model $j_{\rm H_2}\equiv j_{\rm HI}$), as dotted lines. 
We show observations from  \citet{Obreschkow14b,Chowdhury17,Posti18,Murugeshan19,Mancera-Pina21,DiTeodoro21,Hardwick22a,Hardwick22b}, as labelled. We highlight the measurements of \citet{Obreschkow14b} (OG14) in all panels using coloured symbols - this is the only sample where measurements of the sAM of stars, atomic and molecular gas are presented for the same galaxies (shown with different colours in the right-hand panel, as labelled).}
\label{AMz0}
\end{center}
\end{figure*}
\subsubsection{Structural properties of galaxies at $z=0$}\label{structuresec}

One of the major modifications in \shark\ v2.0 is the more sophisticated modelling of galaxy angular momentum, and the exchange of angular momentum between the ISM and stars ($\S$~\ref{newjmodel}). Hence, it is natural to explore what the effect of that is on structural scaling relations of galaxies. Here, we focus on the size-mass and the specific angular momentum-mass relations at $z=0$.

We start with the size-mass relation presented in Fig.~\ref{Sizesz0}. 
The latter is shown for the disk and bulge components of galaxies at $z=0$ for \shark\ v2.0 and \citetalias{Lagos18c}, compared with observations from \citet{Lange16,Robotham22,Bellstedt23}, in the top two panels. The bottom panel shows the half-stellar mass radius, $r_{\star}$, vs galaxy stellar mass compared with the stellar-mass weighted half-mass radius vs stellar mass of \citet{Robotham22,Bellstedt23}. 
Note that \citet{Lange16,Robotham22,Bellstedt23} use the GAMA survey. In Appendix~\ref{SecSizesGAMA} we explain why these measurements are different and the process of measuring the size-mass relation from the \citet{Bellstedt23} catalogue.
Note that for both \shark\ and the observations we use the same Eq.~(\ref{EqRstar}) to calculate $r_{\star}$.
%the latest GAMA photometry \citep{Bellstedt20} and stellar mass estimates \citep{Bellstedt20b}. 

The disk sizes in \shark\ v2.0 are smaller than \citetalias{Lagos18c}, especially at the low-mass end. This is the regime where we expect our new angular momentum treatment to make a difference. As these galaxies have molecular gas that is more concentrated than the atomic gas, the new model would produce new stars that form at overall smaller radii than the old model, which assume all disk components to have the same angular momentum distribution. The lower disk sizes in \shark\ v2.0 agree better with the observations 
%of \citet{Robotham22,Bellstedt23} 
than v1.1 (\citetalias{Lagos18c}).

The bulge size-mass relation (middle panel in Fig.~\ref{Sizesz0}) in \shark\ v2.0 is steeper than in \citetalias{Lagos18c} with a plateau around a bulge mass of $10^{10}\,\rm M_{\odot}$, which is roughly when galaxies transition from being disk- to bulge-dominated. The steeper size-mass relation agrees better with observations than the prediction of \citetalias{Lagos18c}, especially on the regime of massive spheroids. Note that there is a clear systematic difference between the bulge-size mass relations in the observations shown. This is because defining a bulge and its size is more difficult than for disks in observations (see discussion in \citealt{Robotham22}).
%, and hence there is more ambiguity  

The galaxy size-mass relation in the bottom panel of 
Fig.~\ref{Sizesz0} shows that \shark\ v2.0 has a clear transition at a stellar mass of $10^{9.7}-10^{10}\,\rm M_{\odot}$. 
At lower stellar masses, the disk size dominates $r_{\star}$,  while at higher masses it is the bulge size that determines $r_{\star}$. The transition zone was not clearly present in \citetalias{Lagos18c}, in which the size-mass relation was better described by a single power law. Compared with observations, \shark\ v2.0 is in better agreement, although the transition region in observations starts at lower masses.
%and extends for a wider mass range. 
Note, however, that this is not part of the diagnostics used to tune the parameters of the model, so the improvement in the agreement with observations is a success of the new angular momentum model.

%specific AM - mass relations

%size-mass-age relations for disks and bulges

%supplementary material: HI, H2 scaling relations, and gas metallicity scaling relations as shown in Lagos+2018

Fig.~\ref{AMz0} shows the stellar specific angular momentum (sAM) vs stellar mass, atomic gas and all ISM sAM vs gas mass, baryon sAM vs baryon mass, and the sAM of stars, atomic and molecular gas as a function of stellar mass. For the first three cases, we show the relation for galaxies of varying disk-to-total stellar mass ratio, D/T, as labelled in the third panel, while for the fourth panel we show the sAM-stellar mass relations for galaxies with a $\rm D/T > 0.5$. We compare with a large compilation of observations as labelled. 

The stellar sAM-mass relation (left panel of Fig.~\ref{AMz0}) shows a strong dependence of the zero-point of the relation on D/T, so that galaxies with lower D/T have a lower sAM at fixed stellar mass. This is the well known morphological dependence of the sAM-mass relation, first discussed in \citet{Fall83}. Overall, \shark\ v2.0 agrees very well with the observations, and even the scatter for disk-dominated galaxies ($\rm D/T>0.5$) is similar to the scatter reported in observations, which for the most part contain disk galaxies only. The baryon sAM-mass relation (middle-right panel of Fig.~\ref{AMz0}), shows a similar behavior, but the scatter increases with increasing baryon mass in a way that resembles what observations report. The difference in baryon sAM between galaxies with $\rm D/T>0.75$ and $\rm D/T > 0$ is larger than the differences obtained in stellar sAM. 
The atomic (or total ISM) sAM-mass relation (middle-left panel of Fig.~\ref{AMz0}) is the tightest of all, in agreement with what has been reported in observations. However, we see that \shark\ v2.0 produces a slightly too shallow relation. This slope directly depends on what we assume for the halo spin parameter. \citetalias{Lagos18c} assumed a halo spin distribution with a mode that was independent of halo mass. In \shark\ v2.0, we instead assume a very weak dependence of the mode of the spin distribution on halo mass as reported in \citet{Kim15AM}. A slightly steeper relation would help the model reproduce the observed slope, however, we decide not to change the spin-halo mass dependence arbitrarily. Ideally, we would want to use the spin parameter directly inferred from the $N$-body simulation. However, such measurements are very noisy and only reliable for halos resolved with several hundred particles, which prohibits the use of that in SAMs, in which we use halos sampled with a number of particles as low as $20$. 

The right panel of Fig.~\ref{AMz0} illustrates the power of the new angular momentum exchange model ($\S$~\ref{newjmodel}) by showing the difference between the sAM of the atomic and molecular gas, and the stars at fixed stellar mass. The offset between the sAM of atomic and molecular gas increases slightly as the stellar mass decreases. We also show observations from \citet{Obreschkow14b}, which include $j_{\star}$, $j_{\rm HI}$, and $j_{\rm H_2}$ for the same sample of galaxies (THINGS, The HI Nearby Galaxy Survey). The difference between $j_{\star}$ and $j_{\rm HI}$ in THINGS is $\approx 0.3$~dex, while in \shark\ v2.0 is closer to $0.5$~dex. In \shark\ v2.0, $j_{\rm H_2}$ is $\approx 0.15$~dex higher than $j_{\star}$ at fixed stellar mass, while in THINGS this difference is $\approx 0.05$~dex. Nevertheless, within the scatter of the observations, our predictions agree well. We also show the relations for \shark\ v1.1 (\citetalias{Lagos18c}) in dotted lines, and find clear disagreements with the observations, which are most evident for $j_{\rm HI}$, which in \shark\ v1.1 is $\approx 0.2-0.6$~dex lower than \citet{Obreschkow14b}. Note that $j_{\star}$ and $j_{\rm HI}$ in \shark\ v1.1 differ only by $0-0.15$~dex, while observations consistently prefer $\approx 0.3$~dex across the whole stellar mass range probed. By definition, $j_{\rm H_2}=j_{\rm HI}$ in \shark\ v1.1, while in \shark\ v2.0 $j_{\rm HI}$ is higher than $j_{\rm H_2}$ by $0.2-0.5$~dex (with the difference increasing at lower stellar masses), which is similar to the differences seen in THINGS. The improvements we see in the sAM-mass scaling relations are a direct result of the new angular momentum exchange model ($\S$~\ref{newjmodel}).
%The scatter of the gas sAM-stellar mass relations also increases as the stellar mass decreases.

Overall we see that \shark\ v2.0 with the default parameters adopted (Table~\ref{tab:parameters}) produces structural scaling relations that are in broad agreement with observations in the local Universe. We can attribute that in large part to the new  angular momentum exchange model. {  Other SAMs that have implemented more sophisticated treatment for the angular momentum build up of galaxies (e.g. \citealt{Stevens16,Zoldan18}) are also able to reproduce the mass-size and angular momentum-mass relations of galaxies relatively well.}
In the future we also plan to explore other structural scaling relations, such as the size-mass relation across cosmic time, the Tully-Fisher relation, and the relation between sizes-mass and galaxy age reported in \citet{Robotham22}.

\subsubsection{The BH-stellar mass relation and morphological dependence}

\begin{figure}
\begin{center}
\includegraphics[trim=9mm 8mm 12mm 15mm, clip,width=0.49\textwidth]{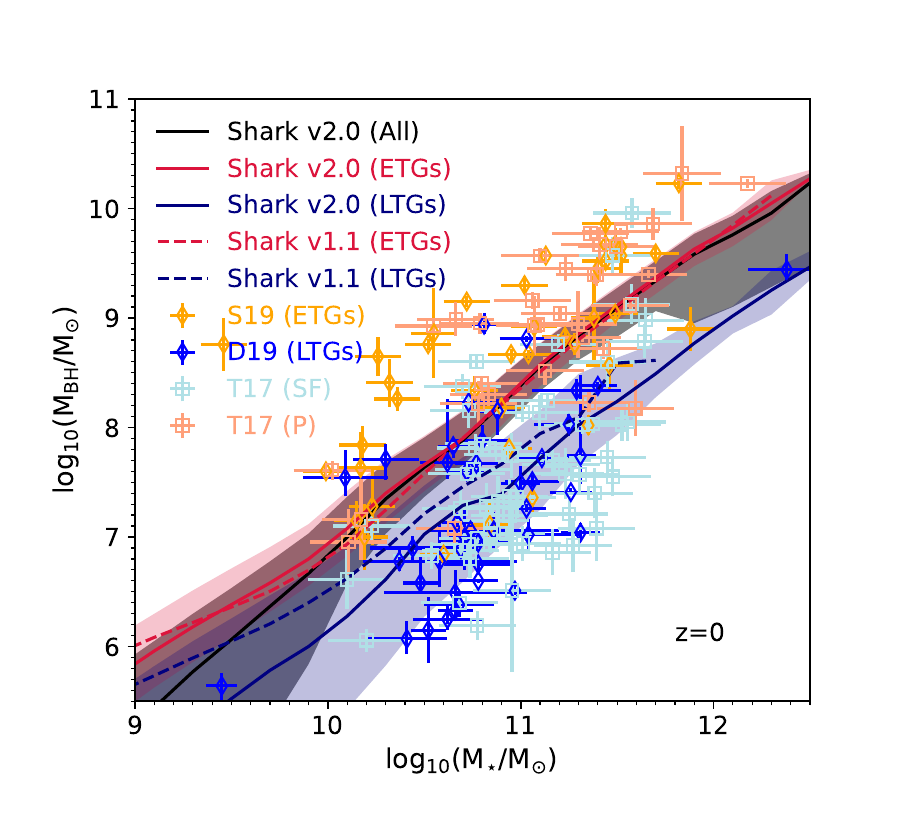}
\caption{BH-stellar mass relation of all galaxies in \shark\ v2.0 at $z=0$ and the subsample of galaxies classified as ETGs (those with a D/T$<0.5$) and LTGs (those with D/T$\ge 0.5$), as labelled. We also show the relation for ETGs and LTGs in \shark\ v1.1 (\citetalias{Lagos18c}) as dashed lines. Observations of ETGs and LTGs from \citet{Sahu19} and \citet{Davis19}, respectively, and from passive (P) and star-forming galaxies (SF) from \citet{Terrazas17} are shown with symbols, as labelled. The BH-stellar mass relation of P and SF galaxies in \citet{Terrazas17} agree well with those of ETGs and LTGs, respectively, of \citet{Sahu19} and \citet{Davis19}.}
\label{BHMSz0}
\end{center}
\end{figure}

Fig.~\ref{BHMSz0} shows the BH-stellar mass relation at $z=0$ for \shark\ v2.0 and v1.1 (\citetalias{Lagos18c}), split by galaxy morphology. We do the latter classification based on D/T, with D/T$<0.5$ corresponding to early-type galaxies (ETGs) and D/T$\ge 0.5$ to late-type galaxies (LTGs). Both \shark\ v2.0 and v1.1 predict LTGs to have systematically lower BH masses than ETGs at fixed stellar mass. The offset in the zero-point of the BH-stellar mass relation between the LTG and ETG populations is larger in \shark\ v2.0 ($\approx 0.5-1.5$~dex) than in v1.1 ($\approx 0.3-0.6$~dex), especially clear at $M_{\star}\lesssim 10^{10.5}\,\rm M_{\odot}$. Another interesting difference is that \shark\ v2.0 has LTGs as massive as $10^{12.2}\,\rm M_{\odot}$ while in v1.1 they at most reach stellar masses of $\approx 10^{11.5}\,\rm M_{\odot}$. 

Observations show that the typical BH mass difference between LTGs and ETGs is $\approx 1$~dex at fixed stellar mass. \shark\ v2.0 predicts a difference of $\approx 0.8-1$~dex in agreement with the observations, while v1.1 prefers smaller differences of $\approx 0.3-0.6$~dex. Both \shark\ versions predict the BH-stellar mass relation of LTGs to have a larger scatter than that of ETGs, however, this difference in scatter between morphological types is larger in \shark\ v2.0 than v1.1. Observations suggest a qualitatively similar difference, with the scatter of the BH-stellar mass relation of LTGs or star-forming galaxies being larger than for ETGs or passive galaxies. We also find that overall the BH-stellar mass relation becomes tighter for both galaxy populations with increasing stellar mass. More observations are needed to establish whether that is also the case in the real Universe.

\section{Quenching of massive galaxies}\label{basicresults2}

One of the major developments in \shark\ v2.0 compared with \citetalias{Lagos18c} is the new AGN feedback model ($\S$~\ref{newagnmodel}). As such, we focus here on analysing the effect this has on the quenching of galaxies, especially massive galaxies. This is also an area that has attracted a lot of attention due to recent results from the JWST, which point to massive-quiescent galaxies being more common than previously thought of at $z>3$ \citep{Carnall23,Nanayakkara22,Valentino23,Long23}. 

This section is organised as follows: $\S$~\ref{sec:quench1} focuses on the quenching of galaxies in the local universe, and $\S$~\ref{sec:quench2} analyses how quenching develops across cosmic time.

\subsection{Quenching in local universe galaxies}\label{sec:quench1}
\begin{figure}
\begin{center}
\includegraphics[trim=3mm 4mm 3mm 3mm, clip,width=0.47\textwidth]{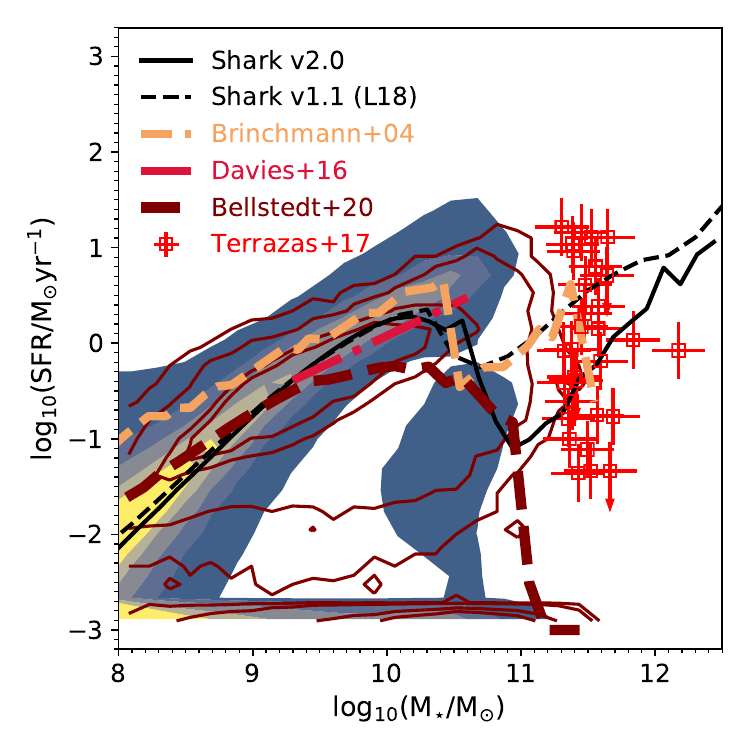}
\caption{SFR as a function of stellar mass for galaxies at $z=0$. The black solid and dashed lines show the medians for the galaxies in \shark\ v2.0 and v1.1 (\citetalias{Lagos18c}), respectively, for bins with $\ge 10$ objects. The filled contours show percentile ranges ranging from $99^{\rm th}$ to $10^{\rm th}$, from the outer to the inner regions, for \shark\ v2.0. The brown, dashed line shows the median SFR-stellar mass relation of SDSS galaxies as reported by \citet{Brinchmann04}; the blue dot-dashed line shows the main sequence as measured from GAMA by \citet{Davies16}; the pink dashed line and thin solid lines show the median and the  $99^{\rm th}$, $95^{\rm th}$, $68^{\rm th}$ and $50^{\rm th}$ percentile ranges of GAMA galaxies from \citet{Bellstedt20b}. Observations of massive galaxies from \citet{Terrazas17} are also shown as red symbols as a reference of typical SFRs of massive galaxies that are above the dynamical range sampled by GAMA. Some of the latter measurements are upper limits, and are indicated by down-pointing arrows.}
\label{MSz0}
\end{center}
\end{figure}

\begin{figure}
\begin{center}
\includegraphics[trim=3mm 8mm 8mm 15mm, clip,width=0.49\textwidth]{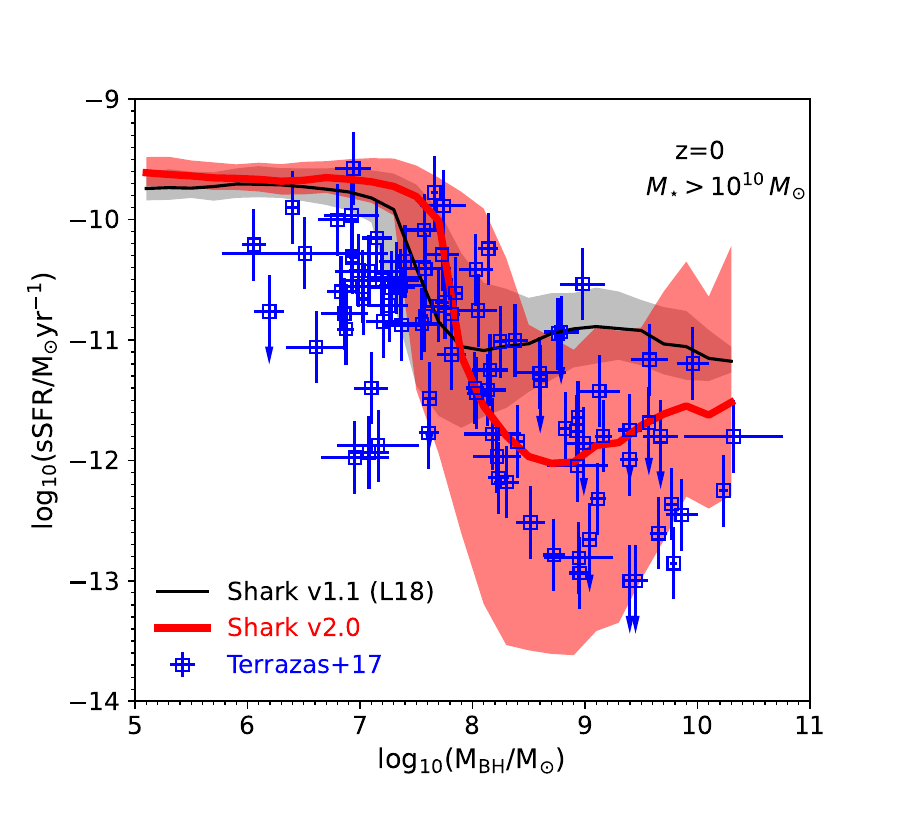}
\caption{Specific SFR as a function of BH mass for galaxies at $z=0$ with stellar masses $>10^{10}\,\rm M_{\odot}$. The results for \shark\ v1.1 (\citetalias{Lagos18c}) and v2.0 are shown as a black line with the grey shaded region, and red line with the faded red shaded region, respectively. Lines show medians and the shaded regions show the $16^{\rm th}-84^{\rm th}$ percentile ranges for bins with $\ge 10$ objects. Observational estimates from \citet{Terrazas17} are shown as symbols.}
\label{BHSSFR}
\end{center}
\end{figure}

\begin{figure}
\begin{center}
\includegraphics[trim=5mm 4.5mm 2mm 4mm, clip,width=0.49\textwidth]{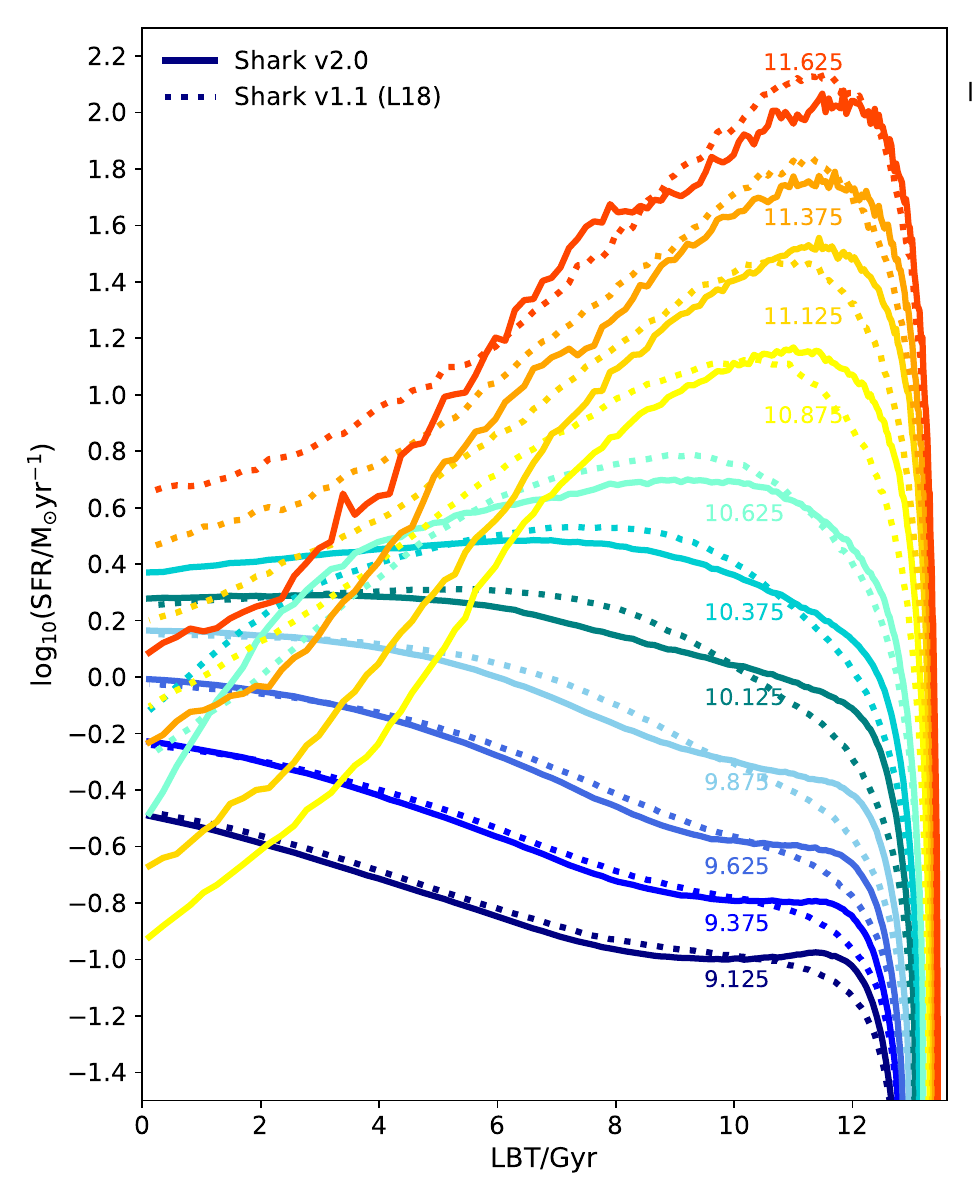}
\caption{The median SFR as a function of lookback time (LBT) of galaxies at $z=0$ in different bins of stellar mass (of width $0.25$~dex) for \shark\ v2.0 (solid lines) and v1.1 (dotted lines). The mean stellar mass of each bin are labelled next to the corresponding lines.} 
%
%as labelled. Stellar mass bins have a width of $0.25$~dex. Lines with shaded regions show the medians and the $16^{\rm th}-84^{\rm th}$ percentile ranges for bins with $\ge 10$ objects. The top and bottom panels show galaxies in \shark\ v1.1 (\citetalias{Lagos18c}) and v2.0, respectively.}
\label{SFH}
\end{center}
\end{figure}

Fig.~\ref{MSz0} shows the SFR-stellar mass plane at $z=0$ in \shark\ v2.0 and \citetalias{Lagos18c}. Observational measurements of the median SFR-stellar mass relation of SDSS galaxies from \citet{Brinchmann04}, the main sequence in GAMA from \citet{Davies16} and the distribution of all GAMA galaxies at $z\le 0.06$ and their median SFR-stellar mass plane distribution from the catalogue of \citet{Robotham20,Bellstedt20b}, are also shown. For the latter, galaxies both on and off the main sequence (MS) are included.

The median SFR at fixed stellar mass is very similar between \shark\ v2.0 and \citetalias{Lagos18c} in galaxies with $M_{\star}<10^{10}\,\rm M_{\odot}$. Above that stellar mass, the models diverge in a way that galaxies in \shark\ v2.0 appear more quenched than those in \citetalias{Lagos18c} by roughly an order of magnitude, at least for galaxies with stellar masses $10^{10.5}-10^{11.5}\,\rm M_{\odot}$. The most massive galaxies, $M_{\star}>10^{12}\,\rm M_{\odot}$, in both versions of \shark\ have similar SFRs, about an order of magnitude below the MS.  

Compared with observations, we see that the MS in both \shark\ versions agree well with the measurements of \citet{Davies16}. There is evidence though that galaxies in GAMA as analysed in \citet{Bellstedt23} start to, on average, deviate from the MS at $M_{\star}\gtrsim 10^{9.5}\,\rm M_{\odot}$. This in \shark\ v2.0 and \citetalias{Lagos18c} happens at higher stellar masses, $M_{\star}\approx 10^{10}\,\rm M_{\odot}$. The deviations from MS in the early SDSS analysis of \citet{Brinchmann04} happens at higher stellar masses, closer to what we get in \shark. Because GAMA's survey area is too small to have a representative sample of very massive galaxies, we also show individual measurements of the SFR and stellar mass of a sample of massive galaxies from \citet{Terrazas17}. Note that some of these measurements are upper limits.

The quenching of massive galaxies is linked to AGN feedback. It is thus natural to explore the connection between quenching and BH mass. Fig.~\ref{BHSSFR} shows the specific SFR ($\equiv \rm SFR/M_{\star}$; sSFR) as a function of the central BH mass of galaxies with $M_{\star}>10^{10}\rm \, M_{\odot}$ at $z=0$ for \shark\ v2.0 and \citetalias{Lagos18c}, compared with the observations of \citet{Terrazas17}. This plane clearly shows a BH mass scale at which quenching happens in \shark. This transition mass for \shark\ v2.0 is $M_{\rm BH}\approx 10^{7.5}\,\rm M_{\odot}$ and for \citetalias{Lagos18c} is $M_{\rm BH}\approx 10^{7.25}\,\rm M_{\odot}$ {  (with the difference in the threshold mass between the two models becoming increasingly larger at high redshift; see Fig.~\ref{BHSSFRevo})}. What is interesting is that above that mass threshold, \shark\ v2.0 galaxies reach much lower sSFRs (i.e. are more quenched) than \citetalias{Lagos18c} galaxies, but also the scatter of sSFR at fixed BH mass is much larger in \shark\ v2.0 than \citetalias{Lagos18c}. The relation obtained by \shark\ v2.0 resembles much better what the observations of \citet{Terrazas17} suggest. \citet{Terrazas20} used this plane to diagnose the way AGN feedback acts in the Illustris-TNG simulations, and found that the transition region in Illustris-TNG, which happens at $M_{\rm BH}\approx 10^{8}\,\rm M_{\odot}$,  was much sharper than observations suggest, and even bimodal, with galaxies below the transition mass being star-forming, and right above, being quenched. \shark\ v2.0 displays a much smoother transition that agrees better with observations. We came back to the sSFR-$M_{\rm BH}$ plane in $\S$~\ref{sec:quench2}.

Lastly, we analyse the SFR histories (SFHs) of galaxies at $z=0$ in different bins of stellar mass for \shark\ v2.0 and \citetalias{Lagos18c} in Fig.~\ref{SFH}. Galaxies in both \shark\ v2.0 and \citetalias{Lagos18c} have on average raising SFHs in the stellar mass range $10^9-10^{9.8}\,\rm M_{\odot}$. Galaxies in the next stellar mass bin, $10^{10}-10^{10.25}\,\rm M_{\odot}$, in both models have a median SFH that is close to constant over the last $6$~Gyrs. At higher stellar masses there is a clear peak at lookback times $>9$~Gyrs, followed by a decline in SFR. 
%The main difference between \shark\ v2.0 and \citetalias{Lagos18c} is in the level of ``downsizing'' experienced by massive galaxies. 
In \citetalias{Lagos18c}, galaxies with $M_{\star}\gtrsim 10^{10.5}\,\rm M_{\odot}$ have relatively self-similar SFHs, scaled up or down in the overall normalisation. In \shark\ v2.0, the SFHs of massive galaxies present a much larger variations at fixed stellar mass, and on average end up more quenched towards $z=0$. In \shark\ v1.1, once quenching kicks in for galaxies with $M_{\star}>10^{10.5}\,\rm M_{\odot}$, we see that the SFR continues to increase with increasing stellar mass, while in \shark\ v2.0, galaxies with $10^{10.5}\,\rm M_{\odot}<M_{\star}<10^{10.75}\,\rm M_{\odot}$ have higher SFRs at lookback time $<4$~Gyr than galaxies with $10^{10.75}\,\rm M_{\odot}<M_{\star}<10^{11.5}\,\rm M_{\odot}$, on average.
%this is very different, with the SFHs of galaxies being more skewed to higher lookback time with increasing stellar mass. This yields a stronger ``downsizing'' signal, in which more massive galaxies formed their stellar masses earlier and on shorter timescales than lower mass galaxies. 
This agrees qualitatively with the ``downsizing'' signal observed in local Universe galaxies (e.g. \citealt{Thomas10,Bellstedt20b}), and shows that this new version of \shark\ overcomes one of the shortcomings of \citetalias{Lagos18c} identified in \citet{Bravo22,Bravo23}.

\subsection{The onset of galaxy quenching}\label{sec:quench2}

\begin{figure}
\begin{center}
\includegraphics[trim=3mm 3mm 3mm 3.5mm, clip,width=0.262\textwidth]{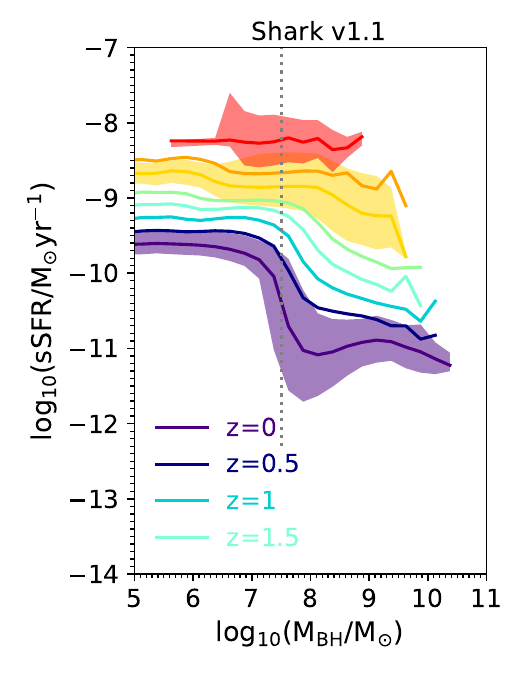}
\includegraphics[trim=21mm 3mm 2mm 3.5mm, clip,width=0.208\textwidth]{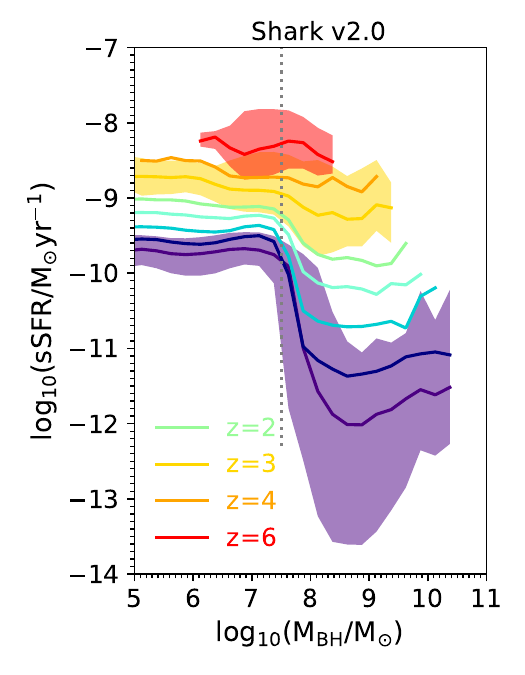}
\caption{The specific SFR vs BH mass for galaxies with stellar masses $\ge 10^9\,\rm M_{\odot}$ from $z=0$ to $z=6$, as labelled, for \shark\, v1.0 (right) and v2.0 (left). The applied stellar mass limit is to avoid the emergence of quenched satellite galaxies producing a visible decrease in specific SFR as the focus here is on AGN feedback. Lines show the medians, and the shaded regions show the $16^{\rm th}-84^{\rm th}$ percentile ranges. The latter are only shown for $z=0$, $z=3$ and $z=6$ to avoid crowding the figure. The vertical dotted line shows for reference a BH mass of $10^{7.5}\rm \, M_{\odot}$.}
\label{BHSSFRevo}
\end{center}
\end{figure}

To study the onset of quenching in galaxies, we start by exploring the sSFR-BH mass plane across cosmic time in Fig.~\ref{BHSSFRevo} for \shark\ v2.0 and \citetalias{Lagos18c}. We show how the median relation for galaxies with $M_{\star}>10^9\,\rm M_{\odot}$ evolves from $z=0$ to $z=6$ in both models. 

In \citetalias{Lagos18c}, the BH mass at which galaxies transition from being primarily star-forming to displaying signs of quenching (i.e. a decrease in sSFR) increases with increasing redshift. By $z=3$, this happens at $M_{\rm BH}\approx 10^{8.3}\,\rm M_{\odot}$, $\approx 1.2$~dex higher than the transition mass at $z=0$ for that model. At $z=4$ there are only weak signs of quenching happening in galaxies with $M_{\rm BH}\gtrsim 10^9\,\rm M_{\odot}$. There is a stark contrast with \shark\ v2.0, which predicts that the BH transition mass is relatively redshift independent at $M_{\rm BH}\approx 10^{7.5}\,\rm M_{\odot}$. The reason for this difference comes down to how radio mode feedback was implemented in \citetalias{Lagos18c} vs \shark\ v2.0. As discussed in $\S$~\ref{newagnmodel}, the heating power of AGN feedback in the radio mode of \citet{Croton06} was calculated using only the accretion rate coming from the hot-halo mode instead of the total accretion rate of the BH. The hot-halo mode contributes very little to the total BH accretion rate at high-z, hence we need to move to very high BH masses to start seeing a larger contribution from that mode, which can then lead to quenching. In \shark\ v2.0 we instead are agnostic to the accretion channel which makes more physical sense, and simply calculate the total jet power that can be used to offset gas cooling {  in the jet mode of AGN feedback}. This leads to an approximately constant BH mass above which enough jet power is produced to lead to quenching (or at least to significant deviations from the MS). Combining robust BH mass and sSFR measurements in observations across cosmic time would greatly help to constrain the predictions here. {  In the future we plan to quantify how the changes in sSFR are tracked by changes in BH mass, to understand how BH growth is leading to galaxies falling off the MS.}

\begin{figure}
\begin{center}
\includegraphics[trim=5mm 4mm 3mm 3mm, clip,width=0.499\textwidth]{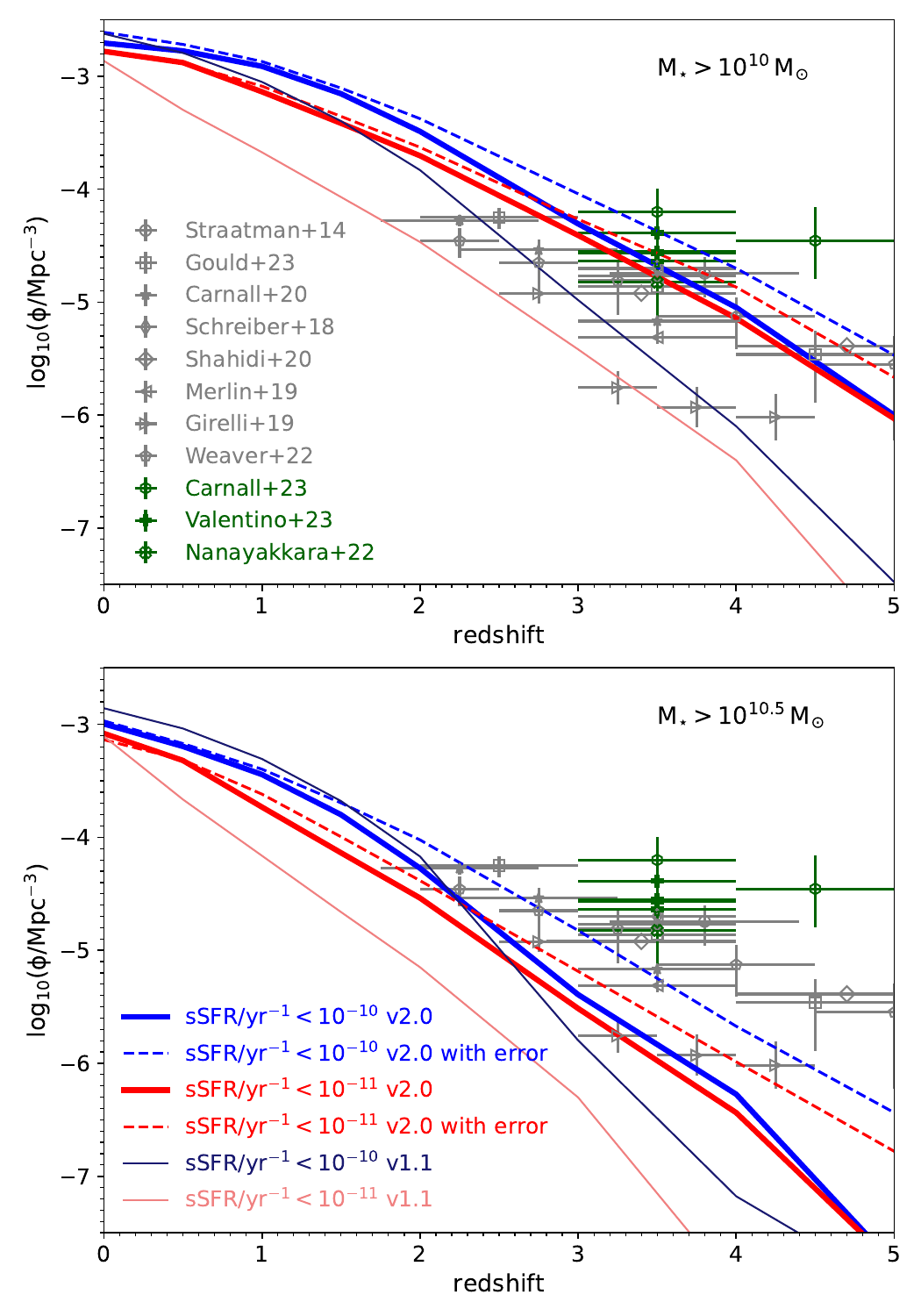}
\caption{Number density of passive massive galaxies selected based on two stellar mass thresholds, $M_{\star}>10^{10}\,\rm M_{\odot}$ (top) and $M_{\star}>10^{10.5}\,\rm M_{\odot}$ (bottom), as labelled. To select passive galaxies we use two different thresholds in specific SFR as labelled, which approximate typical values adopted in the literature. For \shark\ v2.0 we show the intrinsic predictions (i.e. taking stellar masses and SFRs without assuming any errors; thick solid lines), and the prediction after convolving stellar masses and SFRs with errors that are Gaussian-distributed with a width $0.3$~dex but no bias (i.e. Gaussian centred at $0$; dashed lines). \shark\ v1.1 intrinsic predictions are also shown in both panels (thin solid lines, as labelled in the bottom panel). Observational estimates are also shown: in grey symbols (pre-JWST results) are from \citet{Straatman14,Schreiber18,Merlin19,Girelli19,Carnall20,Weaver22,Gould23}, and green symbols (JWST results) are from \citet{Nanayakkara22,Carnall23,Valentino23}, as labelled at the top panel. }
\label{NumDensityPassiveGals}
\end{center}
\end{figure}

One of the key motivations to study the quenching of massive galaxies comes from the recent discoveries of a sizeable population of massive-quiescent galaxies in the early Universe (e.g. \citealt{Schreiber18,Carnall20,Weaver22, Gould23, Nanayakkara22, Carnall23, Valentino23,Long23}). These observations have revealed that these galaxies exist in number densities $\gtrsim 10^{-5}\,\rm Mpc^{-3}$. We hence investigate what \citetalias{Lagos18c} and \shark\ v2.0 predict in Fig.~\ref{NumDensityPassiveGals} for the number density of massive-quiescent galaxies at $0\le z\le 5$. We do this by using two definitions of sSFR, $<10^{-10}\,\rm yr^{-1}$ (as adopted by \citealt{Long22}) and $<10^{-11}\,\rm yr^{-1}$ (a more conservative cut). These two thresholds represent deviations from the MS at $z=3$ of $\approx 1.3-1.5$~ and $2.3-2.5$~dex, respectively. 
We also investigate two stellar mass selections, $\gtrsim 10^{10}\,\rm M_{\odot}$ (top panel in Fig.~\ref{NumDensityPassiveGals})  and $\gtrsim 10^{10.5}\,\rm M_{\odot}$ (bottom panel in Fig.~\ref{NumDensityPassiveGals}), as they represent typical mass thresholds being adopted in observations. 

\begin{figure*}
\begin{center}
\includegraphics[trim=4.5mm 4mm 4mm 4mm, clip,width=0.99\textwidth]{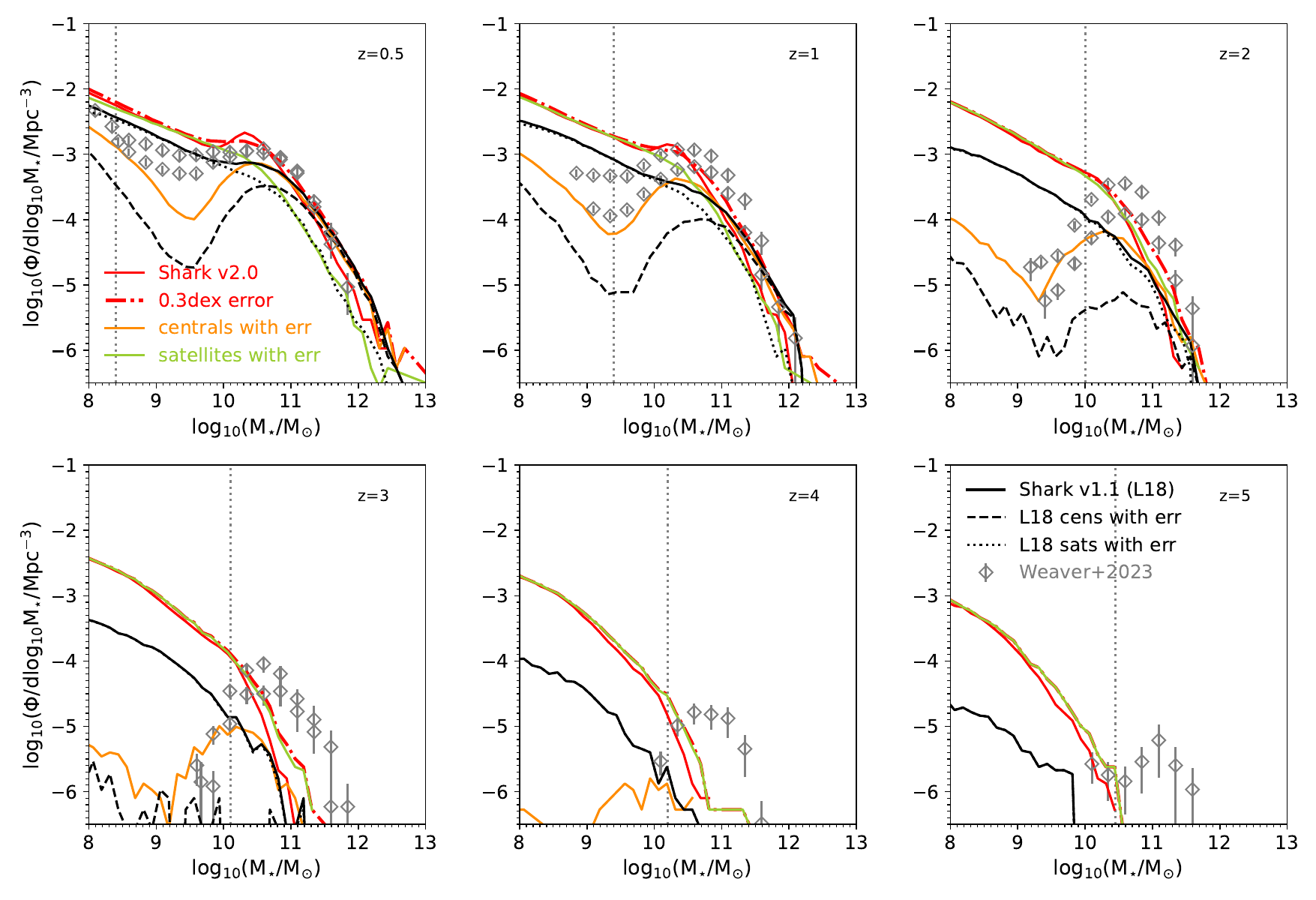}
\caption{Galaxy SMF in \shark\ v2.0 (solid red line) and the default model in \citetalias{Lagos18c} (solid black line), for galaxies classified at passive with a simple threshold of $\rm sSFR < 10^{-10.75}\,\rm yr^{-1}$ from $z=0.5$ to $z=5$, as labelled. 
We also show for \shark\ v2.0 the SMF of passive galaxies after applying 
a random error to the stellar masses and SFRs of $0.3$~dex (dot-dashed line) and then splitting those galaxies between centrals (orange) and satellites (green). We show the same for central and satellite galaxies in \citetalias{Lagos18c} as dashed and dotted black lines, respectively. Observations from \citet{Weaver22} from the COSMOS survey are also shown. {  Two sets of data points for \citet{Weaver22} are shown in the panels where their adopted redshift bins do not fully coincide with ours, corresponding to the closest adopted redshift bins in their work.}. Vertical dotted lines mark the approximate stellar mass above which which observational estimates of the SMF are reliable according to \citet{Weaver22}.}
\label{SMFpass}
\end{center}
\end{figure*}

\begin{figure*}
\begin{center}
\includegraphics[trim=4.5mm 4mm 4mm 4mm, clip,width=0.99\textwidth]{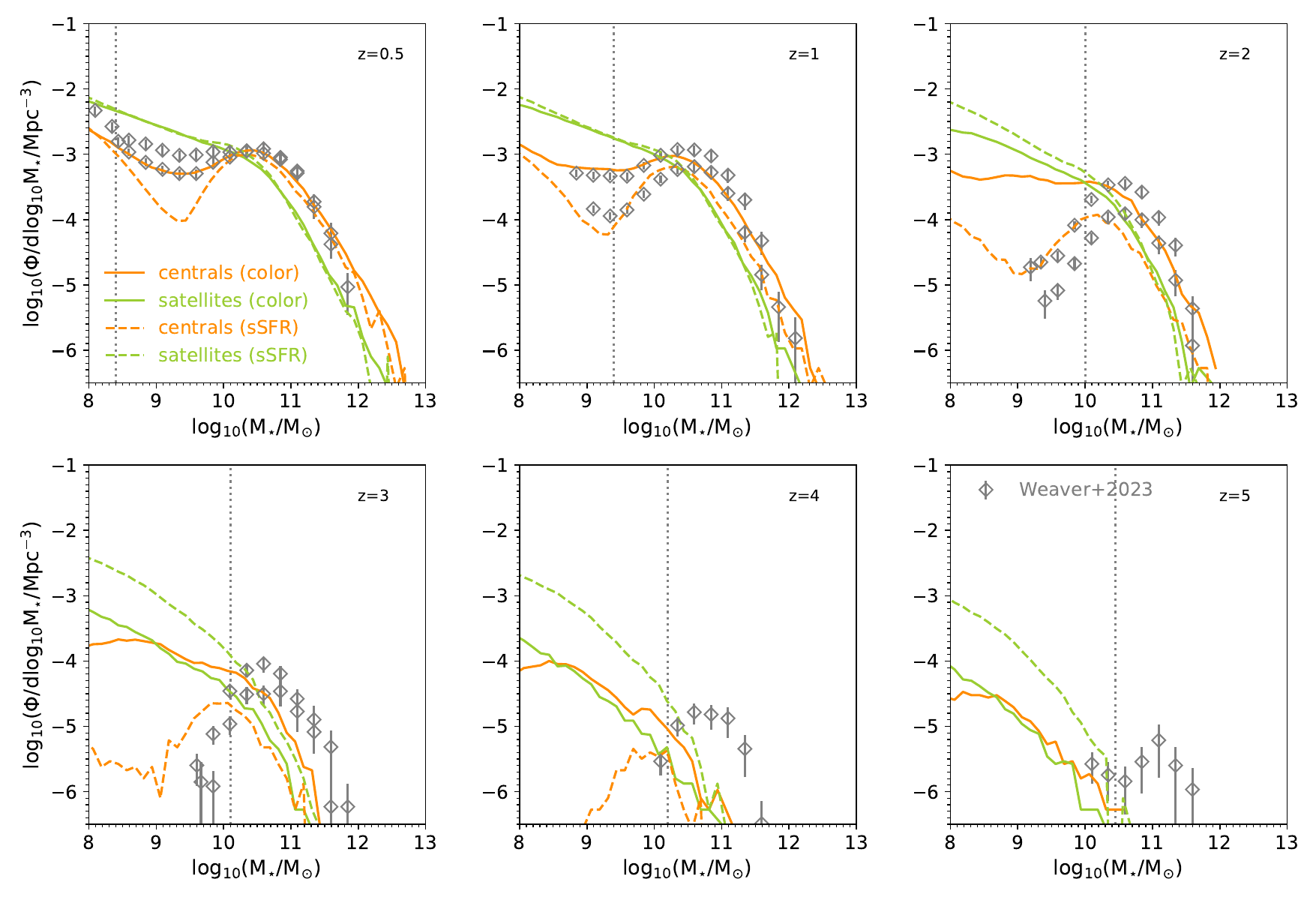}
\caption{As Fig.~\ref{SMFpass} but for centrals and satellites in \shark\ v2.0 selected to have $\rm sSFR < 10^{-10.75}\,\rm yr^{-1}$ (dashed lines), or to comply with the colour-colour selection of Eq.~(\ref{colselec}) (solid lines). The latter is what was used by \citet{Weaver22} to select passive galaxies and construct the SMF.}
\label{SMFpassCol}
\end{center}
\end{figure*}

Fig.~\ref{NumDensityPassiveGals} shows two important differences between \citetalias{Lagos18c} and \shark\ v2.0: (1) overall the number density of massive-quiescent galaxies in \shark\ v2.0 is $\gtrsim 1$~dex higher than \citetalias{Lagos18c}, with differences being largest for galaxies with $\rm sSFR<10^{-11}\, yr^{-1}$, $\approx 2$~dex; (2) there is little difference in number density between selecting galaxies with $\rm sSFR<10^{-11}\, yr^{-1}$ or $\rm sSFR<10^{-10}\, yr^{-1}$ in \shark\ v2.0 ($\lesssim 0.2$~dex), compared to the large differences seen in \citetalias{Lagos18c} ($\gtrsim 0.5$~dex). This tells us that quenched galaxies in \shark\ v2.0 have on average sSFR $\gtrsim 1$~dex smaller than \citetalias{Lagos18c}. We find that the large difference between \shark\ v2.0 and \citetalias{Lagos18c} is due to the new {  jet} mode AGN feedback model rather than the inclusion of {  a wind} mode AGN feedback. When we turn off the latter we only see a slightly lower number density of massive-quiescent galaxies at $z\ge 4$ of $0.2$~dex, and almost no differences at $z<4$.

Overall the number density of massive-quiescent galaxies predicted by \shark\ v2.0 are in good agreement with current observational constraints within the uncertainties. Most of the scatter between current observational estimates comes from the different employed methods to define ``quenched'' (e.g. colour-colour selection vs post-starburst features in the spectrum; see \citealt{Valentino23} for a discussion). 
The comparison between observations with simulations is further complicated by the unconstrained systematic and random uncertainties in the estimation of stellar mass and SFRs in these galaxies. To get a sense of what the impact of this could be, we convolve the stellar masses and SFRs in \shark\ galaxies with a random Gaussian distribution centred around $0$ and with a width of $0.3$~dex. We select then massive-quiescent galaxies based on sSFRs and stellar masses after including errors. This effect is shown as dashed lines in Fig.~\ref{NumDensityPassiveGals}. The consequence of adding errors can be quite large, especially for the rarest objects (bottom panel in Fig.~\ref{NumDensityPassiveGals} shows $\approx 0.8$~dex difference between having errors or not in the population with $\rm sSFR < 10^{-10}\,yr^{-1}$). We should note that errors of the order of $0.3$~dex for stellar masses and SFRs are likely lower limits, as even at $z=0$ with better data quality and multi-wavelength coverage, stellar mass errors are typically $0.2$~dex \citep{Robotham20}. A full understanding of the level of agreement/tension between observations and simulations likely requires full forward modelling to select galaxies in the same colour-colour space or with the same spectral features as in observations. We leave that for future work.

We now investigate the evolution of the SMF of passive galaxies in Fig.~\ref{SMFpass} from $0.5\le z\le 5$. We select passive galaxies as those with a $\rm sSFR < 10^{-10.75}\,\rm yr^{-1}$ following Shuntov et al. (in preparation). We discuss the effect of the passive selection criteria when discussing Fig.~\ref{SMFpassCol}.
%Shuntov et al. (in preparation) show that such criterion produces a similar SMF of passive galaxies as the colour-colour selection applied in \citet{Weaver22}, so we choose to apply the sSFR cut above for simplicity. 
Comparing \shark\ v2.0 with \citetalias{Lagos18c} first (solid red and dashed black lines, respectively), we find that at all redshifts \shark\ v2.0 produces more passive galaxies across the whole mass range than \citetalias{Lagos18c}. The differences become larger with increasing redshift, from $\approx 0.3$~dex at $z=0.5$ to $1.5$~dex at $z=5$. The difference in number density between the two models is largest for central galaxies (orange solid vs dotted black lines), with \citetalias{Lagos18c} producing a number density of central, passive galaxies $\lesssim 10^{-5.5}\, \rm Mpc^{-3}$ at $M_{\star}\approx 10^{10.5}\,\rm M_{\odot}$, while \shark\ v2.0 predicts number densities of $10^{-4}\rm \, M_{\odot}$. At $z\ge 3$, \citetalias{Lagos18c} predicts virtually no quenched, central galaxies, with number densities $<10^{-6}\,\rm M_{\odot}$, in contrast with \shark\ v2.0 which predicts peak number densities of $\approx 10^{-4.5}\,\rm Mpc^{-3}$ and $\approx 10^{-5.5}\,\rm Mpc^{-3}$ at $z=3$ and $z=4$, respectively. At $z\le 3$ a population of passive, low-mass centrals emerges in both \shark\ v2.0 and \citetalias{Lagos18c} with masses $M_{\star}\lesssim 10^9\,\rm M_{\odot}$. These galaxies corresponds to those inhabiting very low-mass halos that suffer from photo-ionisation from the UV background, whose efficiency depends on a halo's circular velocity and redshift \citep{Sobacchi13}, as described in $\S$~4.4.9 in \citetalias{Lagos18c}.

The difference between the SMF of passive, satellite galaxies between \shark\ v2.0 and \citetalias{Lagos18c} comes in part from the new dynamical friction timescale model we adopted ($\S$~\ref{poultonmodel}) in v2.0 producing longer dynamical friction timescales leading to a longer survival of satellite galaxies than the \citet{Lacey93} model adopted in \citetalias{Lagos18c}. Satellite galaxies make up most of the population of passive galaxies at $M_{\star}\lesssim 10^{10.2}\,\rm M_{\odot}$ at $z=0.5$ and at increasingly higher stellar masses with increasing redshift. For example, at $z\ge 4$, passive, satellite galaxies dominate the number density across the whole mass range, showing that environment (i.e. RPS and tidal stripping) can effectively quench galaxies at very early cosmic times.  

Fig.~\ref{SMFpass} shows the observational results of \citet{Weaver22}, in which a colour-colour selection was used to classify galaxies as passive or star-forming using the near-UV (NUV), r-band (r), and J-band (J) magnitudes, with passive galaxies being those with

\begin{equation}
    \rm (NUV - r) > 3 \, (r - J)\, +\, 1;\, (NUV - r) > 3.1,\label{colselec}
\end{equation}

\noindent and star-forming galaxies are those that do not comply with the selection above.
Compared with the observations of \citet{Weaver22}, \shark\ v2.0 performs overall better than \citetalias{Lagos18c}, reproducing well the population of passive, central galaxies {  (at least up to $z=2$)} which likely dominate the high-mass end in the observations. At $z\ge 3$, \shark\ v2.0 struggles to reproduce the high-mass end of the passive SMF, but this is in part due to systematic uncertainties in the observations. For example, Shuntov et al. (in preparation) using COSMOS-Web \citep{Casey22COSMOSWeb} show that the high-mass end they recover at those high-redshift for passive galaxies is about $0.5$~dex lower than that reported by \citet{Weaver22} and closer to the \shark\ v2.0 predictions. Another important limitation to mention in observations is that the number densities are unconstrained for galaxies {  with masses to the left of} the vertical lines shown in Fig.~\ref{SMFpass}, 
%close to unconstrained for galaxies with stellar masses $<10^{10}\,\rm M_{\odot}$ at $z\ge 1.5$, 
and hence the onset of environmental quenching in the very early Universe as predicted by \shark\ cannot be clearly studied with current observations of the SMF. 

We assess the effect the selection of passive galaxies has on the resulting SMF, we compare the SMF of galaxies with a sSFR$<10^{-10.75}\,\rm yr^{-1}$ and that comply with the colour-colour selection of Eq.~(\ref{colselec}). We compute galaxy spectral energy distributions following the method described in \citet{Lagos19} (referred to as ``{\sc EAGLE}-$\tau$~RR14'' in that paper - see Section~6 in the supplementary material for a short description). The supplementary material shows that \shark\ v2.0 produces galaxy luminosities that reproduce reasonably well the observed luminosity functions of the local Universe from the NUV to the FIR, the K-band luminosity function evolution up to $z=3$, and the far-UV luminosity function up to $z=10$.

The SMF of the passive galaxy populations selected by the methods above are shown in Fig.~\ref{SMFpassCol}. Focusing first on central galaxies, we see that at $z\le 1$ Eq.~(\ref{colselec}) is effective in selecting galaxies of low sSFR and stellar masses $\gtrsim 10^{10}\,\rm M_{\odot}$, however, it tends to overestimate the number of low sSFR central galaxies at lower stellar masses by up to $0.8$~dex. The problem gets worse at increasing redshift, and by $z\ge 3$ most of the centrals classified as passive by their colour have sSFR$>10^{-10.75}\,\rm yr^{-1}$. By $z=5$ there are $>1000$ times more galaxies with colours consistent with being passive but with  sSFR$>10^{-10.75}\,\rm yr^{-1}$. Interestingly, for satellite galaxies we see the opposite. At $z\le 1$ galaxies with sSFR$<10^{-10.75}\,\rm yr^{-1}$ comply with the colour selection of Eq.~(\ref{colselec}), but at higher redshifts, the colour selection {\it underestimates} the number of satellites with low sSFRs. By $z=5$ this underestimation is of $\approx 1$~dex. To understand where these disparate effects on centrals and satellite galaxies come from, we studied the median properties of centrals/satellites with $M_{\star}\ge 10^{10}\,\rm M_{\odot}$. 
We found that centrals primarily correspond to star-bursting, dusty galaxies (i.e. most of their SFR is associated to a burst with super-solar metallicities) at $z>2$. Satellites at $z\ge 2$ on the other hand, have metallicities that are $0.7-2$~dex below centrals and have their star formation taking place in the galaxy disk.
%their overall different gas metallicities: satellite galaxies tend to have a lower dust attenuation than centrals at high-z by XXX. 
Hence, Eq.~(\ref{colselec}) appears to be effective in selecting passive galaxies when they have gas metallicities close to solar metallicity and they are not dusty star-forming galaxies (see \citealt{Lagos20} for a detailed analysis of the contamination of dusty star-forming galaxies in the selection of passive galaxies based on colour-colour diagrams). 

We overall see that the SMF of galaxies selected to be ``passive'' following the colour-colour selection of Eq.~(\ref{colselec}) is in much better agreement with the observational estimates of \citet{Weaver22}, coming from the same colour-colour selection (see for example high-mass end at $z=2$ and $z=3$). In the future we plan to investigate other more modern methods to select passive galaxies based on colours using independent magnitudes and Bayesian methods to assign probabilities of being passive \citep{Gould23,Long23}.
{  Other SAMs and cosmological hydrodynamical simulations have also pointed to the large contamination that adopted colour-colour selections have (e.g. \citealt{Lustig23}), highlighting the need to compare with observations carefully.}

%XXXX aca va la parte de la seleccion de passive.

%FINISH DISCUSSION AFTER UPDATING THE OBSERVATIONS.
%NOTE THAT OBSERVATIONS FROM SHUNTOV+ FROM COSMOS-WEBB LOOK MUCH BETTER COMPARED TO SHARK AT Z=4 AND Z=5 WITH LOWER STELLAR MASSES. 

\begin{figure}
\begin{center}
\includegraphics[trim=0.5mm 4mm 4mm 3mm, clip,width=0.46\textwidth]{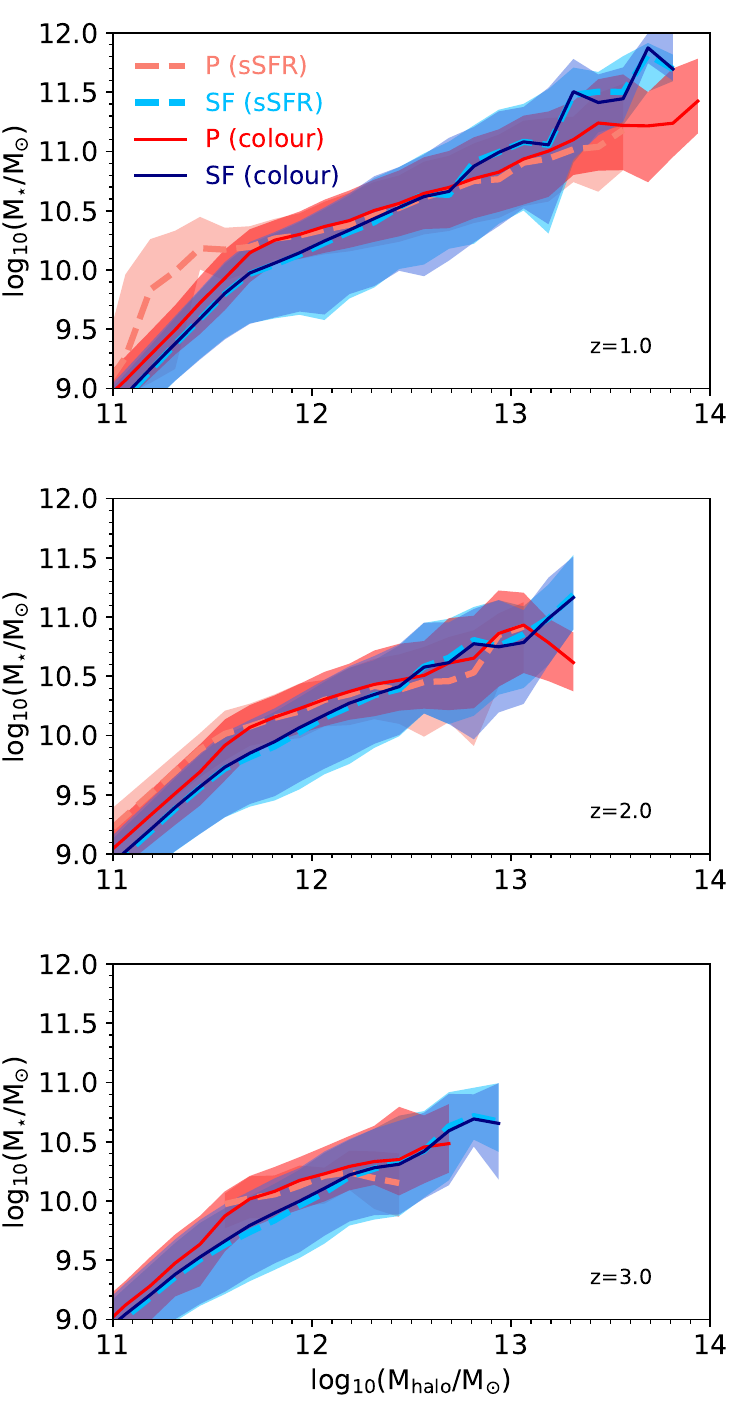}
\caption{Stellar vs halo mass relation for central galaxies in \shark\ v2.0 selected to be passive by either assuming $\rm sSFR<10^{-10.75}\,\rm yr^{-1}$ (salmon dashed line), or by selecting them based on their NUV-optical colour (see text for details; red solid line); and star-forming, by either selecting them to have sSFRs that are $>-0.3$~dex from the MS (light blue dashed line) or based on their NUV-optical colour (blue solid line). Lines with shaded regions show the medians and $16^{\rm th}-84^{\rm th}$ percentile ranges, respectively. 
%We also show for reference the median stellar-halo mass relation of all central galaxies in \shark\ v2.0. 
Only bins with $\ge 10$ galaxies are shown. 
This is presented for $z=1,\,2,\,3$, as labelled. Overall, passive galaxies display a flatter stellar-halo mass relation than star-forming galaxies, with the difference being larger when galaxies are selected based on their sSFR rather than colour.}
\label{SMHMPass}
\end{center}
\end{figure}

Finally, we present predictions for the stellar-to-halo mass relation of passive and star-forming central galaxies in \shark\ v2.0 in Fig.~\ref{SMHMPass}. We again test two definitions of passive galaxies as described above (i.e. based on sSFR and colour), while star-forming galaxies are defined based on their sSFR distance to the MS or by not complying with the colour-colour selection of Eq.~(\ref{colselec}). These two selections are referred to as ``sSFR'' and ``colour'' in Fig.~\ref{SMHMPass}. We measure the MS in \shark\ v2.0 by doing a linear fit to the relation between $\rm log_{10}(sSFR)$ and $\rm log_{10}(M_{\star})$ for central galaxies with stellar masses in the range $7\times 10^8-10^{10}\,\rm M_{\odot}$ at each of the shown redshifts. These mass limits are chosen conservatively to avoid resolution effects at the lower end, and the regime where AGN feedback becomes efficient at the higher end. Star-forming galaxies, in the case of the ``sSFR'' selection, are those with a distance to the main sequence $\ge -0.3$~dex. 

Fig.~\ref{SMHMPass} shows that passive galaxies display a weaker dependence of their stellar mass on halo mass at $M_{\rm halo}\gtrsim 10^{11.6}\rm \, M_{\odot}$ than star-forming galaxies. The difference is larger when we use the sSFR to select galaxies than with the colour-colour selection. Interestingly, the latter misses most of the central galaxies with low sSFRs that are hosted by halos with $M_{\rm halo}<10^{11.6}\,\rm M_{\odot}$ at $z=1$ and $z=2$. At $z=3$ the colour-colour selection picks centrals in those low-mass halos that are not passive according to their sSFR. 
%
%, in contrast with star-forming central galaxies, which follow better the empirical stellar-halo mass relations derived from abundance matching. 
The overall different stellar-halo mass relation of passive and star-forming galaxies should translate into different clustering signals for these two populations. The flatness of the stellar-halo mass relations of passive central galaxies leads to their stellar masses being lower than that of star-forming galaxies at fixed halo mass in halos with masses $\gtrsim 10^{12.5}\,\rm M_{\odot}$, on average.
At halo masses $10^{11.6}\lesssim  M_{\rm halo}/\rm M_{\odot}\lesssim10^{12.5}$, the flatness of the stellar-halo mass relation of passive galaxies leads to them having {\it higher} stellar masses than star-forming galaxies. This is very clear at $z=1$ and apparent at $z=2$. At $z=3$ we have too few high-mass halos to establish the continuation of such a difference. We will revisit this using larger cosmological volume simulations in the future.

At $M_{\rm halo}\lesssim 10^{11.6}\,\rm M_{\odot}$, the stellar-halo mass relation of passive galaxies is very different between the two methods employed to select them, with the colour-colour selection leading to a relation that is similar to that of star-forming galaxies (especially clear 
%leading to a steepening of the relation at higher halo masses than the sSFR selection (specially visible 
at $z=1$ and $z=2$).
%there is a sudden steepening of the stellar-halo mass relation fo.r passive galaxies that is visible at $z=2$ and $z=1$. 
%This steepening at low halo masses is in part due to stellar feedback becoming very efficient and photoionisation from the UV background kicking in. 

One of the only observational measurements we can compare Fig.~\ref{SMHMPass} with were presented by \citet{Cowley19Pass}, who derived a stellar-to-halo mass relation of passive and star-forming galaxies from the clustering of these populations at $z\approx 2-3$. Note that \citet{Cowley19Pass} use colour selections similar to Eq.~(\ref{colselec}) to isolate passive galaxies. \citet{Cowley19Pass}
 found that at $M_{\rm halo}\gtrsim 10^{12.7}\,\rm M_{\odot}$ passive and star-forming galaxies followed similar relations, but below that mass, passive galaxies were more massive than star-forming galaxies at fixed halo mass by $\approx 0.5$~dex. This difference is larger than what we find at $z=2,\,3$ and $M_{\rm halo}\lesssim 10^{12.5}\,\rm M_{\odot}$ in \shark\ v2.0 (which is closer to $0.3$~dex), but overall consistent within the scatter. At higher halo masses, the colour selection indeed leads to stellar-halo mass relations between star-forming and passive galaxies in \shark\ v2.0 that are similar (and even indistinguishable at $z=3$), in agreement with the conclusions of \citet{Cowley19Pass}. However, if galaxies were selected by sSFR we predict that different stellar-halo mass relations of passive and star-forming galaxies should be seen at these high halo masses at least up to $z=2$. 
 \citet{Magliocchetti23}, also using clustering measurements, presented constraints on the host halo mass of massive-quiescent galaxies (those with $M_{\star}\approx 10^{10}-10^{11}\,\rm M_{\odot}$) and show that the host halo masses of these galaxies are likely $\gtrsim 10^{12}\,\rm M_{\odot}$ across a wide redshift range, $0\le z \le 5$. Although we broadly see something similar in \shark\ v2.0, a careful comparison requires selecting galaxies using the same colour-colour selection employed in  \citet{Magliocchetti23} and adding photometric redshift uncertainties, which we leave for future work.
 %used to isolate passive galaxies.
 %than what is found for the sSFR selection, where the difference between the two galaxy populations is clear in all panels.
 %, and at $z=3$  even though the sSFR selection 
 %happens similar to the difference we see in the middle panel of Fig.~\ref{SMHMPass} at $M_{\rm halo}\approx 10^{11.5}\,\rm M_{\odot}$. 
 %There are several caveats in comparing our results directly to the measurements of \citet{Cowley19Pass}, which include the exact way passive and star-forming galaxies were selected being different to what we have applied here. A one-to-one comparison with the clustering measurements of \citet{Cowley19Pass} will be presented in future work.

The predictions of Fig.~\ref{SMHMPass} will be testable in the near future thanks to JWST programs focused on obtaining large statistical samples of massive-quiescent galaxies across cosmic time.

\subsection{\shark\ predictions in the context of other galaxy formation simulations}\label{sec:quench3}

{  The topic of massive quiescent galaxies has generated a lot of interest given the recent JWST results. Our results point to the treatment of AGN and its feedback as a critical aspect behind the rise of the first massive quenched galaxies. \citet{Xie24} show that in the GAEA semi-analytic model this is also the case. Their model agrees well with the number density of passive galaxies up to $z\approx 3$, in great part thanks to the new AGN feedback model implemented in GAEA (which was described in \citealt{Fontanot20}). In the {\tt Magneticum} cosmological hydrodynamical simulations, \citet{Kimmig23} show that AGN feedback is also the culprit of the quenching of massive galaxies in the very early Universe. \citet{Pandya17} show that in the hydrodynamical simulations of \citet{Choi17b}, AGN feedback produces a population of massive quenched galaxies at $z\approx 2$, but the quenching appears to be too slow to produce enough quenched galaxies. This also seems to be the case on other cosmological hydrodynamical simulations, as discussed in \citet{Valentino23}.
AGN feedback is thus a major, but still enigmatic quenching mechanism around and before cosmic noon, hence extending the well-established importance of AGN quenching at late times \citep{Croton06,Bower06} to the entire star-forming era.

Some simulations on the market have been able to produce a number density of quenched galaxies at $z>2$ that agrees with observations, such as {\tt Illustris-TNG} (see \citealt{Valentino23} for details), GAEA \citep{Xie24} and \shark\ v2.0 (this work). However, the problem of massive galaxy quenching is more complex than that. 
%needs to be taken in how well the simulation reproduces the local universe observation. 
For example, if a high number density of massive quiescent galaxies at high-z is accompanied by an over-prediction of the number density of massive galaxies at $z=0$, then such a model may not be optimal. \citet{Valentino23} show that {\tt Illustris-TNG} produces a higher number density of massive quenched galaxies at $z=3$ than the {\tt EAGLE} simulations; however, the same difference can be seen in the number density of massive galaxies at $z=0$, where {\tt EAGLE} agrees better with observations than {\tt Illustris-TNG}. We argue here that massive galaxy quenching needs to be assessed throughout cosmic time in any one model to understand if and how AGN feedback may be effective. 
Hence our focus here was in quenching of massive galaxies at $z=0$ in $\S$~\ref{sec:quench1}, followed by quenching at high z in $\S$~\ref{sec:quench2}.}

\section{Discussion and conclusions}\label{conclusions}

We have introduced a new version of the \shark\ SAM (v2.0) after a number of improvements to the physics included. These changes comprise: (i) a model describing the exchange of angular momentum between the interstellar medium and stars that results in different angular momenta for the atomic and molecular ISM and stars in galaxies; 
(ii) an updated dynamical friction timescale of satellite galaxies; 
(iii) a new AGN feedback model which includes two modes, a {  wind} and a {  jet} mode, with the {  jet} mode directly tied to the jet power production and the {  wind} mode consisting of a radiation pressure-driven outflow; (iv) a model that tracks the development of BH spins, which together with the mass and accretion rate, are used to define different BH accretion disk states; (v) a model for the gradual ram-pressure stripping of the hot and cold gas of satellite galaxies; (vi) a model for the tidal stripping of gas and stars of satellite galaxies; 
(vii) a method for automatic parameter exploration of the model using particle swarm optimisation. The model parameters were chosen to fit the $z<0.1$ SMF of \citep{Li09} only. No high redshift constraints were used.

We showed that \shark\ v2.0 provides predictions that agree better with observations than \shark\ v1.1 (\citetalias{Lagos18c}). Those include: the evolution of the SMF up to $z=7$ (Fig.~\ref{SMF}), the halo-mass conditional baryon mass function at $z=0$ and the contribution from satellite and central galaxies (Fig.~\ref{BMF}); the stellar size-mass relation  (Fig.~\ref{Sizesz0}) and specific angular momentum-mass relation of different baryon components of galaxies (Fig~\ref{AMz0}) at $z=0$; the BH-stellar mass relation of late- and early-type galaxies at $z=0$ (Fig.~\ref{BHMSz0}). We show that these improvements in large part relate to the new physics included in \shark. Specifically the conditional baryon mass function of satellite galaxies agrees better with observations in part due to the new dynamical friction timescale; and the stellar size-mass and specific angular momentum-mass relations improved significantly thanks to the new angular momentum treatment of galaxy components. 

We presented a detailed analysis of galaxy quenching in \shark\ v2.0 and improvements over v1.1 (\citetalias{Lagos18c}). These included:
\begin{itemize}
    \item Massive galaxies at $z=0$, $M_{\star}\gtrsim 10^{10.5}\,\rm M_{\odot}$ have $\approx 1$~dex lower SFRs in \shark\ v2.0 than v1.1. This allows the model to better reproduce the SFRs of massive galaxies observed  (Fig.~\ref{MSz0}) and the sSFR-BH mass relation of galaxies with $M_{\star}> 10^{10}\,\rm M_{\odot}$ (Fig.~\ref{BHSSFR}) at $z=0$. The latter shows that \shark\ v2.0 also produces a much larger sSFR scatter at fixed BH mass for BH masses $>10^{8}\,\rm M_{\odot}$ than v1.1, in much better agreement with the observations of \citet{Terrazas17}. 
    \item \shark\ v2.0 predicts that the transition of galaxies from being star-forming (i.e. main sequence) to displaying a clear decrease in sSFR relative to the main sequence happens at roughly the same BH mass at all redshifts ($\approx 10^{7.5}\,\rm M_{\odot}$), while v1.1 requires an increasingly massive BH to transition to quenched with increasing redshift (Fig.~\ref{BHSSFRevo}). By $z=4$, in \shark\ v1.1 only galaxies with a BH mass $\gtrsim 10^9\,\rm M_{\odot}$ show signs of a decreased sSFR. 
    \item \shark\ v2.0 produces $\approx 1$~dex higher number density of massive-quiescent galaxies at $z\gtrsim 2$ than v1.1, and those that are classified as being quiescent display lower sSFRs in \shark\ v2.0 than v1.1. Our new results agree well with current observational constraints on the number density of massive-quiescent galaxies coming from the JWST \citep{Carnall23,Nanayakkara22,Valentino23,Long23} (see Fig.~\ref{NumDensityPassiveGals}). We highlight that the overall abundance of massive galaxies in \shark\ v2.0 is similar to that of v1.1 (\citetalias{Lagos18c}), but the key difference is in the fraction of those that are quenched.
    \item We analyse the SMF of passive galaxies from $z=0.5$ to $z=5$ and show that at $z\ge 2$ \shark\ v2.0 produces $\gtrsim 100$ times more central-passive galaxies than v1.1, and this difference increases with increasing redshift. Similarly, \shark\ v2.0 predicts $\approx 10-80$ times more passive satellite galaxies at $z\ge 3$ than v1.1. These differences tend to disappear towards $z=0$ with both models predicting similar SMF of passive galaxies (Fig.~\ref{SMFpass}). Our new model is in better agreement with current observational constraints of the passive SMF (especially when we select galaxies using the same colour-colour criterion employed in observations), though there are likely still too few passive galaxies with masses $>10^{11}\,\rm M_{\odot}$ at $z\ge 4$ in \shark\ v2.0.
    \item We present predictions for the stellar-halo mass relation of star-forming and passive galaxies at $z\ge 1$ and find clear differences between the two populations. Passive galaxies tend to display a flat stellar-halo mass relation at $M_{\rm halo}\gtrsim 10^{11}\,\rm M_{\odot}$, so that at $M_{\rm halo}\approx 10^{11}-10^{12.5}\,\rm M_{\odot}$ they are more massive than their star-forming counterparts, while the opposite happens at $M_{\rm halo}\gtrsim 10^{12.5}\,\rm M_{\odot}$ (Fig.~\ref{SMHMPass}). The exact magnitude of this effect depends on the criteria used to select passive and star-forming galaxies.
\end{itemize}

The differences listed above between \shark\ v1.1 and v2.0 can almost exclusively be associated with the implementation of the new AGN model presented in $\S$~\ref{newagnmodel}, although other processes, such as the new dynamical friction timescale, also play a (secondary) role. This work thus {\it demonstrates the power of SAMs in quickly assessing the possible physical mechanisms behind current tensions with observations}.

In accompanying papers we present a detailed analysis of the BH and AGN population across cosmic time (Bravo et al. in preparation); and a thorough analysis of the parameter space of \shark\ v1.1 and v2.0 (Proctor et al. in preparation). In the future, we plan to explore different definitions of what makes a galaxy passive and the impact that has on the number density, SMF evolution and clustering of passive galaxies; and assess the derived properties of massive-quiescent galaxies to understand the effect of systematic uncertainties. 

%Following the philosophy of \shark, we will continue to improve the physics included in the model. 
%also incorporate new extensions to the physics of the model that have been developed by 

\section*{Data Availability}
The \surfs\ halo and subhalo catalogue and corresponding merger trees used in this work can be accessed from \url{https://tinyurl.com/y6ql46d4}. \shark\ is a public code and the source and python scripts used to produce the plots in this paper can be found at \url{https://github.com/ICRAR/shark/}.

\section*{Acknowledgements}
We thank the anonymous referee for their constructive feedback. 
We thank Sabine Bellstedt, Kate Gould, John Weaver, Jen Hardwick and Nandini Sahu for sharing observational data that has been used in this work. 
CL has received funding from the ARC Centre of
Excellence for All Sky Astrophysics in 3 Dimensions (ASTRO 3D), through project number CE170100013, and is a recipient of the ARC Discovery Project DP210101945. 
MB has received funding from McMaster University through the William and Caroline Herschel Fellowship. 
 DO and ASGR acknowledge support from the ARC Future Fellowship scheme (FT190100083 and FT200100375, respectively).
 KP and ACG acknowledge Research Training Program and ICRAR scholarships.
This work was supported by resources provided by The Pawsey Supercomputing Centre with funding from the 
Australian Government and the Government of Western Australia.

%%%%%%%%%%%%%%%%%%%%%%%%%%%%%%%%%%%%%%%%%%%%%%%%%%



%\bibliographystyle{mnras}

\bibliographystyle{mn2e_trunc8}
\bibliography{SharkAM}

%%%%%%%%%%%%%%%%%%%%%%%%%%%%%%%%%%%%%%%%%%%%%%%%%%

%%%%%%%%%%%%%%%%% APPENDICES %%%%%%%%%%%%%%%%%%%%%

\appendix
\section{Minor modifications to physical models}\label{AppendixMinorModifications}

In addition to the major changes in physical models introduced in $\S$~\ref{LargeModifications}, we also make small modifications to existing models in \shark, as described below.

\subsection{Evolving metal yield}\label{evolvingyield}

In \shark\ v1.1, a constant yield from newly formed stars was assumed. In v2.0, we allow alternative assumptions. In particular, we include a model that assumes a constant ejecta metallicity regardless of
the metallicity of the gas that formed the stars. This in practice assumes an effective yield that depends on the initial metallicity of the star-forming gas:

\begin{equation}
p_{\rm eff}  = {p} - \frac{Z_{\rm cold}}{4},
\end{equation}

\noindent where $p=0.029$ and $Z_{\rm cold}$ is the metallicity of the gas from which stars are being formed. The value of $p$ was computed in \citetalias{Lagos18c} using \citet{Conroy09} for a \citealt{Chabrier03} IMF, and assuming a minimum and maximum masses of stars formed, $m_{\rm min} = 1\,\rm M_{\odot}$ and $m_{\rm max} = 120\, \rm M_{\odot}$, respectively. This model can be adopted using {\tt evolving\_yield = true} in the parameter file.

\subsection{Stellar feedback}\label{StellarFeedbackChanges}

The stellar feedback model adopted in the default \shark\ model in \citetalias{Lagos18c}, is based on that from \citet{Lagos13}, where the mass loading, the ratio between the outflow and instantaneous SFR, $\dot{M}_{\rm out}/\psi$, is a function of both redshift and the maximum circular velocity of the DM halo where the galaxy resides, $f=f(z,v_\mathrm{circ})$ in the case of central galaxies. For satellite galaxies, we use the maximum circular velocity the host halo of the galaxies had right before becoming satellite subhalos. In \shark\ v2.0, we include 
%o correct this, we have added the ability to set 
a minimum value for the mass loading, $\beta_\mathrm{min}$, such that $f=\max(f(z,v_\mathrm{circ}),\beta_\mathrm{min})$. If $\beta_{\rm min}=0$, then we recover the \citetalias{Lagos18c} model. The parameters adopted for the stellar feedback model in this paper and \citetalias{Lagos18c} are shown in Table~\ref{tab:parameters}.
%For Figure \ref{fig:ml_vcirc} this has been set to $\beta_\mathrm{min}=1$.

\begin{figure}
    \centering
    \includegraphics[trim=7mm 12mm 1.5mm 7mm, clip,width=0.5\textwidth]{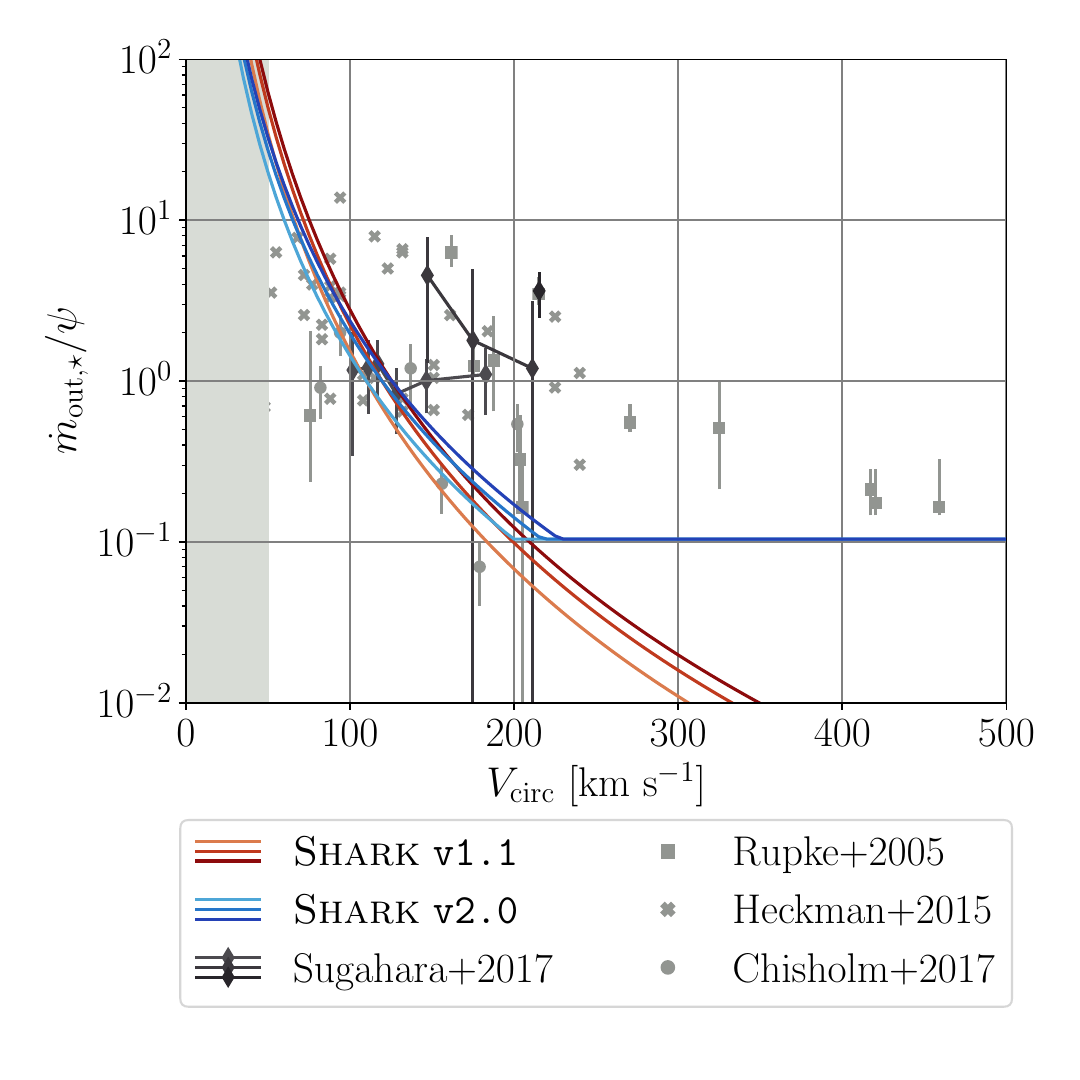}
    \caption{Mass loading vs circular velocity at $z=0,\,1,\,2$ for \shark\ v2.0 (blue lines) and v1.1 (\citetalias{Lagos18c}; red lines). Redshifts are ordered from darker to lighter colours from $z=0$ to $z=2$. %The model adopted in \citetalias{Lagos18c} is shown in lines if red shading, while the new model, which adopts a floor mass loading value is shown in dashed blue lines, as labelled. For this visualisation, we adopted a $\beta_{\rm min}=1$. 
    Observational estimates from \citet{Rupke05,heckman2015,chisholm2017,sugahara2017} are shown as symbols, as labelled.}
    \label{fig:ml_vcirc}
\end{figure}

A comparison of the modelled mass loading-$v_\mathrm{circ}$ relation with 
observations is shown in 
%\placeholder{predictions form hydrodynamic simulations and} measurements from observations is shown in 
Fig.~\ref{fig:ml_vcirc}. Note that these observations have been selected to avoid AGN galaxies, which can show higher mass loading but due to AGN activity rather than stellar-driven winds. Also note that the uncertainty in these measurements is larger than the errorbars show as systematic effects are typically not accounted for when computing errors. 
Overall the introduction of a floor in the mass loading allows the \shark\ v2.0 to agree better with observations, and the slightly higher value adopted for $v_{\rm hot}$ (see Table~\ref{tab:parameters}) also allows for better agreement in galaxies with $V_{\rm circ}<200\,\rm km\,s^{-1}$.

%While in good agreement below a halo mass of $\sim10^{10.5}$ $M_\odot$ with \placeholder{simulations like FIRE \citneed\ and EAGLE \citneed\ and} observations, at higher halo masses the default model in \shark\ underestimates the mass loading.

%This means that for massive galaxies un-physically strong SFRs would be required to produce significant gas outflow from stellar-driven winds, so this galaxies lack a mechanism to deplete their gas reservoirs.

\subsection{Halo spin parameter dependence on halo mass}

\shark\ v1.1 assumed that the distribution of the halo spin parameter was a log-normal distribution of mean $0.03$, $\langle \lambda\rangle$, and width $0.5$. In \shark\ v2.0, we assume that the width is independent of halo mass, but the mean is weakly dependent on halo mass so that $\langle \lambda\rangle = 0.009\,\,{\rm log_{10}} (M_{\rm halo}/\rm M_{\odot}) \,-\, 0.076$ which represents the $\langle \lambda\rangle$ dependence on halo mass reported in \citet{Kim15AM}.

\section{The {\tt griffin19} BH spin model}\label{AppendixSpin}

In \shark\ v2.0, the spin of BHs is determined by the history of BH-BH mergers and gas accretion. Below we summarise the models introduced in \shark\ v2.0 for these processes.

\subsection{BH-BH mergers}

During galaxy mergers, the central supermassive BHs of each galaxy are also merged. We assume that
the mass of the final BH is the sum of the merging BH masses, which assumes that the mass lost to gravitational radiation is negligible. For the BH spin, we follow \citet{Lagos09} and calculate the resulting spin following two different models, which are applied depending on the initial spins. We remind the reader that we define the norm of the BH spin $a=|{\bf a}|$.

If the two BHs merging are non-rotating, which can happen due to our BH seed assumption of non-rotating BHs, we follow \citet{Berti08}, and calculate the resulting spin, $a_{f}$, from the BH mass ratio, $q=m_2/m_1$, where $q \le 1$:

\begin{equation}
    a_f = \frac{2\,\sqrt{3}\,q}{(1+q)^2} - \frac{2.029\, q^2}{(1+q)^4}, \label{spinBHmerg}
\end{equation}

\noindent The first term of Eq.~\ref{spinBHmerg} comes from the orbital angular momentum of a particle at the innermost stable circular orbit of a non-rotating BH, while the second term in the equation accounts
for the angular momentum radiated in the final plunge.

If at least one of the BHs involved in the merger has a non-zero spin, we instead apply 
 the semi-analytic fitting formulae presented in \citet{Rezzolla08}, where the authors fit an extensive compilation of numerical results on BH mergers. Here, the two BHs with masses $m_1$ and $m_2$ have spins ${\bf a}_{1}$ and  ${\bf a}_{2}$, respectively, and the resulting spin is given by 
 
\begin{eqnarray}    
 a_{f} = \frac{1}{(1+q)^2}\, \Big[ a^2_1 + a^2_2\, q^4 + 2\, a_1\,a_2\, q^2 \cos\phi\,+\\
    2\,(a_1\, \cos\theta + a_2\,q^2\,\cos\xi)\,|{\bf l}|\,q +|{\bf l}|^2\,q^2 \Big]^{1/2},\nonumber
\end{eqnarray}

\noindent where $\mu$ is the symmetric mass ratio $q/(q + 1)^2$, and ${\bf l}$ is the contribution of the orbital angular momentum to $a_f$. Following \citet{Griffin19}, we assume that the direction of ${\bf l}$ is that of the initial orbital angular momentum, while its magnitude is given by:

\begin{eqnarray}
 |{\bf l}| = \frac{s_{4}}{(1+q^2)^2}\, (a^2_1 + a^2_1\, q^4 + 2\, a_1\,a_2\,q^2\,\cos\phi) \,+ \\ 
 \left( \frac{s_5\,\mu + t_0 + 2}{1+q^2} \right)\, 
(a_1\, \cos\theta + a_2\,q^2\,\cos\xi) \, + \nonumber \\ 2\,\sqrt{3} + t_2\,\mu + t_3\,\mu^2, \nonumber
\end{eqnarray}

\noindent where  $s_{4}=-0.129$, $s_{5}=-0.384$, $t_{0}=-2.686$, $t_{2}=-3.454$, $t_{3}=2.353$ are the values obtained in \cite{Rezzolla08}. The angles $\phi$, $\theta$ and $\xi$ are the angles between the spins
of the two BHs and their orbital angular momentum, and are given by:

\begin{eqnarray}
     \cos\phi &=& {\bf \hat{a}_1} \cdot {\bf \hat{a}_1},\\
     \cos\theta &=&  {\bf \hat{a}_1} \cdot \hat{\bf{l}},\nonumber\\ 
      \cos\xi &=& {\bf \hat{a}_2} \cdot \hat{\bf{l}}.\nonumber
\end{eqnarray}

\noindent During a BH-BH merger, we randomly select the directions for $\bf{a_1}$, $\bf{a_2}$ and $\bf{l}$. 

\subsection{BH spin development during accretion events}

In our implementation, an accretion event is caused by either of the three processes bringing in gas to the central BH: hot-halo gas cooling, galaxy mergers or disk instabilities. To calculate the BH spin after a gas accretion episode, $a_f$, we follow \cite{Bardeen70}:

\begin{equation}
 a_f = \frac{1}{3} \,\sqrt{\hat{r}_{\rm lso,i}}\, \frac{m_{\rm BH,i}}{m_{\rm BH,f}} \,\left( 4 - \left[ 3\,\hat{r}_{\rm lso,i} \,\left( \frac{m_{\rm BH,i}}{m_{\rm BH,f}} \right)^2 - 2\right]^{1/2} \right), \label{af_acc}
\end{equation}

\noindent where $\hat{r}_{\rm lso}$ is defined in Eq.~\ref{riso_eq}, and $m_{\rm BH,i}$ and $m_{\rm BH,f}$ are the BH masses before and after an accretion event. The latter are related by:

\begin{equation}
 m_{\rm BH,f} = m_{\rm BH,i}\, +\, (1-\epsilon_{\rm TD})\,\Delta m,
\end{equation}

\noindent where $\Delta m$ is the mass accreted from the accretion disk in this accretion episode and $\epsilon_{\rm TD}$ is defined in Eq.~\ref{epsilonTD}. 

We follow \citet{Griffin19} and describe the BH accretion disk as having three regions: an outer disk at $r>r_{\rm in}$ and angular momentum ${\bf J}_{\rm out}$; an inner disk at $r_{\rm warp}<r<r_{\rm in}$, which has angular momentum ${\bf J}_{\rm in}$, where $r_{\rm warp}$ is the warp radius; and the warped disk internal to $r_{\rm warp}$ with angular momentum ${\bf J}_{\rm warp}$. 
If ${\bf J}_{\rm BH}$ (defined in $\S$~\ref{BHnohairproprs}) and ${\bf J}_{\rm in}$ have different directions, a spinning BH induces a Lense-Thirring precession in the misaligned disk elements. As the precession rate falls off as $r^{-3}$, the expectation is that at smaller radii $\bf J_{\rm BH}$ and ${\bf J}_{\rm in}$ become perfectly aligned or anti-aligned \citep{Bardeen75}. At sufficiently large radii, however, there will still be a misalignment between the vectors. The point at which the disk transitions from being (anti-)aligned to misaligned happens at $r_{\rm warp}$. 

We assume that at the beginning of a gas accretion event, ${\bf J}_{\rm warp}$ and  ${\bf J}_{\rm in}$ are aligned. Due to torques, ${\bf J}_{\rm BH}$ aligns with ${\bf J}_{\rm tot} = {\bf J}_{\rm BH} + {\bf J}_{\rm warp}$; ${\bf J}_{\rm tot}$ does not change during the torquing process. We follow \citet{King05} and assume that ${\bf J}_{\rm warp}$ completely aligns or anti-aligns with ${\bf J}_{\rm BH}$. The gas internal to $r_{\rm warp}$ is then accreted onto the BH. As this is an iterative process, and accretion from the internal disk continues to happen,  ${\bf J}_{\rm BH}$ eventually aligns with ${\bf J}_{\rm in}$.

In \shark\ v2.0 we implement the same three options introduced in \citet{Griffin19} for the relation between the angular momentum of the outer and inner disks, which are controlled by the parameter {\tt accretion$_{-}$disk$_{-}$model}:

\begin{itemize}
    \item {\tt prolonged}: assumes that ${\bf J}_{\rm out}$ is aligned with ${\bf J}_{\rm in}$. This model generally leads to BHs that are maximally rotating (e.g. \citealt{Lagos09}).
    \item {\tt selfgravitydisk}: assumes the chaotic mode introduced in \citet{King08}, which assumes that ${\bf J}_{\rm in}$ is randomly oriented with respect to ${\bf J}_{\rm out}$, and $r_{\rm in}$ to be equal to the disk self-gravitating radius (described below).
    \item {\tt warpeddisk}: also assumes the chaotic mode introduced in \citet{King08}, but follows the \citet{Volonteri07} model for the warped disk, which assumes $r_{\rm warp}$ is where the timescale for radial diffusion of the warp due to viscosity is equal to the local Lense-Thirring precession timescale (described below).
\end{itemize}

\subsubsection{The \citet{King08} self-gravitating accretion disk}

In this mode, $r_{\rm in}=r_{\rm sg}$, where

\begin{equation}
    r_{\rm sg} = 4790 \, \alpha^{14/27}_{\rm TD} \,\dot{m}^{-8/27}\,\left(\frac{m_{\rm BH}}{10^8\,\rm M_{\odot}}\right)^{-26/27}\, 2\,r_{\rm G}.
    \label{rsg}
\end{equation}

\noindent Here, $\dot{m}$ is the BH accretion rate divided by the Eddington accretion rate (Eq.~\ref{eddmacc}), $r_{\rm G}$ is the gravitational radius of the BH (defined in $\S$~\ref{BHnohairproprs}; note that $2\,r_{\rm G}$ is known as the Schwarzschild radius) and $\alpha_{\rm TD}$ is the dimensionless \citet{Shakura73} thin-disk viscosity parameter. Following \citet{Griffin19} we adopt $\alpha_{\rm TD}=\alpha_{\rm ADAF}=0.1$.

Integrating the mass surface density profile of the accretion disk within $r=[0,r_{\rm sg}]$ gives the self-gravity mass for the thin disk:

\begin{equation}
    m_{\rm sg} = 1.35\, {\rm M_{\odot}} \, \alpha^{-4/5}_{\rm TD} \,\dot{m}^{3/5}\,\left(\frac{m_{\rm BH}}{10^8\,\rm M_{\odot}}\right)^{11/5}\, \left(\frac{r_{\rm sg}}{2 r_{\rm G}}\right)^{7/5}.
    \label{msg}
\end{equation}

\noindent We refer to Section~$2.5$ in \citet{Griffin19} for the full derivation of Eqs.~\ref{rsg}~and~\ref{msg}. 

Eqs.~\ref{rsg}~and~\ref{msg} are used to define $J_{\rm in}$,
\begin{equation}
    J_{\rm in} = \frac{m_{\rm sg}}{\sqrt{2}\,a\,m_{\rm BH}}\, \sqrt{\frac{r_{\rm sg}}{2\,r_{\rm G}}}. \label{jindef}
\end{equation}
\noindent In the first BH accretion episode of a timestep, we randomise the angle between ${\bf J}_{\rm in}$ and ${\bf J}_{\rm BH}$, $\theta_{\rm i}$, between $[0,\pi]$. The  angle between ${\bf J}_{\rm in}$ and ${\bf J}_{\rm BH}$ after the accretion event, $\theta_{\rm f}$, is calculated assuming conservation of ${\bf J}_{\rm tot}$,

\begin{equation}
    {\rm cos}(\theta_{\rm f}) = \frac{J_{\rm in}\, + \,J_{\rm BH}\,{\rm cos}(\theta_{\rm i})}{\sqrt{J^2_{\rm BH}\, + \,J^2_{\rm in} + 2\,J_{\rm BH}\,J_{\rm in}\,{\rm cos}(\theta_{\rm i})}}.
\end{equation}

\noindent The BH spin $a_{\rm f}$ (Eq.~\ref{af_acc}) is calculated using $\Delta m=m_{\rm sg}$ and calculating $\hat{r}_{\rm lso}$ using the initial spin of the BH prior to the accretion of $m_{\rm sg}$. The direction of $a_{\rm f}$ depends on the angular momenta ratio between the self-gravitating disk and the BH, $J_{\rm in}/2\,J_{\rm BH}$:
\begin{eqnarray}
    {\rm cos}(\theta_{\rm i}) &\ge & -J_{\rm in}/2\,J_{\rm BH}\, \, \, \, \, \, \, {\text  {\bf J}_{\rm in}\,\text{and}\,{\bf J}_{\rm BH}\,\text{align},}\\
    &<& -J_{\rm in}/2\,J_{\rm BH}\, \, \, \, \, \, \,  {\text {\bf J}_{\rm in}\,{\text{and}}\,{\bf J}_{\rm BH}\,{\text{anti-align}}.}\nonumber
\end{eqnarray}

\noindent Once a fraction of $m_{\rm sg}$ is accreted by the BH, a new $\theta_{\rm i}$ is randomly chosen; the process continues until the whole accretion disk mass is consumed. 

\subsubsection{The \citet{Volonteri07} warped accretion disk}

In this mode, we calculate $r_{\rm warp}$ and $m_{\rm warp}$ following \citet{Volonteri07}, 

\begin{equation}
        r_{\rm warp} = 3410\,a^{5/8} \, \alpha^{-1/2}_{\rm TD}\, \dot{m}^{-1/4}\,\left(\frac{m_{\rm BH}}{10^8\,\rm M_{\odot}}\right)^{1/8}\, \left(\frac{\nu_2}{\nu_1}\right)^{-5/8}\,2\,r_{\rm G},\label{rwarp}
\end{equation}

\noindent where $\nu_1$ and $\nu_2$ are the horizontal and vertical viscosities of the accretion disk, respectively. We follow \citet{Papaloizou83} and assume $\nu_1/\nu_2 \approx \alpha^2_{\rm TD}$, which is valid in the thin disk approximation. This leads to $m_{\rm warp}$ that is defined like Eq.~\ref{msg} but using $r_{\rm warp}$ instead of $r_{\rm sg}$. 
We refer to Section~$2.6$ in \citet{Griffin19} for the full derivation of Eq.~\ref{rwarp}. 
We define $J_{\rm warp}$ following Eq.~\ref{jindef} but using $m_{\rm warp}$ and $r_{\rm warp}$ instead of $m_{\rm sg}$ and $r_{\rm sg}$, respectively.

In this model, $r_{\rm in}=r_{\rm sg}$, which is generally $>r_{\rm warp}$. Thus, the BH accretes in chunks of $m_{\rm warp}$ until it consumes $m_{\rm sg}$. After every accretion of $m_{\rm warp}$, we assume the new ${\bf J}_{\rm warp}$ is aligned with ${\bf J}_{\rm in}$. 

As with the {\tt selfgravitydisk} model, the initial direction of $J_{\rm in}$ is chosen randomly. Once $m_{\rm sg}$ is consumed, a new inner disk is formed with updated values of $m_{\rm sg}$, $r_{\rm sg}$ and $J_{\rm in}$, and with a new random direction for $J_{\rm in}$.

Note that because usually $m_{\rm warp}<m_{\rm sg}$, this model requires more iterations to calculate the resulting spin than the {\tt selfgravitydisk} model. In principle, if enough iterations are taken, the BH spin would converge. We carried out numerical tests and found that if $\gtrsim 50$ small chunks of either $m_{\rm sg}$ or $m_{\rm warp}$ are accreted, $a_{\rm f}$ converges. Hence to minimise the computing time, we introduce a maximum number of iterations {\tt loop$_{-}$limit$_{-}$accretion} that is by default set to $50$.

\section{Formation of hot halos}\label{sec:hothalo}

\begin{figure}
\begin{center}
\includegraphics[trim=0mm 1mm 3mm 4mm, clip,width=0.49\textwidth]{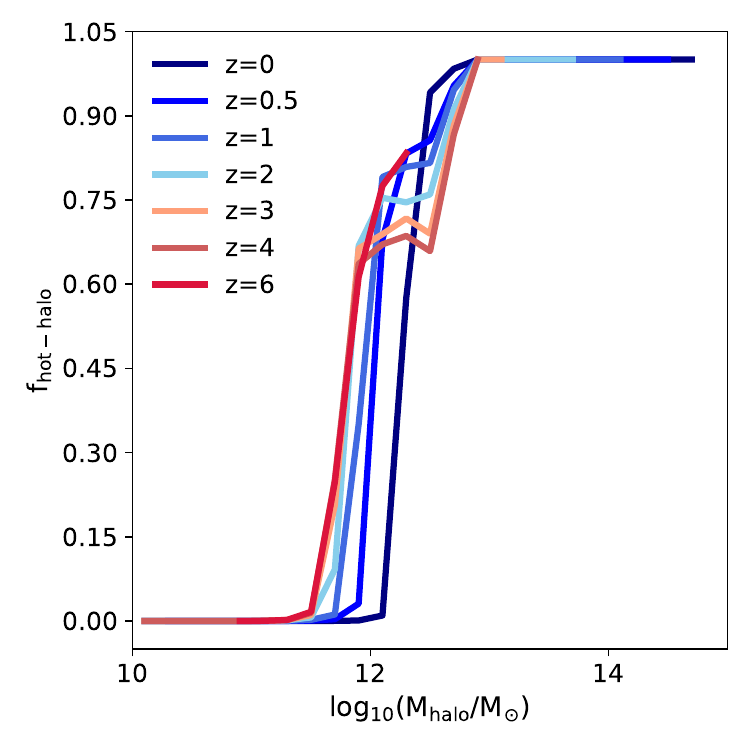}
\caption{Fraction of halos that have developed a hot gaseous halo as a function of halo mass in \shark\ v2.0 and following the model of \citet{Correa18}. This is shown for several redshifts, as labelled.}
\label{hothalo}
\end{center}
\end{figure}

$\S$~\ref{radiomodemodel} described how \shark\ v2.0 assumed the AGN radio-mode feedback to act on halos that have developed a hot gaseous halo, following the description of \citet{Correa18}. Fig.~\ref{hothalo} shows the fraction of halos that have developed a hot gaseous halo as a function of halo mass at different redshifts from $z=6$ to $z=0$. 

The halo mass scale associated to the transition to most halos having developed a hot gaseous halo is usually slightly below $10^{12}\,\rm M_{\odot}$, and moves to slightly higher halo masses with time. The trend here is similar to that reported in \citet{Correa18} for the {\sc EAGLE} simulations (see their Fig.~$13$). 

\section{Computing the size-mass relation from GAMA data}\label{SecSizesGAMA}

Fig.~\ref{Sizesz0} presents observations from 
 \citet{Lange16,Robotham22,Bellstedt23}, all of which use the GAMA survey. However, their results can be very different, especially for the bulge components, so it is worth explaining where these differences come from and how we compute the size-mass relation from the data of \citet{Robotham22,Bellstedt23}. 
 
 The differences between the studies above are many, with \citet{Robotham22,Bellstedt23} using the latest GAMA photometry, which includes VST and VISTA observations \citep{Bellstedt20} and more sophisticated methods to simultaneously model the light profile and SED of galaxies \citep{Robotham22}. The latter results from the combination of VST ugri and VISTA ZYJHKs photometry. \citet{Lange16} used an older version of the photometry from SDSS data (which is shallower) and fitted only the r-band light with a single or two Sersic components without considering mass-to-light ratio variations. 
 
 \citet{Bellstedt23} classified their galaxies as single Sersic fits (split between pure disks, pure spheroids and ambiguous), or double Sersic fits (split between bulge+disk, point source+disk, and disk+disk).
 For the single Sersic fit galaxies, the distinction between the three populations was: a Sersic index $\le 1.5$ corresponds to a pure disk; a Sersic index $\ge 2.5$ corresponds to a pure Spheroid; Sersic indices in the range $1.5-2.5$ are considered ``ambiguous''. Note that disk+disk galaxies are very few and we ignore them here.
 
 The stellar-mass size relation of galaxy disks in the top panel of Fig.~\ref{Sizesz0} is computed in two different ways: by combining pure disks, the disks of bulge+disk and point source+disk galaxies (referred to as ``disks''); and adding ambiguous sources (referred to as ``disks+ambig''). For the middle panel of Fig.~\ref{Sizesz0}, we again compute the size-mass relation in two ways: including pure spheroids and bulges of bulge+disk galaxies (empty symbols) and also adding the ambiguous sources (filled symbols). 
 %We show the two combinations because in this case it does make a big difference to the resulting relation. 
 For the bottom panel of Fig.~\ref{Sizesz0} and for single Sersic galaxies, we simply take the best fit half-stellar mass radius and derived SED-fitting stellar mass. For bulge+disk and point source+disk we measure a stellar mass weighted half-mass radius as: 

 \begin{equation}
     r_{\star} = \frac{\left(r_{\star,\rm bulge}\,M_{\star,\rm bulge} + r_{\star,\rm disk}\,M_{\star,\rm disk}\right)}{M_{\star}},\label{EqRstar}
 \end{equation}
 
 \noindent with $M_{\star} = M_{\star,\rm bulge} + M_{\star,\rm disk}$. Note that in the case of bulge+disk galaxies, $r_{\star,\rm bulge}$ is well defined, but in the case of point source+disk galaxies it is not as the bulge is effectively unresolved. For the latter we tested that the resulting relation remained unchanged if we assumed $r_{\star,\rm bulge} < 1$~kpc.
%%%%%%%%%%%%%%%%%%%%%%%%%%%%%%%%%%%%%%%%%%%%%%%%%%

% Don't change these lines
\bsp	% typesetting comment
\label{lastpage}
\end{document}